\documentclass[12pt]{article}
\title{Universal BPS structure of stationary supergravity solutions}

\global\arraycolsep=1pt
\usepackage[colorlinks=true,backref=true,linkcolor=black,anchorcolor=black,citecolor=black,filecolor=black,menucolor=black,pagecolor=black,urlcolor=black]{hyperref}

\usepackage{amsmath}
\usepackage{mathrsfs}
\usepackage[Symbol]{upgreek}
\usepackage{bbm}
\usepackage{dsfont}
\usepackage{amssymb}
\usepackage[vcentermath]{youngtab}
\usepackage{caption}
\usepackage{graphicx}
\usepackage{wasysym}

\newcommand{\ba}{/ \hspace{-1.2ex}} 
\newcommand{\baa}{/ \hspace{-1.4ex}}
\newcommand{\baaa}{\, / \hspace{-1.6ex}}

\def\baQ{ \hspace{.3mm} /  \hspace{-1.7mm}   Q  \hspace{0.2mm}}
\def\batQ{\hspace{0.5mm} /  \hspace{-1.9mm}   Q  \hspace{.4mm}}
\def\baaQ{ /  \hspace{-2.0mm}   Q  \hspace{1.0mm}}
\def\baP{ \hspace{.3mm} /  \hspace{-1.7mm}   P  \hspace{0.2mm}}
\def\batP{\hspace{0.5mm} /  \hspace{-1.9mm}   P  \hspace{.4mm}}
\def\baaP{ /  \hspace{-2.0mm}   P  \hspace{1.0mm}}
\def\bap{ /  \hspace{-1.5mm}   p  \hspace{0.5mm}}

\def\baaaP{ \baaa \hspace{.2mm}P}
\def\baaU{ \hspace{.3mm} \baa U}
\def\baupsilon{/  \hspace{-1.5mm}   \upsilon  \hspace{0.5mm}}
\newcommand{\Scal}[1]{\Bigl ({#1} \Bigr )}
\newcommand{\scal}[1]{\bigl ({#1} \bigr )}
\def\bea{\begin{eqnarray}}
\def\eea{\end{eqnarray}}
\def\be{\begin{equation}}
\def\ee{\end{equation}}
\def\ie{{\it i.e.}\ }
\def\eg{{\it e.g.}\ }
\def\viz{{\it viz.}\ }
\newcommand{\CR}{\nonumber \\*}

\newcommand{\trace}{\hbox {Tr}~}

\DeclareMathAlphabet{\mathpzc}{OT1}{pzc}{m}{it}

\DeclareMathOperator{\ad}{ad}

\newcommand{\gra}[2]{{\scriptscriptstyle (#1 , #2 )}}
\newcommand{\ord}[1]{{\scriptscriptstyle (#1)}}

\def\L{{\cal L}}

\newcommand{\sfrac}[2]{{\scriptstyle \frac{#1}{#2}}}

\def\e{{\boldsymbol{e}}}
\def\f{{\boldsymbol{f}}}
\def\h{{\boldsymbol{h}}}
\def\x{{\boldsymbol{x}}}
\def\y{{\boldsymbol{y}}}

\def\z{{\boldsymbol{z}}}
\def\upsi{{{\boldsymbol \upsilon}}}

\def\nn{\nonumber}

\def\N{\mathcal{N}}
\def\Nquatre{{\mathcal{N}=4}}
\def\ft#1#2{\tfrac{#1}{#2}}

\def\C{{\mathscr{C}}}
\def\V{{\mathcal{V}}}
\def\w{{\scriptstyle W}}
\def\Z{\mathds{Z}}

\def\G{{\mathfrak{G}}}
\def\H{{\mathfrak{H}}}
\def\P{{\mathfrak{P}}}
\def\I{{\mathfrak{I}}}
\def\J{{\mathfrak{J}}}
\def\j{{\mathfrak{j}}}

\def\m{{\mathpzc{m}}}
\def\n{{\mathpzc{n}}}

\def\un{{\mathpzc{1}}}
\def\deux{{\mathpzc{2}}}
\def\trois{{\mathpzc{3}}}

\def\bras{\hspace{0.5mm} \left<^{\hspace{-2.8mm}*\hspace{1.2mm} }}
\def\invo{{\APLstar}}
\def\convo{{\mbox{\tiny$\invo$}\hspace{-0.5mm}}}
\def\Ic{{I\hspace{-0.6mm}c}}
\def\bulletup{ {\scriptstyle  \bullet} }
\def\yngd{ \Yboxdim8pt {\young(\bulletup)}} 
\def\DJo{$\;$\kern-.4em \hbox{D\kern-.8em\raise.15ex\hbox{--}\kern.35em okovi\'c}}

\newcommand{\eprint}[1]{{\href{http://arxiv.org/abs/#1}{\texttt{[#1}]}}}
\newcommand{\eprintN}[1]{{\href{http://arxiv.org/abs/#1}{\texttt{#1 [hep-th]}}}}

\begin{document}
\allowdisplaybreaks[1]
\renewcommand{\thefootnote}{\fnsymbol{footnote}}
\numberwithin{equation}{section}
\def\corr{$\spadesuit \, $}
\def\trefle{$ \, $}
\def\kscorr{$\diamondsuit \, $}
\begin{titlepage}
\begin{flushright}
\
\vskip -2.5cm
{\small AEI-2009-024}\\
{\small Imperial/TP/09/KSS/02}\\
\vskip 1cm
\end{flushright}
\begin{center}
{\Large \bf
Universal BPS structure of \\[2mm] stationary supergravity solutions}
\\
\lineskip .75em
\vskip 3em
\normalsize
{\large  Guillaume Bossard\footnote{email address: bossard@aei.mpg.de},
Hermann Nicolai\footnote{email address: Hermann.Nicolai@aei.mpg.de} and
K.S. Stelle\footnote{email address: k.stelle@imperial.ac.uk}}\\
\vskip 1 em
$^{\ast\dagger\ddagger}${\it AEI, Max-Planck-Institut f\"{u}r Gravitationsphysik\\
Am M\"{u}hlenberg 1, D-14476 Potsdam, Germany}
\\
\vskip 1 em
$^{\ddagger}${\it Theoretical Physics Group, Imperial College London\\
Prince Consort Road, London SW7 2AZ, UK}

\vskip 1 em
\end{center}
\begin{abstract}
{\footnotesize
We study asymptotically flat stationary solutions of four-dimensional 
supergravity theories via the associated $\G/ \H^*$ pseudo-Riemannian 
non-linear sigma models in three spatial dimensions. The Noether charge 
$\C$ associated to $\G$ is shown to satisfy a characteristic equation 
that determines it as a function of the four-dimensional conserved charges. 
The matrix $\C$ is nilpotent for non-rotating extremal solutions. The nilpotency degree of 
$\C$ is directly related to the BPS degree of the corresponding solution when they are BPS. 
Equivalently, the charges can be described in terms of a Weyl spinor 
$|\C\rangle$ of $Spin^*(2\N)$, and then the characteristic equation 
becomes equivalent to a generalisation of the Cartan pure spinor constraint 
on $|\C\rangle$. The invariance of a given solution with respect to 
supersymmetry is determined by an algebraic `Dirac equation' on the 
Weyl spinor $|\C\rangle$. We explicitly solve this equation for all 
pure supergravity theories and we characterise the stratified structure 
of the moduli space of asymptotically Taub--NUT black holes with respect 
to their BPS degree. The analysis is valid for any asymptotically 
flat stationary solutions for which the singularities are protected 
by horizons. The $\H^*$-orbits of extremal solutions are identified as Lagrangian submanifolds of nilpotent orbits of $\G$, and so the moduli space of extremal spherically symmetric black holes is identified as a Lagrangian subvariety of the variety of nilpotent elements of $\mathfrak{g}$. We also generalise the notion of active duality transformations 
to an `almost action' of the three-dimensional duality group $\G$ 
on asymptotically flat stationary solutions.
}
\end{abstract}

\end{titlepage}
\renewcommand{\thefootnote}{\arabic{footnote}}
\setcounter{footnote}{0}

\tableofcontents

\section{Introduction}

Black hole solutions of supergravity theories have been extensively 
studied in the literature \cite{Maison,Maison1}. This applies in 
particular to BPS solutions, that is, to supersymmetric solutions admitting 
Killing spinors (see \eg \cite{KillingSpinor} and references therein).
In this publication, we focus on theories for which the scalar fields 
lie in a symmetric space, which can be represented as the quotient
of the global hidden symmetry by a maximal subgroup \cite{Maison,CJ,JuliaL}.
Stationary solutions can then be identified as solutions of a  
three-dimensional non-linear sigma model over a symmetric space 
$\G / \H^*$ coupled to Euclidean gravity, such that the maximal subgroup
$\H^*$ is {\em non-compact}. The group $\G$ extends the global `hidden 
symmetry' group $\G_4$ of four-dimensional supergravity; while 
the latter acts only on matter degrees of freedom, the larger group
$\G\supset\G_4$ also incorporates gravity and the Ehlers $SL(2,\mathds{R})$ 
symmetry \cite{Ehlers}.

For a large class of theories, the duality group $\G$ is simple, and 
then all the non-extremal black hole solutions are in $\H^*$-orbits of solutions 
of pure Einstein theory; indeed, all of these solutions can be obtained 
through group theoretical methods \cite{Maison,Maison1}. The Noether 
current associated to the duality symmetry gives rise to a charge $\C$ 
lying in the Lie algebra $\mathfrak{g}\equiv$ Lie$(\G)$. We display the 
physical content of this charge in terms of four-dimensional 
quantities.  Imposing regularity conditions on the solution requires that $\C$ satisfy 
a characteristic equation which determines $\C$ non-linearly as a function 
of the conserved charges of the four-dimensional theory, \ie the Komar mass, the NUT charge and the electromagnetic charges. Consequently, they transform all together in a non-linear representation of the group $\H^*$. 

Our results are based on an extension of the general classification of 
three-dimensional supergravity theories \cite{3Dclass} to theories with 
Euclidean signature which characterise the stationary solutions of 
supergravity theories in four dimensions. For $\N$-extended supergravity, 
the group $\H^*$ is the product of the group $Spin^*(2\N)$ (for $\N>1$, 
a non-compact real form of the group $Spin(2\N)$ appearing for Lorentzian 
supergravities \cite{3Dclass}) with a symmetry group determined by the matter 
content of the theory. The charge matrix $\C$ can then be associated to a 
charge state vector $|\C\rangle$ which transforms as a $Spin^*(2\N)$ 
chiral spinor. For asymptotically flat solutions (flat in the sense of Misner \cite{Misner}, \ie including the 
asymptotically Taub--NUT ones) the BPS condition is equivalent to an 
algebraic `Dirac equation' (see (\ref{susy}) for a precise formulation)
\be\label{Dirac}
\quad  \ba \epsilon \, |\C\rangle = 0  \quad 
\ee
where the `momentum' $\epsilon$ is the asymptotic supersymmetry parameter 
(Killing spinor) transforming in the (pseudo-real) vector representation 
of $SO^*(2\N)$. This equation follows from the dilatino variation, which 
in three dimensions carries all the essential information about residual 
supersymmetries. Equally important, (\ref{Dirac}) determines the charge 
$\C$  for BPS solutions as a function of the conserved charges of the four-dimensional theory 
in terms of a simple rational function, whereas the generic non-BPS solution of the 
characteristic equation is generally a non-rational function. For $\N\leq 5$ 
pure supergravities, the characteristic equation for the charge matrix $\C$ 
simply reduces to the $Spin^*(2\N)$ pure spinor equation for the charge 
spinor $|\C\rangle$, and can be solved in full generality. 

In order to identify the general conditions on charges for various
BPS degrees, we solve equation (\ref{Dirac}) systematically for all 
pure supergravity theories (the $\N=4$ theory coupled to $n$ vector multiplets 
is also considered in the last section). In this way we are able to provide a systematic
classification of BPS solutions for all supergravities. Our analysis 
encompasses previous work on BPS solutions, such as, for instance, the
BPS asymptotically Minkowskian stationary black holes solutions in $\N=2$
supergravity  \cite{Tod,BPSsolutions}, whose classification became possible
through the discovery of the so-called attractor mechanism \cite{attractors}
(see \cite{lectures} for an introduction to the subject, and \cite{R2} for an extension 
of these results including $R^2$ corrections). In addition (and amongst other results) we are able to prove
the conjecture of \cite{E7entropy} on the vanishing of the horizon area 
for $\frac14$ and $\frac12$ BPS solutions to $\N=8$ supergravity. We then 
conjecture an expression for the horizon area of asymptotically Taub--NUT 
BPS black holes, which turns out not to be $E_{7(7)}$ invariant in general. 
Moreover, neither the horizon area, nor the surface gravity are invariant 
under the action of $Spin^*(16)$, but the product of these two 
is proportional to the square root of the $E_{8(8)}$ invariant $\trace \C^2$. \footnote{For rotating solutions this also involves the angular momentum per unit of mass, which is also left invariant by the action of $Spin^*(16)$.}  

The moduli space of stationary single-particle solutions admits a stratified 
structure whose filtration corresponds to the BPS degree in pure supergravity
theories with $\N\le5$. The strata of BPS degrees $(n/{\N})$ then 
can be given as coset spaces $\H^*/\J_n$ which we identify explicitly 
in terms of the isotropy subgroups $\J_n\subset \H^*$ of the given 
charges. We also describe the moduli space of stationary single-particle solutions of $\N=6$ and $\N=8$ supergravities. In these cases, the stratification is slightly more involved. We show that the BPS degree is characterised in a $\G$-invariant way by the nilpotency degree of the charge matrix $\C$ in both the 
fundamental and the adjoint representation of $\mathfrak{g}$. Another main new result of this work is the
demonstration that these BPS strata, initially obtained as orbits of
the asymptotic-structure-preserving group $\H^*$, are diffeomorphic to {\em Lagrangian 
submanifolds of nilpotent orbits under the adjoint action of $\G$.}

We also generalise the notion of active duality transformations \cite{Active} 
to the case of three-dimensional Euclidean theories. Unlike in four dimensions,
this procedure fails to define a Lie group action because of the singular 
behaviour of the action on the BPS solutions. This failure is directly 
related to the failure of the Iwasawa decomposition when the maximal subgroup
$\H^*\subset \G$ is non-compact: the elements of $\G$ mapping non-BPS to
BPS solutions are precisely the ones for which the Iwasawa decomposition 
breaks down. We will explain in some detail how the BPS strata are related 
to the `Iwasawa failure sets' in $\G$.

A chief motivation for the present work was provided by the general 
conjecture of \cite{HT} (see also \cite{OP}) according to which the global 
hidden symmetries $\G$ of supergravity become replaced by certain arithmetic 
subgroups $\G(\Z)\subset\G$ in the quantum theory,\footnote{The main 
  example here is maximal $\N=8$ supergravity \cite{CJ} whose global 
  symmetry $\G_4 = E_{7(7)}$ is broken to 
  $E_7(\Z) = E_{7(7)}\cap Sp(56,\Z)$ upon quantisation,
  where the symplectic group $Sp(56,\Z)$ encodes the 
  Dirac--Schwinger--Zwanziger quantisation condition for the 
  electromagnetic charges.} and to explore whether and in what sense this 
claim can remain valid as one descends to three dimensions. This case  
cannot be simply extrapolated from higher dimensional examples, because 
it differs from those in two crucial respects: (1) unlike in dimensions 
$d\geq 4$, the central charges of the superalgebra no longer combine into 
representations of the global hidden symmetry group $\G$ \cite{dWN}, and 
(2) the quantisation condition would now also apply to the gravitational 
charges (mass and NUT charge) \cite{Henneaux}. For maximal supergravity, our analysis 
leads us to the conclusion that the quantum moduli space of maximal 
supergravity solitons is {\em not} described by a lattice in the adjoint 
representation of an arithmetic subgroup of $E_{8(8)}$. 

We will argue that the singular behaviour of the duality transformations
on the subset of BPS solutions within the space of all stationary solutions
might be resolved at the quantum level. This conjecture is based on our
description of the $\H^*$-orbits as Lagrangian submanifolds of the $\G$-adjoint
orbits of the corresponding charge matrix: if there is no representation of the 
duality group $\G$ on the moduli space of asymptotically flat stationary solutions, 
there might nevertheless exist a unitary representation of $\G$ 
on the space of functions defined on this moduli space. 
The action of $\G$ on the adjoint orbits induces a unitary representation on the spaces of functions supported on these Lagrangian submanifolds, that is on the moduli space of solutions. We speculate on an interpretation of the formula for the Eisenstein series obtained from the minimal unitary representation of $\G$ 
\cite{minimalEisenstein,GKN} in terms of observables of the quantum mechanics 
of a particle living in the moduli space of $\ft12$ BPS black holes. 
kscorr Whereas such a construction of the Eisenstein series seems meaningful 
in the study of the moduli space of particle solutions, the na\"{i}ve 
generalisation of higher dimensional formulas for Eisenstein 
series as sums over a lattice representation of  $\G(\Z)$ 
should not be interpreted as a sum over the quantised charges. 

An interesting problem for future study will be the extension of the present results to solutions 
with a lightlike Killing vector \cite{NullReduction}, which are plane-wave when they are BPS (see \eg 
\cite{NullBPS} and references therein). Here we only remark that, in four spacetime dimensions, 
$pp$-type plane wave solutions cannot be asymptotically flat because the solutions of the transverse Laplace equation decay only logarithmically.


\section{Duality groups of stationary solutions}
In Einstein theory coupled to matter, one generally knows exact solutions 
only when the corresponding metric admits a certain number of commuting 
Killing vectors, and when these isometries  furthermore leave invariant 
the various gauge fields and matter fields of the theory. The existence 
of $k$ such Killing vectors permits elimination of the dependance of the 
solution on $k$ corresponding coordinates, in such a way that the solution 
can be interpreted in $(d\!-\!k)$ dimensions. Moreover, specific dimensionally 
reduced theories admit enlarged sets of symmetries which are non-linearly 
realised on the solutions. When all of the reduction Killing vectors are spacelike, 
the fields of a dimensionally reduced theory are defined on a 
$(d\!-\!k)$-dimensional spacetime and the Hamiltonian of the theory is 
positively defined. By contrast, when one of the Killing vectors is timelike, 
as a general property of timelike dimensional reductions, the action of 
the dimensionally reduced theory is indefinite \cite{Maison,Maison1}.\footnote{   Timelike dimensional reduction has also been used to generate solutions in \cite{diago}.} This is not a problem
since we are here concerned only with the classical equations of motion in an Euclidean-signature
reduced theory for stationary solutions.

We will consider Einstein theory coupled to abelian vector fields and 
scalar fields living in a symmetric space. We assume that the isometry 
group of the scalar coset space is a semi-simple Lie group $\G_4$ which defines 
a symmetry of the equations of motion, and moreover that each simple or 
abelian group arising in the decomposition of $\G_4$ acts non-trivially 
on the vector fields. The scalar coset space $\G_4 / \H_4$ is a Riemannian manifold 
defined by the quotient of $\G_4$ by its maximal compact subgroup $\H_4$. 
The various Lie groups $\G_4$ satisfying these criteria are listed in 
\cite{Maison}. We denote by $\mathfrak{l}_4$ the representation carried by the Maxwell 
degrees of freedom under $\G_4$. If we consider only stationary 
solutions, we can consider them as solutions of a dimensionally reduced 
Euclidean three-dimensional theory. This dimensional reduction yields one 
scalar from the metric, one scalar for each Maxwell field, together with all the original
scalars of the four-dimensional theory together with one vector field coming 
from the metric and one vector field from each Maxwell field. For a timelike
Killing vector, the kinetic terms of tensor fields whose rank has been reduced 
by an odd number come with a negative sign, while the remaining fields' kinetic terms are positive  
(for a spacelike Killing vector, they would all appear with a positive
sign). This holds, in particular, for the scalars arising from 
Maxwell fields ($1 \rightarrow 0$) and for the vector field arising from 
the metric  ($2 \rightarrow 1$). After dualisation of the vector fields 
to scalars, the vector field arising from the metric turns into a scalar 
field with a {\em positive} kinetic term. The Maxwell vectors become 
scalars after dualisation, with negative kinetic terms similarly to the Maxwell 
scalars. We will call the latter `electric' fields, and we will call the scalars 
arising from vectors upon dualisation `magnetic' fields. 

The stationary solutions of pure gravity admit the so-called Ehlers group 
\cite{Ehlers} as a symmetry, yielding a formulation 
of the theory as an $SL(2,\mathds{R})/ SO(2)$ non-linear sigma model 
coupled to three-dimensional gravity. This property generalises 
to Einstein theory with matter in such a way that we get an 
$\mathfrak{sl}(2,\mathds{R})\oplus \mathfrak{g}_4$ set of symmetry 
generators. The scalars arising from the Maxwell fields admit a 
shift symmetry, since the Maxwell fields can be shifted by constants
(global gauge transformations) already in four dimensions, so 
the $\G_4$ symmetry is enhanced to the non-semi-simple group 
$\G_4 \ltimes \mathfrak{l}_4$. After dimensional reduction to three 
dimensions, the magnetic scalars obtained by dualisation from the vectors
also admit shift symmetries, since each dualisation leaves an undetermined
integration constant. Altogether, these shift symmetries still transform 
in the $\mathfrak{l}_4$ of $\G_4$. Moreover, the commutators of the Ehlers 
generators with these shift symmetries generate new generators which also 
belong to the $\mathfrak{l}_4$ representation of $\G_4$, and which are 
themselves nonlinearly realised on the fields. The whole duality group then becomes a simple Lie group \cite{Maison}, for which the algebra admits 
a five-graded decomposition with respect to the diagonal generator of 
the Ehlers $SL(2,\mathds{R})$
\be 
\mathfrak{g} \cong \mathfrak{sl}(2,\mathds{R}) \oplus \mathfrak{g}_4 
\oplus ({\bf 2} \otimes \mathfrak{l}_4) \cong  {\bf 1}^{\ord{-2}} 
\oplus \mathfrak{l}_4^{\ord{-1}} \oplus ( {\bf 1} \oplus 
\mathfrak{g}_4 )^\ord{0}  \oplus \mathfrak{l}_4^\ord{1} 
\oplus {\bf 1}^\ord{2} \label{five}\ee
The maximal compact subgroup $\H$ of this group is generated by the $\mathfrak{so}(2) 
\oplus \mathfrak{h}_4$ subalgebra of $\mathfrak{sl}(2,\mathds{R}) 
\oplus \mathfrak{g}_4$, together with the compact combination of the two 
$\mathfrak{l}_4$.  Because the scalar fields arising from the Maxwell 
fields have negative kinetic terms, however, the maximal subgroup $\H^*$ 
for a timelike dimensional reduction is a {\em non-compact real form} 
of this maximal compact subgroup \cite{Maison}, in contrast to spacelike reductions, for which $\H$ is fully compact. 

The resulting three-dimensional theory is described in terms of a coset representative 
$\V\in\G/\H^*$ which contains all the propagating (scalar) degrees of freedom
of the theory, plus the three-dimensional metric\footnote{We 
 denote curved spacetime indices by Greek letters $\mu,\nu,\dots$ in both 
 four and three dimensions.} $\gamma_{\mu\nu}$ which, however, carries no physical degrees of 
freedom in three dimensions. The Maurer--Cartan form $\V^{-1} d \V$ 
decomposes as
\be
\V^{-1} d\V = Q + P \quad, \qquad Q\equiv Q_\mu dx^\mu\in \mathfrak{h}^* 
  \;\; , \;\;\; 
 P\equiv P_\mu dx^\mu\in \mathfrak{g}\ominus\mathfrak{h}^* 
\ee
The Bianchi identity then gives 
\be 
d Q + Q^2 = - P^2\ ,  \hspace{10mm} 
d_Q P \equiv d  P + \{ Q , P \} = 0 \ .\label{Bianchi}
\ee
The equations of motion of the scalar fields are 
\be 
d_Q  \star P \equiv d \star P + \{ Q , \, \star P \} = 0 \label{EMP} 
\ee
and the Einstein equations are 
\be 
R_{\mu\nu}  = \trace P_\mu P_\nu \label{EinsteinE}
\ee

We will consider in this paper solutions which are asymptotically flat in the sense 
of Misner \cite{Misner} (which are not generally to be confused with asymptotically Minkowskian solutions). 
Strictly speaking,  by this we mean that four-dimensional spacetime admits 
a function $r$ that tends to infinity at spatial infinity and which defines a 
proper distance in this limit, $g^{\mu\nu}Ê\partial_\mu r \partial_\nu r 
\rightarrow 1$, and such that all the components of the Riemann tensor in any 
vierbein frame tend to zero as $\mathcal{O}(r^{-3})$ as $r \rightarrow + 
\infty$. In the same way, the Maxwell field strengths are required to 
tend to zero as $\mathcal{O}(r^{-2})$ in this limit in any vierbein frame.\footnote{ More generally, one could
consider solutions tending asymptotically to an arbitrary $\G_4 / \H_4$ constant matrix, but this can be standardised to the unit matrix by making a $\G_4$ transformation.}
The four-dimensional coset elements $\in \G_4 / \H_4$ are required to 
tend asymptotically to the unit matrix as $\mathds{1} + \mathcal{O}(r^{-1})$. In order for charges 
to be well defined, we also require that the timelike Killing vector 
$\kappa \equiv \kappa^\mu \partial_\mu$ leaves invariant the function $r$, 
that it satisfies asymptotic hypersurface orthogonality 
$\varepsilon^{\mu\nu\sigma\rho} \kappa_\nu \partial_\sigma \kappa_\rho 
= \mathcal{ O}(r^{-2})$, and that its squared norm $-H$ tends to negative unity 
as $H = 1 + \mathcal{O}(r^{-1})$ in the asymptotic region. We assume 
that the action of the timelike isometry on the domain $M_+$ of the 
four-dimensional manifold $M$ on which $H$ is positively defined  
(\ie outside possible horizons and ergospheres) is free and proper. 
$M_+$ then admits an abelian principal bundle structure whose fibres are the 
timelike isometry orbits and whose base is a three-dimensional Riemannian 
manifold $V$ which is asymptotically Euclidean. For such specific 
solutions, all the fields of the four-dimensional theory can be defined 
from pull-backs of the scalar fields living in the symmetric space $\G/\H^*$ 
defined throughout $V$, and the asymptotic condition in the four-dimensional 
theory is equivalent to the requirement that the coset representative 
$\V$ goes to the unit matrix as $\V = \mathds{1} +  \mathcal{O}(r^{-1})$ in the 
asymptotic region.

\subsection{Conserved charges}
We define the Komar two-form $K \equiv  \partial_\mu  \kappa_\nu dx^\mu \wedge
dx^\nu$ \cite{Komar}, which is invariant under the action of the timelike 
isometry and which is asymptotically horizontal. The latter condition is equivalent 
to the requirement that the scalar field $B$, dual to the Kaluza--Klein 
vector arising from the metric by dimensional reduction, vanishes as 
$\mathcal{O}(r^{-1})$ as $r \rightarrow + \infty$.  We define a set of 
local sections of the principal bundle on each open set of an atlas of the three-dimensional manifold $V$, 
which we denote collectively $s$. Then we can define the Komar mass 
and the Komar NUT charge as follows \cite{nous}
\be 
m \equiv \frac{1}{8 \pi }Ê\int_{\partial V}  s^* \star K  \hspace{10mm}  n 
\equiv \frac{1}{8 \pi }Ê\int_{\partial V} s^* K \label{def}Ê
\ee
The Maxwell fields also define conserved charges. The Maxwell equation 
$d \star \mathcal{F}  = 0$, where $\mathcal{F }\equiv\delta{\cal{L}}/\delta F$ 
is a linear combination of the two-form field strengths $F$ depending on the four-dimensional 
scalar fields, permits one to define electric charges, and the Bianchi 
identities $d F = 0$ permits one to define magnetic charges, as follows:
\be 
q \equiv  \frac{1}{2 \pi }Ê\int_{\partial V}  s^* \star \mathcal{F}  
\hspace{10mm} p \equiv  \frac{1}{2 \pi }Ê\int_{\partial V}  s^* F \ .
\ee
These transform together in the representation $\mathfrak{l}_4$ of $\G_4$. 
Finally, the rigid $\G_4$ invariance of the four-dimensional theory gives rise to 
an associated conserved current such that the associated three-form $J_3$ 
transforms in the adjoint representation of $\G_4$, and satisfies $d J_3 = 0$ if 
the scalar field equations are obeyed. However, $J_3$ cannot generally 
be written as a local function of the fields and their derivatives 
in four dimensions. 

We now wish to analyse these conserved charges from the point of 
view of the three-dimensional theory defined on $V$, and to clarify their
transformation properties under the action of the three-dimensional
duality group $\G$. In consequence of the invariance of the three-dimensional 
action under this group, there exists an associated Noether current in three dimensions. 
Indeed, the equations of motion (\ref{EMP}) can be rewritten as 
\be  d \star \V P  \V^{-1} = 0 \ . \ee
Therefore, the $\mathfrak{g}$-valued Noether current is $\star \V P \V^{-1}$. 
Since the three-dimensional theory is Euclidean, we cannot properly talk 
about a conserved charge. Nevertheless, since $\star \V P \V^{-1}$ is $d$-closed, 
the integral of this $2$-form on a given homology cycle does not depend on 
the representative of the cycle. As a result, for stationary solutions, 
the integral of this three-dimensional current over any spacelike closed surface, containing 
in its interior all the singularities and topologically non-trivial subspaces of the 
solution, defines a $\mathfrak{g}$-valued charge matrix $\C$:
\be\label{Charge} 
\C \equiv \frac{1}{4 \pi} \int_{\partial V} \star \V P \V^{-1} \ .
\ee
This transforms in the adjoint representation of $\G$ by the standard non-linear action and it 
can easily be computed by looking at the asymptotic 
value of the current if (as we assume to be the case)
\be\label{Pasympt}
P = \C \, \frac{ dr }{r^2} + \mathcal{O}(r^{-2}) \;\; .
\ee 
For asymptotically flat solutions, $\V$ goes to the identity matrix 
asymptotically and the charge matrix $\C$ in that case is given by the asymptotic value 
of the one-form $P$. $\C$ then lies in $\mathfrak{g} \ominus \mathfrak{h}^*$
and can thus be decomposed into three irreducible representations with 
respect to $\mathfrak{so}(2) \oplus \mathfrak{h}_4$ according to 
\be\label{g-h*}
\mathfrak{g} \ominus \mathfrak{h}^* 
\cong \scal{  \mathfrak{sl}(2,\mathds{R})\ominus \mathfrak{so}(2) }  
\oplus \mathfrak{l}_4 \oplus \scal{Ê\mathfrak{g}_4 \ominus \mathfrak{h}_4 }
\ee 
We stress once again that the metric induced by the Cartan-Killing metric 
of $\mathfrak{g}$ on the coset (\ref{g-h*}) is positive definite on the 
first and last summand, and negative definite on $\mathfrak{l}_4$.

The decomposition (\ref{g-h*}) is in precise accord with the structure 
of the conserved charges in four dimensions as described above. Namely, the
computation of $\C$ permits one to identify its $\mathfrak{sl}(2,\mathds{R}) 
\ominus \mathfrak{so}(2)$ component as the Komar mass and the Komar NUT 
charge, and its $\mathfrak{l}_4$ components with the electromagnetic charges. 
The remaining $\mathfrak{g}_4 \ominus \mathfrak{h}_4$ charges come from the
$\G_4$ Noether current of the original four-dimensional theory, which transforms in 
the adjoint of $\G_4$.  For a stationary 
solution, $\L_\kappa J_3 = 0 $ and $i_\kappa J_3$ then defines a conserved 
two-form which is furthermore manifestly invariant and horizontal with 
respect to the timelike isometry. Although $J_3$ in general is not a local function 
of the fields and their four-dimensional derivatives, $i_\kappa J_3$ can be 
written in terms of the pull-backs of the scalar fields of the 
three-dimensional model for stationary field configurations. One thus obtains that the integral 
of the pull-back 
\be 
\frac{1}{4 \pi} \int_\Sigma  s^* i_\kappa J_3 
\ee
on any homology two-cycle $\Sigma$ of $V$, does not depend on the 
representative of that cycle. An important fact is that {\em the scalar 
charges, that is, the $\mathfrak{g}_4 \ominus \mathfrak{h}_4$ components
of $\C$, will not constitute independent integration parameters.} This was demonstrated in 
full generality in \cite{Maison}.  We will see that it is natural to impose characteristic equations on the charges, with the consequence that the scalar
charges become functions of the gravitational and electromagnetic 
charges in the case of pure supergravity theories. We note also
that the contribution of the angular momentum in 
(\ref{Pasympt}) is subleading (that is, it belongs to the 
$\mathcal{O}(r^{-2})$ part of (\ref{Pasympt})); hence the conserved 
charge $\C$ will be insensitive to the angular momentum parameter $a$.

Defining the usual generators of $\mathfrak{sl}(2,\mathds{R})$, $\h, \e$ 
and $\f$ by
\be 
[Ê\h , \e ] = 2 \e \hspace{10mm} [\h,\f] = - 2 \f \hspace{10mm} [\e ,\f ] 
= \h \ee
we can summarise what has been said above in the equation 
\be  
\star \V P \V^{-1} =  4 s^* \star K \h  - 4 s^* K ( \e + \f )  + 
s^* \star \mathcal{F}  +  s^* F + s^* i_\kappa J_3 + \mathcal{O}(r^{-2}) 
\ee
where the electromagnetic current $s^* \star \mathcal{F}  +  s^* F$, which 
transforms under $\G_4$ in the representation $\mathfrak{l}_4$, is understood 
to be valued in the corresponding generators of $\G$ with the appropriate
normalisation. For example, in $\N=8 $ supergravity the $28$ Maxwell fields 
$F^{ij}$  transform under  $SO(8) \subset SU(8)$ as antisymmetric tensors. 
The compact generators of $\mathfrak{e}_{8(8)} \ominus \mathfrak{so}^*(16)$ transform in the $\boldsymbol{56}$ of $E_{7(7)}$. 
They are  conveniently represented by a complex antisymmetric tensor 
$\boldsymbol{Z}^{ij}$ of $SU(8)$ and its Hermitean conjugate 
$\boldsymbol{Z}_{ij}$. Then 
\be 
\star \mathcal{F} =   \star \mathcal{F}_{ij} \boldsymbol{Z}^{ij} -  
\star \mathcal{F}^{ij} \boldsymbol{Z}_{ij} \hspace{10mm} 
F = i F_{ij}  \boldsymbol{Z}^{ij} + i F^{ij}  \boldsymbol{Z}_{ij} \ .
\ee 
Note that only the sum $ \star \mathcal{F}  + F$ transforms covariantly 
under the action of $E_{7(7)}$. 

The charge matrix $\C$ is associated to ``instantons'' of the three-dimensional 
Euclidean theory. The single-point instantons correspond to single-particle 
like solutions of the four-dimensional theory. Naively, one would thus expect 
these solutions to appear in multiplets transforming in the linear 
representation of $\H^*$ defined by $\mathfrak{g} \ominus \mathfrak{h}^*$.  
However, matters are not so simple, because the charge matrix $\C$ is restricted to satisfy $\H^*$ invariant constraints in general, so that the number of independent parameters describing these solutions is much smaller --
as was to be anticipated in view of the dependence of the charges
associated to the four-dimensional scalar fields on the gravitational 
and electromagnetic charges. This is because the charges parametrising 
the solutions are the conserved charges {\em in four
dimensions}, that is the mass, the NUT charge and the electromagnetic 
charges. This, in turn, is due to the fact that the particle-like solutions 
are supported by vector fields through Gauss's law. A useful analogy
here is that of a free particle in Minkowski space. When the momentum 
of this particle is timelike, it can be rotated to the rest frame. Here,
the role of momentum is played by the charge matrix, while the non-compact
group $\H^*$ plays the role of the Lorentz group. The electromagnetic 
charges belong to the $\mathfrak{l}_4$ representation of $\G_4$, just like 
the non-compact generators of $\mathfrak{h}^*$. The action of these 
generators on the Maxwell charges is linear in the scalar and the gravity 
charges, in such a way that for a non-zero value of $m^2 + n^2$ one can 
always find a generator that acts on the Maxwell charge as a shift 
parallel to it. This generator of $\mathfrak{h}^*$ defines an $SO(1,1)$ 
subgroup of $\H^*$ which mixes the electromagnetic charges with the others. 
For any charge matrix satisfying $\trace \C^2 > 0$ the action of this 
abelian subgroup of $\H^*$ permits one to cancel the electromagnetic charges. 
It then follows from the five-graded decomposition of $\mathfrak{g}$ that 
one can find an element of the compact subgroup of the Ehlers 
group that cancels the NUT charge without modifying the electromagnetic and 
the scalar charges. It has been proven in \cite{Maison} that a static 
solution without electromagnetic charges will have singularities outside 
the horizons if the scalar fields are not constant throughout spacetime. 
In this way, a theorem was proved that all static solutions regular 
outside the horizon with a charge matrix satisfying $\trace \C^2 > 0$ 
lie on the $\H^*$-orbit of the Schwarzschild solution. This also led to a generalisation of
Mazur's theorem, obtaining that all non-extremal axisymmetric stationary 
and asymptotically Minkowskian black holes lie on the $\H^*$-orbit of the Kerr 
solution (with some angular momentum parameter $a$).\footnote{For the reader's convenience we recall that Mazur's theorem states that an asymptotically Minkowskian axisymmetric stationary non-extremal black hole solution with a non-degenerate horizon is 
uniquely determined by its mass, its angular momentum and its electromagnetic charges \cite{Mazur}.}

Although the Mazur proof is more difficult to generalise to the case of asymptotically Taub--NUT solutions, it 
is reasonable to conjecture that all non-extremal axisymmetric 
stationary particle-like solutions lie on the $\H^*$-orbit of some
Kerr solution.  

It follows, as a corollary, that any $\H^*$ invariant equation satisfied by 
the charge matrix $\C$ of a Kerr solution is also satisfied by the charge 
matrix of any non-extremal axisymmetric stationary particle-like solution. 
Although there is no general proof that all the extremal axisymmetric 
stationary particle-like solutions can be obtained by taking the appropriate 
limit of a non-extremal solution, so far all known such solutions can 
be obtained in this way. By continuity, any $\H^*$ invariant equation satisfied 
by the charge matrix $\C$ is also valid for such extremal solutions. Using
Weyl coordinates \cite{ExactSolutions}, the coset representative $\V$ associated to the 
Schwarzschild solution with mass $m$ and its associated charge can be 
written in terms of the non-compact generator $\h$ of 
$\mathfrak{sl}(2,\mathds{R})$ only, \viz
\be
\V = \exp \left(\frac12 \ln \frac{r-m}{r+m} \, \h \right) 
\quad\Rightarrow\qquad \C = m \h
\ee
where we have used (\ref{Charge}). According to the five-graded decomposition 
(\ref{five}), the generator $\h$ in the adjoint representation acts as 
the diagonal matrix ${\rm diag} \, [2,1,0,-1,-2]$, where $1$ is the identity 
on $\mathfrak{l}_4$ and $0$ acts on $\mathfrak{g}_4 \oplus \{Ê\h \}$. This 
implies that 
\be {\ad_\h}^5 = 5 {\ad_\h}^3 - 4 \ad_\h \ .\ee
The four-dimensional theories leading to coset models associated to 
simple groups $\G$ after timelike dimensional reduction have been classified 
in \cite{Maison}. They correspond to models for which the four-dimensional 
scalars parametrise a symmetric space whose isometry group acts 
non-trivially on the vector fields. In particular, the list of \cite{Maison}
includes two theories for which the three-dimensional duality group is a 
real form of $E_8$, namely $\N=8$ supergravity \cite{CJ}, and the 
exceptional `magic' $\N=2$ supergravity \cite{magic} with real forms 
$E_{8(8)}$ and $E_{8(-24)}$, respectively. Since the fundamental 
representation of $E_8$ is the adjoint representation, 
we have, for these two theories,
\be 
\h^5 = 5 \h^3 - 4 \h \label{N8} \ .
\ee
However, $\h$ turns out to satisfy a lower order polynomial equation in general. Indeed, for all 
the other groups listed in \cite{Maison}, the fundamental representation 
of $\G$ admits a three-graded decomposition with respect to the generator 
$\h$, in such a way that the latter takes the form ${\rm diag}\,[1,0,-1]$. 
The three-graded decomposition of the groups listed in \cite{Maison}Ê is 
displayed in appendix~\ref{groups}. It follows that in these cases one has the stronger relation
\be 
\h^3 = \h \label{N}\ .
\ee
We then define the BPS parameter $c^2$ by 
\be 
\boxed{\quad c^2 \equiv \frac{1}{k} \, \trace \C^2 \quad} 
\label{bpsparam}
\ee 
with $ k \equiv \trace  \h^2 > 0$, where the normalisation is chosen such that
$c^2 = m^2$ for the Schwarzschild solution. Owing to the indefinite metric 
on the coset space $\mathfrak{g}\ominus\mathfrak{h}^*$, the trace 
$\trace \C^2$ and thus the square of the BPS parameter $c^2$ can assume either 
sign.\footnote{This is in contradistinction to spacelike reductions, for which 
  the metric on the coset is positively defined, whence $\trace \C^2 =0$
  would imply $\C =0$. We thus recover the well-known result that, in order for
  BPS solutions to exist, the Killing vector must be non-spacelike \cite{KillingSpinor}.} 
  However, negative values of $c^2$ correspond to hyper-extremal solutions 
which we will not consider (such as e.g. the Reissner-Nordstr\"om solution 
with $c^2 = m^2-e^2 <0$, which has a naked singularity). Hence, the BPS 
parameter will always be assumed to be non-negative in the following. 
Equation (\ref{N8}) then implies that for any solution in the $\H^*$-orbit 
of the Kerr solution, the charge matrix $\C$ satisfies
\be 
\boxed{ \biggl . \quad \C^5 = 5 c^2 \C^3 - 4 c^4  \C  \quad \biggr . } 
\label{polynom}
\ee
For all but two exceptional cases with $E_{8(8)}$ and 
$E_{8(-24)}$, we have the stronger constraint
\be\label{cubic} \boxed{ \biggl . \quad 
\C^3 = c^2 \C  \quad \biggr . } 
\ee
from (\ref{N}), in which case the fundamental representation admits a three-graded decomposition. 
Then, using the theorem of \cite{Maison}, it follows that these equations are 
satisfied by the charge matrix of any asymptotically flat non-extremal 
axisymmetric stationary single-particle solution. Furthermore, it follows that {\em non-rotating extremal solutions (like the BPS solutions), for which $c=0$, are characterised by nilpotent charge matrices $\C$.} Note however that the BPS parameter is non-zero for extremal rotating solutions. The extremality parameter $\varkappa$ is defined as
\be \varkappa \equiv \sqrt{ c^2 - a^2}  \label{extremalityP} \ee
where $a$ is the angular momentum by unit of mass.\footnote{Note that the definition of angular momentum is slightly more subtle in asymptotically Taub--NUT spacetimes, nevertheless one can define it  unambiguously by requiring the corresponding Komar integral to be independent of the local section of $U(1) \rightarrow M_+ \rightarrow V$ \cite{Komar}.}  For an asymptotically Taub--NUT black hole, the extremality parameter is equal to the product of the horizon area and the surface gravity divided by a factor of $4\pi$. Neither the horizon area nor the surface gravity is left invariant by the action of $\H^*$, but nevertheless $\varkappa$ is an invariant.

The current $\star \V P \V^{-1} $ is the representative of a cohomology class of $V$, and as such it defines a linear map from the second homology group of $V$ to $\mathfrak{g}$. 
\bea \star \V P \V^{-1}\   :  H^2(V) \hspace{2mm} &\longrightarrow&  \hspace{2mm}  \mathfrak{g} \CR
\Sigma \hspace{6mm} &\longrightarrow&  \hspace{2mm}   \C_{|\Sigma} \ .\eea
The algebraic structure of $\mathfrak{g}$ then permits one to define a non-linear map from $H^2(V)$ into the universal enveloping algebra of $\mathfrak{g}$ for any polynomial $\Upsilon$ as follows
\bea \Upsilon\  :  H^2(V)   \hspace{2mm}  &\longrightarrow&  \hspace{2mm}   U(\mathfrak{g}) \CR
\Sigma \hspace{6mm} &\longrightarrow& \hspace{2mm}   \Upsilon(\C_{|\Sigma}) \ .\eea
For any stationary asymptotically flat solution regular outside the horizon of any non-exceptional model of the list \cite{Maison}, we conjecture that this map vanishes identically on $H^2 (V)$ for $\Upsilon = \C^3 - c^2 \C$, and moreover that it vanishes for $\Upsilon = \C^5 - 5 c^2  \C^3 + 4 c^4 \C $ in the exceptional cases of $\N=8$ supergravity and the exceptional $\N=2$ magic supergravity.

\subsection{Supergravity and BPS conditions}

When considering stationary solutions in supergravity theories, 
the Euclidean three-dimensional point of view is very convenient for obtaining an
understanding of the BPS properties of stationary solutions. $\N$-extended supergravity in four dimensions
admits $U(\N)$ as an R symmetry group for $\N<8$, and $SU(8)$ for $\N=8$. Upon dimensional 
reduction from four to three dimensions, the compact R symmetry $U(\N)$ 
is enlarged to $SO(2\N)$ if the Killing vector is spacelike \cite{3Dclass},
and to the group $SO^*(2\N)$ (non-compact for $\N>1$) if it is timelike 
\cite{Maison}. It is the latter case that is relevant to the stationary 
solutions considered here. In this case, the group of automorphisms of 
the  $2\N$-extended superalgebra in three dimensions is the product 
of the three-dimensional rotation group $SU(2)$ and the R symmetry group
$Spin^*(2\N)$.

The fields of pure $\N$-extended supergravity are in one-to-one 
correspondence with the $p$-form representations of $U(\N)$, where $p$ is 
even for the boson fields, and odd for the fermionic fields. For $\N=8$, 
there is no $U(1)$ factor, and the $p$-form representations are related 
by duality to the complex conjugates of the $(8-p)$-form representations,  
while scalar fields are complex self-dual (\ie pseudo-real). As we will explain, there 
is a similar pattern for the conserved charges of the stationary solutions:
the mass and the NUT charge correspond to the trivial representation of 
$SU(\N)$ while the electromagnetic charges correspond to the $2$-form and the $6$-form 
representations of $SU(\N)$ and the scalar charges correspond to the $4$-form 
representation of $SU(\N)$. After a timelike dimensional reduction, these then
combine to form the full charge matrix $\C$ of pure $\N$-extended supergravity, 
which will be shown to be equivalently described by a state $|\C\rangle$
transforming in the Weyl spinor representation of $Spin^*(2\N)$. Likewise, 
and in analogy with the spacelike reduction of \cite{3Dclass}, the bosonic 
and fermionic fields are assigned to spinor representations of $Spin^*(2\N)$, 
and are transformed into one another by the action of $2\N$ extended 
supersymmetry, with the supersymmetry parameter belonging to the (pseudo-real)
vector representation of $SO^*(2\N)$.

Following \cite{3Dclass} one can now in principle classify all possible 
locally supersymmetric theories systematically by studying the restrictions 
that supersymmetry imposes on the target space geometries. Here we 
will not work out the complete Lagrangians, but will concentrate on the relevant
supersymmetry variations. Furthermore, we will
limit attention to the smaller class of theories obtainable
by dimensional reduction from four dimensional supergravities, and whose 
scalar sectors are governed by irreducible symmetric spaces. A list of such theories can be obtained by matching the tables of \cite{Maison} with previous results on spacelike reductions of \cite{3Dclass}. The K\"{a}hler symmetric spaces can be found in \cite{Gilmore}, and the special K\"{a}hler symmetric spaces have been classified in \cite{Cremmer}. 

For $\N=1$ supergravity theories, the internal symmetry group is the 
product of $Spin^*(2) \cong U(1) $ and a group associated to the matter 
content of the theory. A list of the relevant theories is given in Table I 
below, with the number of vector and scalar supermultiplets in four
dimensions given in the third and fourth columns, respectively. 

\begin{gather}
\begin{array}{|c|c|c|c|}
\hline
  \,\, \G / \H^* \,\, & \,\,  \G_4 / \H_4 \, \,& \,\, {\rm Vector}  \,\, & \,\, {\rm Scalar} \,\, \\*
\hline
& & &\vspace{-4mm} \\*
\hspace{4mm}\frac{SU(1+m,1+n)}{U(1)\times SU(m,1) \times SU(1,n)}  \hspace{4mm}&\hspace{4mm} \frac{U(m,n)}{U(m) \times U(n)} \hspace{4mm} & \hspace{2mm} m+ n \hspace{2mm}  & \hspace{2mm} mn  \hspace{2mm}  \\*
& & &\vspace{-4mm} \\*
 \hline
& & &\vspace{-4mm} \\*
\hspace{4mm}\frac{SO^*(4+2n)}{U(1)\times SU(2,n)}  \hspace{4mm}&\hspace{4mm} \frac{SO^*(2n)\times SU(2) }{U(n)\times SU(2) } \hspace{4mm} & \hspace{2mm} 2n \hspace{2mm}  & \hspace{2mm} \frac{n(n-1)}{2}  \hspace{2mm}  \\*
& & &\vspace{-4mm} \\*
 \hline
& & &\vspace{-4mm} \\*
\hspace{4mm}\frac{Sp(2+2n,\mathds{R})}{U(1)\times SU(1,n)}  \hspace{4mm}&\hspace{4mm}\frac{Sp(2n,\mathds{R})}{U(n)} \hspace{4mm} & \hspace{2mm} n \hspace{2mm}  & \hspace{2mm} \frac{n(n+1)}{2}  \hspace{2mm}  \\*
& & &\vspace{-4mm} \\*
 \hline
& & & \vspace{-4mm} \\*
\hspace{4mm}\frac{E_{7(-25)}}{U(1)\times E_{6(-14)}}  \hspace{4mm}&\hspace{4mm} \frac{SO(2,10)}{SO(2) \times SO(10)} \hspace{4mm} & \hspace{2mm} 16 \hspace{2mm}  & \hspace{2mm} 10  \hspace{2mm}  \vspace{-4mm} \\*
& & & \\*
 \hline
\end{array} \nonumber 
\end{gather}
\begin{centerline} {\small Table I : Irreducible homogenous spaces of $\N=1$ supergravity}\end{centerline}

For $\N=2$, the internal symmetry is the product of $Spin^*(4)$ with a 
group associated to the matter content of the theory (vector multiplets
or hypermultiplets). Now, $Spin^*(4) \cong SU(1,1) \times SU(2)$, where 
the $SU(2)$ factor acts only on the scalar fields belonging to hypermultiplets, 
and on the fermions. The theories that can be analysed within the 
present framework have vector multiplets but no hypermultiplets
in four dimensions (the number is given in the third column of the table
below). These models are displayed in Table II.

For $\N\geq3$, the possible supergravity theories are much more constrained, 
and the target spaces must be symmetric spaces.\footnote{Whereas this is
not true for the $\N\leq 2$ theories, cf. \cite{3Dclass}.}
When $\N=3,4$, we can still couple in an arbitrary number of matter 
multiplets, whereas for $\N\geq 5$ the theories are uniquely determined.
The complete list is given in Table III below. Note that we need to invoke
the low rank isomorphisms $Spin^*(6) \cong SU(3,1)$ (for $\N=3$) and  
$Spin^*(8) \cong Spin(6,2)$ (for $\N=4$), respectively, in order to match 
the tables with the general theory. 

\begin{gather}
\begin{array}{|c|c|c|}
\hline
  \,\, \G / \H^* \,\, & \,\,  \G_4 / \H_4 \, \,& \,\, {\rm Vector}  \,\,  \\*
\hline 
& & \vspace{-4mm} \\*
\hspace{4mm}\frac{SU(2,1+n)}{SU(1,1)\times  U(1,n)}  \hspace{4mm}&\hspace{4mm} \frac{U(1,n)}{U(1) \times U(n)} \hspace{4mm} & \hspace{2mm} n \hspace{2mm}   \\*
& & \vspace{-4mm} \\*
 \hline
& & \vspace{-4mm} \\*
\hspace{4mm}\frac{Spin(4,2+n)}{SU(1,1)\times SU(1,1)\times   Spin(2,n)}  \hspace{4mm}&\hspace{4mm} \frac{SU(1,1)}{U(1)} \times \frac{SO(2,n)}{U(1) \times SO(n)} \hspace{4mm} & \hspace{2mm} 1+n \hspace{2mm}   \\*
& & \vspace{-4mm} \\*
 \hline
& & \vspace{-4mm} \\*
\hspace{4mm}\frac{G_{2(2)}}{SU(1,1)\times SU(1,1)}  \hspace{4mm}&\hspace{4mm} \frac{SU(1,1)}{U(1) } \hspace{4mm} & \hspace{2mm} 1\hspace{2mm}   \\*
& & \vspace{-4mm} \\*
 \hline
& & \vspace{-4mm} \\*
\hspace{4mm}\frac{F_{4(4)}}{SU(1,1)\times Sp(6,\mathds{R})}  \hspace{4mm}&\hspace{4mm} \frac{Sp(6,\mathds{R})}{U(1)\times SU(3) } \hspace{4mm} & \hspace{2mm} 6\hspace{2mm}   \\*
& & \vspace{-4mm} \\*
 \hline
& & \vspace{-4mm} \\*
\hspace{4mm}\frac{E_{6(6)}}{SU(1,1)\times SU(3,3)}  \hspace{4mm}&\hspace{4mm} \frac{SU(3,3)}{U(1) \times SU(3)\times SU(3)} \hspace{4mm} & \hspace{2mm} 9\hspace{2mm}   \\*
& & \vspace{-4mm} \\*
 \hline
& & \vspace{-4mm} \\*
\hspace{4mm}\frac{E_{7(-5)}}{SU(1,1)\times SO^*(12)}  \hspace{4mm}&\hspace{4mm} \frac{SO^*(12)}{U(1) \times SU(6)} \hspace{4mm} & \hspace{2mm} 15\hspace{2mm}   \\*
& & \vspace{-4mm} \\*
 \hline
& & \vspace{-4mm} \\*
\hspace{4mm}\frac{E_{8(-24)}}{SU(1,1)\times E_{7(-25)}}  \hspace{4mm}&\hspace{4mm} \frac{ E_{7(-25)}}{U(1) \times E_{6(-78)} } \hspace{4mm} & \hspace{2mm} 27\hspace{2mm}  \vspace{-4mm} \\*
& &\\*
 \hline
\end{array} \nonumber 
\end{gather}
\nopagebreak
\begin{centerline} {\small Table II : Irreducible homogenous spaces of $\N=2$ supergravity }\end{centerline}
\pagebreak[2]

\begin{gather}
\begin{array}{|c|c|c|c|}
\hline
 \, \, \N \, \, &  \,\, \G / \H^* \,\, & \,\,  \G_4 / \H_4 \, \,& \,\, {\rm Vector}  \,\,  \\*
\hline 
& & &\vspace{-4mm} \\*
\hspace{2mm}3\hspace{2mm} & \hspace{4mm}\frac{SU(4,1+n)}{SU(3,1)\times  U(1,n)}  \hspace{4mm}&\hspace{4mm} \frac{U(3,n)}{U(3) \times U(n)} \hspace{4mm} & \hspace{2mm} n \hspace{2mm}   \\*
& & &\vspace{-4mm} \\*
 \hline
& & &\vspace{-4mm} \\*
\hspace{2mm}4\hspace{2mm} & \hspace{4mm}\frac{SO(8,2+n)}{SO(6,2)\times SO(2,n)}  \hspace{4mm}&\hspace{4mm}\frac{SU(1,1)}{U(1)} \times \frac{Spin(6,n)}{SU(4) \times Spin(n)}\hspace{4mm} & \hspace{2mm} n \hspace{2mm}   \\*
& & &\vspace{-4mm} \\*
 \hline
& & &\vspace{-4mm} \\*
\hspace{2mm}5\hspace{2mm} & \hspace{4mm}\frac{E_{6(-14)}}{Spin^*(10)\times U(1)}  \hspace{4mm}&\hspace{4mm} \frac{SU(5,1)}{U(5)} \hspace{4mm} & \hspace{2mm} 0 \hspace{2mm}   \\*
& & &\vspace{-4mm} \\*
 \hline
& & &\vspace{-4mm} \\*
\hspace{2mm}6\hspace{2mm} & \hspace{4mm}\frac{E_{7(-5)}}{Spin^*(12) \times SU(1,1)}  \hspace{4mm}&\hspace{4mm} \frac{Spin^*(12)}{U(6)} \hspace{4mm} & \hspace{2mm} 0 \hspace{2mm}   \\*
& & &\vspace{-4mm} \\*
 \hline
& & & \vspace{-4mm} \\*
\hspace{2mm}8\hspace{2mm} & \hspace{4mm}\frac{E_{8(8)}}{Spin^*(16)}  \hspace{4mm}&\hspace{4mm} \frac{E_{7(7)}}{SU(8) } \hspace{4mm} & \hspace{2mm} 0 \hspace{2mm}   \vspace{-4mm} \\*
& & &\\*
 \hline
\end{array} \nonumber 
\end{gather}
\begin{centerline} {\small Table III : Homogenous spaces of $\N\ge3$ supergravity}\end{centerline}

\vspace{3mm}
Let us now discuss the supersymmetry variations relevant to the BPS analysis
in more detail. For the Lorentzian case (\ie for a spacelike reduction), the relevant
(massless) supermultiplets were already described and studied in 
\cite{3Dclass}. As shown there, these superalgebras and their (massless)
representations can be completely characterised in terms of the 
{\it real Clifford algebras}
\be
\{ \Gamma^I , \Gamma^J \} = 2 \delta^{IJ} \qquad
\mbox{for $I,J,... = 1,\dots, 2\N$}
\ee
Here we must perform a similar analysis, but with $SO(2\N)$
replaced by $SO^*(2\N)$ whose maximal compact subgroup is $U(\N)$ (by 
definition). Since the $2\N$-extended Minkowskian superalgebra thus admits 
a Clifford algebra construction from Cl$(2\N,\mathds{R})$, we will look
for an analogous construction for $SO^*(2\N)$ making use of the complex
Clifford algebra Cl$(\N,\mathds{C})$. Because the group $Spin^*(2\N)$ and 
its irreducible spinorial representations are perhaps less familiar, we 
summarise some relevant results in appendix \ref{Spindix}. Besides the use of 
manifestly $U(\N)$ covariant notation, the crucial tool here is the use 
of fermionic oscillators defined by (see e.g. \cite{Georgi} 
for a pedagogical introduction)
\be
a_i := \frac12 \Big(\Gamma_{2i-1} + i\Gamma_{2i}\Big) \quad , \qquad 
a^i \equiv (a_i)^\dagger = \frac12\Big(\Gamma_{2i-1} - i\Gamma_{2i}\Big)
\ee
for $i,j,\dots = 1,\dots , \N$. These obey the standard anticommutation
relations
\be
\{ a_i , a_j\} = \{a^i , a^j \} = 0 \quad , \qquad
\{a_i , a^j \} = \delta^j_i\ .
\ee
As we will see this formalism greatly facilitates the analysis of the 
BPS conditions.

Making use of the fermionic oscillators introduced above we will thus express 
the various fields in the spinor basis generated by the creation operators 
$a^i$ acting on a `vacuum state' $|0\rangle$ (which is annihilated by all the $a_i$).
Accordingly, for $\N=5,6,8$ the coset components $P_\mu dx^\mu$ of the Cartan 
form are represented by the state
\be\label{Pmu}
|P_\mu\rangle = \Big( P_\mu^\ord{0}  + P^\ord{2}_{\mu\,ij} a^i a^j + 
     P^\ord{4}_{\mu\, ijkl} a^i a^j a^k a^l + \dots \Big) |0\rangle \ .
\ee
For $\N\leq 4$, an arbitrary number of matter multiplets can be coupled
and therefore the state $|P_\mu\rangle$ carries an extra label $\mathcal{A}$
to count the matter multiplets. Inspecting the $\H^*$ groups in the
Tables we see this extra label 
is an $SO(2,n)$ index for $\N=4$ (cf. our discussion of matter-coupled 
$\N=4$ supergravities in section~6), an $SU(1,n)$ index for $\N=3$, and 
so on. Furthermore, the state $|P_\mu\rangle$, or the states
$|P_\mu,\mathcal{A}\rangle$, in principle must satisfy an irreducibility 
(reality) constraint as explained in appendix \ref{Spindix}. However, 
when the group $\H^*$ contains an extra $U(1)$ factor besides the
R symmetry group $Spin^*(2\N)$, the representation (\ref{Pmu}) 
becomes complexified, so we only need to impose a reality constraint 
when no $U(1)$ is available, such as for instance $\N=2$ supergravity
with exceptional $\G$ or $\G=Spin(4,2+n)$. The case $\N=6$ can be 
obtained by a consistent truncation from $\N=8$ (see Section~2.2); for 
the latter, there is again no $U(1)$ factor in $\H^*$, and we
need to require $|P\rangle = \invo |P\rangle$, which 
expresses the well known self-duality of the $\N=8$ multiplet. 
Similarly, the physical fermions are represented by the anti-chiral state 
\be
|\chi \rangle_\alpha = \Big( \psi_{\alpha\,i} a^i +
    \chi_{\alpha\, ijk} a^i a^j a^k + \dots \Big) |0\rangle
\ee
(We use the letter $\psi_i$ for degree-one components, because these
originate from the 4-dimensional gravitinos, while the $\chi_{ijk}$
originate from the 4-dimensional spin-$\frac12$ fermions.) Again, this
representation must satisfy an irreducibility constraint for $\N =8$, 
namely\footnote{Note that $\invo$ being an {\it anti}-pseudo-involution, it raises the $SU(2)$ index by complex conjugation.}
\be
( \invo |\chi \rangle)^\alpha = \varepsilon^{\alpha\beta} | \chi \rangle_\beta \ .
\ee
The $\varepsilon_{\alpha\beta}$ here is necessary because the anti-chiral 
representation of $Spin^*(4M)$ is {\em pseudo-real}, with twice as many
components as the {\em real} chiral spinor. It is thus the additional spatial 
$SU(2)$ symmetry that restores the boson fermion balance required 
by supersymmetry. For theories with an arbitrary number of matter 
multiplets the fermionic state acquires an extra label, just like the 
bosonic state.

While the $U(\N)$ transformation properties of these states are manifest, 
they transform as follows under the non-compact generators of $Spin^*(2\N)$ 
(cf. appendix \ref{Spindix}):
\be
\delta \left| P_\mu \right >  = \frac{1}{2} Ê\Scal{Ê\Lambda^{ij} a_i a_j  -  
    \Lambda_{ij} a^i a^j  }Ê\left| P_\mu \right>  \;\;\; , \quad
\delta \left| \chi \right>_\alpha  =  \frac{1}{2} Ê\Scal{Ê\Lambda^{ij} a_i a_j  
   - \Lambda_{ij} a^i a^j  }\left| \chi \right >_\alpha  
\ee
where for $\N\leq 4$ we suppress the extra index $\mathcal{A}$ for simplicity.
On the other hand, the three-dimensional gravitinos $\psi^i_\alpha$ and the
supersymmetry parameters $\epsilon^i_\alpha$ together with their 
complex conjugates $\psi_i^\alpha\equiv(\psi^i_\alpha)^*$ and 
$\epsilon_i^\alpha\equiv(\epsilon^i_\alpha)^*$ transform in the 
pseudo-real vector representation of $SO^*(2\N)$, that is\footnote{In the 
 Minkowski case, the fundamental representation of $SL(2,\mathds{R})$ is real, 
 and the transformation is simply $\delta\epsilon^I_\alpha = 
 \Lambda^{IJ} \epsilon^J_\alpha$ (recall that $I,J,\dots=1,\dots,2\N$).
 In the complex $U(\N)$ basis, this becomes
 $$
 \delta\epsilon^i_\alpha =  {\Lambda^i}_j \epsilon^j_\alpha + \Lambda^{ij} 
 \epsilon_{\alpha\, j}.  
 $$
 The $\varepsilon_{\alpha\beta}$ in (\ref{SOeps}) thus plays the 
 role of an imaginary unit.} 
\be\label{SOeps} 
\delta\epsilon^i_\alpha =  {\Lambda^i}_j \epsilon^j_\alpha + 
\Lambda^{ij}Ê\varepsilon_{\alpha \beta} \epsilon_j^\beta \quad , \qquad
\delta\epsilon_i^\alpha =  {\Lambda_i}^j \epsilon_j^\alpha + 
\Lambda_{ij}Ê\varepsilon^{\alpha \beta} \epsilon^j_\beta
\ee
and similarly for the gravitinos. The commutator of two $Spin^*(2\N)$
transformations with parameters $\Lambda_1$ and $\Lambda_2$ gives a new transformation with parameters
\bea 
{\Lambda_{12}^{\hspace{2mm} i}}_j &=&  
 {\Lambda_{1}^{\hspace{1mm} i}}_k  {\Lambda_{2}^{\hspace{1mm} k}}_j -  
{\Lambda_{2}^{\hspace{1mm} i}}_k  {\Lambda_{1}^{\hspace{1mm} k}}_j  
+ \Lambda_{1}^{\hspace{1mm} ik} \Lambda_{2\, jk} -  
\Lambda_{2}^{\hspace{1mm} ik} \Lambda_{1\, ik} \CR
\Lambda_{12}^{\hspace{1mm} ij} &=&  
- 2 {\Lambda_{1}^{\hspace{1mm} [i}}_k   \Lambda_{2}^{\hspace{1mm} j] k}  
+ 2 {\Lambda_{2}^{\hspace{1mm} [i}}_k   \Lambda_{1}^{\hspace{1mm} j] k}   
\ .
\eea 
With this notation, the supersymmetry transformations of the fermions read 
\be 
\delta \psi_\alpha^i = d_{\omega + Q}  \epsilon_\alpha^i  \quad , \qquad
\delta \psi^\alpha_i = d_{\omega + Q}  \epsilon^\alpha_i  
\ee
for the gravitino components, and
\be
\delta \left| \chi  \right>_\alpha  = e^\mu_a    {{\sigma^a }_\alpha}^\beta 
  \, \Scal{   \epsilon^i_\beta \, a_i +  \varepsilon_{\beta\gamma}    
  \epsilon_i^\gamma   a^i  } \left| P_\mu \right>    
\ee
for the propagating fermions, where $d_{\omega + Q} $ is the covariant 
exterior differential with respect to the $SU(2)$ spin-connection $\omega$ 
and the $\H^*$ connection $Q$ coming from the scalar fields. We note that 
for $\N=8$ this formula is consistent with the representation constraint,
that is, $(\invo\delta|\chi\rangle)^\alpha = \varepsilon^{\alpha\beta} \delta|\chi\rangle_\beta$.
Using the above definitions and the formulas from Appendix \ref{Spindix} 
(and, more
specifically, the fact that conjugation for a $Spin^*(2\N)$ spinor
involves the matrix $\beta$), it is straightforward to compute the
conjugate spinor supersymmetry transformations: 
\be 
\delta \left< \chi \right |^\alpha =  
-  e^\mu_a    {\sigma^{a \, \alpha}}_\beta \, \left< P_\mu \right | 
\Scal{   \epsilon_i^\beta \,a^i +  \varepsilon^{\beta\gamma}    
\epsilon^i_\gamma   a_i   } \ . 
\ee 

The integrability condition for a supersymmetry transformation with 
parameter $\epsilon$ to preserve the vanishing of the gravitino fields is given 
by the algebraic equation
\be 
\delta \psi^\alpha_i = 0 \quad \Rightarrow \qquad
\scal{ \baa R + d Q + Q^2 } \epsilon = 0 
\ee 
for the curvature $2$-forms $\baa R \equiv \frac{1}{4} {R^{ab}}_{\mu\nu} 
\sigma_{ab} dx^\mu \wedge dx^\nu$ and $d Q + Q^2$, valued in 
$\mathfrak{su}(2)$ and $\mathfrak{so}^*(2\N)$, respectively. In three dimensions, the 
curvature $2$-form $\baa R$ is expressible\footnote{The formula is
 $$ R_{\mu\nu ab} = 4 e_{[\mu[a} R_{b]\nu]} - e_{\mu[a} e_{\nu |b]} R.
 $$} in terms of the Ricci tensor 
$R_{ab}$ by
\be 
\baa R =  \frac{1}{2} \sigma^{ab} \Scal{ e_{a}  \wedgeÊe^c \,  
 R_{cb}  - e_{b} \wedge e^c \, R_{ac} - 
\frac{1}{2}  e_{a } \wedge e_b \, R } \ .
\ee 
The equations of motion (\ref{EinsteinE}) give furthermore that
\be 
\baa R =  \frac12 \sigma^{ab} \Scal{ e_{a}  \wedge dx^\mu e_b^\nu - 
  e_{b}  \wedge dx^\mu e_a^\nu - \frac{1}{2} e_{a}  \wedge  e_b g^{\mu\nu} } 
 \trace P_\mu P_\nu \ .
\ee
Then, using the Bianchi identity (\ref{Bianchi}), one can rewrite the 
integrability condition in terms of the one-form $P$ only, 
\be  
\biggl[ \,  \frac{1}{2} \sigma^{ab} \Scal{ e_{a}   \wedge dx^\mu e_b^\nu - 
e_{b}  \wedge  dx^\mu e_a^\nu - \frac{1}{2} e_{a} \wedge  e_b g^{\mu\nu} } 
\trace P_\mu P_\nu - P  \wedge P \biggr] \epsilon = 0 \ .
\ee
For asymptotically flat solutions, $P$ goes to zero as in (\ref{Pasympt})
for $r\rightarrow + \infty$, and the leading order part of this equation is 
given by
\be  
\frac{1}{2 r^4} \sigma^{ab} \Scal{ \delta^3_a dr \wedge  e_b   
-\delta^3_b dr \wedge   e_a  - \frac{1}{2} e_a \wedge e_b    } \, 
\big(\trace \C^2 \big) \, \epsilon = \mathcal{O}(r^{-3}) \ ,
\ee
where $e^3 \sim dr + \mathcal{O}(1)$. In this way, we arrive at the condition 
\be\label{TrC}
\trace \C^2 = 0 \quad \Leftrightarrow \quad c^2 =0 \ .
\ee
If this equation is satisfied, one can then integrate the first order equation 
for the Killing spinors following from the supersymmetry variations 
of the gravitinos, thus justifying the designation of $c^2$ as the `BPS parameter'.
We stress once more that (\ref{TrC}) does {\em not} imply $\C=0$.
For asymptotically Minkowski solutions (that is, without NUT charge), we 
have checked that, for the pure $\N\leq 5$ supergravities, $c^2$ is indeed 
proportional to the determinant of the Bogomolny matrix\footnote{By
 `Bogomolny matrix' we mean the matrix on the right hand side of the  
 superalgebra when acting on the asymptotic free-particle states in
 four dimensions. This matrix is a function of the masses and central 
 charges, and has vanishing determinant for BPS states.} (this claim will
be proved in Section~3.3).
This is no longer true for $\N=6$ and $\N=8$. From equation (\ref{polynom}), 
one then deduces that {\it the charge matrix is nilpotent for BPS solutions}.
More precisely, we have at least $\C^5 = 0$ in the $E_8$ cases and 
$\C^3 = 0$ otherwise. 

The BPS condition also requires the dilatino fields to be left invariant 
by some supersymmetry generators. In order for a Killing spinor to satisfy 
\be 
\delta |\chi \rangle_\alpha = 0 \quad \Rightarrow\qquad
 e^\mu_a    {{\sigma^a }_\alpha}^\beta \, \Scal{   \epsilon^i_\beta \, a_i 
  +  \varepsilon_{\beta\gamma}    \epsilon_i^\gamma   a^i  } 
     \left| P_\mu \right>   = 0 \label{dilatinoT}
\ee 
the charge state vector must satisfy 
\be\label{susy} \boxed{ \Biggl . \quad
\Big( \epsilon^i_\alpha a_i + 
\varepsilon_{\alpha\beta} \epsilon_i^\beta a^i \Big) |\C\rangle = 0  
\quad \Biggr . }
\ee 
where $(\epsilon^i_\alpha,\epsilon_i^\alpha)$ is the asymptotic 
(for $r\rightarrow\infty$) value of the Killing spinor. As before, for
$\N\leq 4$ the state $|\C\rangle$ may require an extra label, such
that (\ref{susy}) gets replaced by
\be\label{susy1}
\Big(\epsilon^i_\alpha a_i + \varepsilon_{\alpha\beta} \epsilon^\beta_i a^i\Big)
|\C,\mathcal{A}\rangle = 0
\ee

{\em The simple equation (\ref{susy}) (or \ref{susy1})) is a key result of this paper:} 
it encapsulates all the information about solutions of the equations of motion with 
residual supersymmetry and allows {\em a complete analysis of the BPS 
sector} (as we will see below, (\ref{susy}) is a stronger condition than 
the Killing spinor equation). Furthermore, we will show how the analysis 
of the  BPS conditions can be reduced to simple calculations with 
fermionic oscillators by means of (\ref{susy}). Since $\C$ does not
involve the angular momentum parameter $a$, we recover the (known)
result that the BPS analysis is not sensitive to angular momentum.
We note that (\ref{susy}) takes the form of a `Dirac equation' for the 
$\H^*$ spinor $|\C\rangle$, with the `$\gamma$-matrices' $(a_i,a^i)$ and 
the supersymmetry parameter $(\epsilon^i,\epsilon_i)$ as the `momentum'.

Multiplying equation (\ref{susy}) by its conjugate equation, and 
contracting the antichiral Weyl $Spin^*(2\N)$ indices, one gets the 
integrability condition
\be  
\left( \begin{array}{cc} \ \delta_\alpha^\beta \, \Scal{ \left< \C |  
\C \right>  \delta^i_j + \frac{1}{2} \left<   \C | [ a^i , a_j ] 
| \C \right> }Ê \ & \   \varepsilon_{\alpha\beta} \, 
\left< \C | a^i a^j | \C \right>  \ \\  
\  - \varepsilon^{\alpha\beta} \,\left< \C | a_i a_j | \C \right>  
\ &  \ \delta^\alpha_\beta\, \Scal{Ê \left< \C |  \C \right>  \delta_i^j -
\frac{1}{2} \left<   \C | [ a^j , a_i ] | \C \right> }  \ \end{array} \right) 
\left( \begin{array}{c} \epsilon_\beta^j \\  \epsilon_j^\beta \end{array} \right)  = 0 
\ee
necessary for a supersymmetry parameter to correspond to an unbroken supersymmetry
generator. This equation decomposes into two inequivalent representations 
upon $SO^*(2\N)$: first of all, we recover the condition
$c^2 \equiv  \left< \C |  \C \right> = 0$; secondly, we get  
\be
\mathcal{Z} (\epsilon) \equiv \left( \begin{array}{cc} \  \frac{1}{2} \ 
\delta_\alpha^\beta \ \left<   \C | [ a^i , a_j ] | \C \right>   
\ & \  \varepsilon_{\alpha\beta} \  \left< \C | a^i a^j | \C \right>   
\ \\  \ -  \varepsilon^{\alpha\beta} \  \left< \C | a_i a_j | \C \right>   
\ &  \ - \frac{1}{2} \  \delta^\alpha_\beta\   \left<   \C | [ a^j , a_i ]
 | \C \right>   \ \end{array} \right)  
\left( \begin{array}{c} \epsilon_\beta^j \\  \epsilon_j^\beta \end{array} 
\right)  = 0 \ .
\ee
The existence of unbroken supersymmetry generators thus requires both 
$c^2 = 0$ and that the matrix  $ \mathcal{Z}$, transforming under 
$SO^*(2\N)$ in the adjoint representation, leaves invariant the associated 
spinor parameter $\epsilon_\alpha^i$. 

In the foregoing section we identified the {\it charge matrix} $\C$ by
means of the conserved charges of the three-dimensional theory, whereas
in this section we have been working with the {\em state} $\left|\C\right >$,
or a multiplet $|\C,{\mathcal{A}}\rangle$ of such states. The two descriptions
are obviously related, as the matrix $\C$ and state $|\C\rangle$ (or the
multiplet $\big\{|\C,{\mathcal{A}}\rangle\big\}$ for $\N\leq 4$) contain
the same number of charge degrees 
of freedom, but writing down a general formula is neither easy nor
really helpful because the most convenient conventions usually depend 
upon the properties of the specific groups $\G$ and $\H^*$.  For $\N=5$ we will 
spell out the relation between the matrix 
$\C$ and the state vector $|\C\rangle$ explicitly in eqn.~(\ref{C})
of Section~3.3.

\section{Solving the BPS conditions}
Let us now proceed to analyse the BPS condition (\ref{susy}) case by case 
for various values of $\N$.  As the conditions are the same even in 
the presence of several matter multiplets  (for $\N\leq 4$), we will suppress the 
extra index $\mathcal{A}$ in this section. The state vector $\left| \C \right > $ 
of charges has the following general form (cf. appendix \ref{Spindix})
\be 
\left| \C \right> \equiv \Scal{Ê\w + Z_{ij} a^i a^j + 
\Sigma_{ijkl} a^i a^j a^k a^l + \cdots }Ê \left| 0 \right> \ .
\ee
Here $\w \equiv m + i n $ is the complex gravitational charge (mass and NUT parameter),
$Z_{ij} \equiv Q_{ij} + i P_{ij} $ are the electromagnetic charges, and $\Sigma_{ijkl}$ are the scalar 
charges (which will turn out to depend on the other charges). Further charge 
components appear for $\N \geq6$. For $Spin^*(2\N)$ the conjugate spinor 
is (see appendix \ref{Spindix})\footnote{Note the plus sign on all terms, which is related to the signature of the $Spin^*(2\N)$ scalar product.}
\be 
\left< \C \right| = \left< 0 \right |  \Scal{Ê\bar \w + Z^{ij} a_i a_j 
+ \Sigma^{ijkl} a_i a_j a_k a_l + \cdots } 
\ee
from which we compute the norm, hence the BPS parameter, as\footnote{With 
  a factor of $\ft12$ in the case of maximal supergravity because the multiplet is self-dual.}  
\be\label{CC} 
c^2 = \langle\C|\C\rangle =
 |\w|^2 - 2 Z_{ij} Z^{ij} + 24 \Sigma_{ijkl} \Sigma^{ijkl}-  \cdots
\ee

\subsection{General discussion and results for $\N\leq 5$}
Equation (\ref{susy}) can now be decomposed with respect to 
$\oplus_p \bigwedge^{2p-1} \mathds{C}^\N$, that is, the oscillator
basis $a^i \left| 0 \right>$, $a^i a^j a^k \left| 0 \right>$,...  
The one-form component reads, for all $\N$, 
\be 
2 Z_{ij} \epsilon^j_\alpha - \varepsilon_{\alpha\beta} \w \epsilon^\beta_i = 0 
\ee
The parameter $\w$ being non-zero for any non-trivial regular solution, we 
obtain that the spinor parameter associated to an unbroken supersymmetry 
generator satisfies 
\be 
\epsilon^\alpha_i = -  \frac{2}{\w}  \varepsilon^{\alpha\beta} Z_{ij} 
\epsilon^j_\beta \label{e1} 
\ee
relating the Killing spinor to its complex conjugate by a kind of
symplectic Majorana condition. Hence, for all $\N$,
\be 
\frac{4}{|\w|^2} Z^{ik} Z_{jk} \epsilon^j_\alpha = \epsilon^i_\alpha \ .
\label{Projsy}
\ee

At this point it is advantageous to switch to a diagonal basis for the 
matrix $Z_{ij}$, which can be reached by conjugating with a suitable
$SU(\N)$ matrix, 
\be Z_{ij} \ \cong \ \frac{1}{2} \left(\begin{array}{ccccc} \ 0 \ & \ z_1 \  & \ 0 \ & \ 0 \ & \ \cdots  \ \\ -z_1\ & \ 0 \ & \ 0 \ & \ 0 \ & \ \ddots \ \\ \ 0 \ & \ 0 \ & \ 0 \ & \ z_2 \ & \ \ddots \ \\ \ 0 \ & \ 0 \ & \ -z_2 \ & \ 0 \ & \ \ddots \ \\ \ \vdots \ &  \ \ddots \ &  \ \ddots \ &  \ \ddots \ &  \ \ddots \ 
 \end{array}\right) 
\label{suitable} \ee
for $\N = 2M $, and 
\be 
Z_{ij} \ \cong \ \frac{1}{2} \left(\begin{array}{cccccc} \ 0 \ &  \ 0 \ &  \ 0 \ &  \ 0 \ &  \ 0 \ &  \ \cdots \ \\  \ 0 \ & \ 0 \ & \ z_1 \  & \ 0 \ & \ 0 \ & \ \ddots  \ \\   \ 0 \ & -z_1\ & \ 0 \ & \ 0 \ & \ 0 \ & \ \ddots \ \\   \ 0 \ & \ 0 \ & \ 0 \ & \ 0 \ & \ z_2 \ & \ \ddots \ \\   \ 0 \ & \ 0 \ & \ 0 \ & \ -z_2 \ & \ 0 \ & \ \ddots \ \\ \ \vdots \ &  \ \ddots \ & \ \ddots \ &  \ \ddots \ &  \ \ddots \ &  \ \ddots \   \end{array}\right) \ee
for $\N= 2M + 1$. Introducing $M$ antisymmetric tensors $\omega^\m_{ij}$ 
satisfying 
\bea 
\omega^\m_{ik} \, \omega^{\n\, jk} &=& 0  \hspace{10mm} {\rm if} 
\hspace{5mm}  \m \ne \n \CR
 \omega^\m_{ik} \, \omega^{\m \, jk} &=& I^{\m \hspace{2mm} j}_{\hspace{2mm}i} \label{Basis2f}
\eea
with the $I^{\m \hspace{2mm} j}_{\hspace{2mm} i}$ being projectors onto 
the orthogonal $2$-dimensional complex subspaces
\bea 
I^{\m \hspace{2mm} k}_{\hspace{2mm} i}  I^{\n 
\hspace{2mm} j}_{\hspace{2mm} k}  &=& 0  \hspace{10mm} {\rm if} \hspace{5mm}  
\m \ne \n \CR
 I^{\m \hspace{2mm} k}_{\hspace{2mm} i}  I^{\m \hspace{2mm} j}_{\hspace{2mm} k} &=& I^{\m \hspace{2mm} j}_{\hspace{2mm} i}  \CR
I^{\m \hspace{2mm} i}_{\hspace{2mm} i}  &=& 2\ ,
\eea
we can re-express $Z_{ij}$ in the form
\be 
Z_{ij} = \frac{1}{2} \sum_\m  z_\m \  \omega^\m_{ij} \ .
\ee
Substituting this expression in equation (\ref{Projsy}) we obtain
\be 
\sum_\m \frac{ |z_\m|^2}{ |\w|^2 }  I^{\m \hspace{2mm} i}_{\hspace{2mm} j} 
\epsilon^j_\alpha = \epsilon^i_\alpha \ .
\ee
Consequently, the spinor parameter can have non-zero components {\em only in 
those subspaces for which $|z_\m |^2 = |\w|^2$.} In accordance with 
established terminology {\em we shall speak of an $(n/\N)$ BPS solution if 
this relation is satisfied for $n$ out of $M$ values $z_\m$.}
For a spinor lying in the subspace $\m$ associated to the projector 
$I^{\m \hspace{2mm} i}_{\hspace{2mm} j}$ for which $|z_\m |^2 = |\w|^2$, 
equation (\ref{e1}) then becomes 
\be  
\epsilon^\alpha_i = -  \frac{z_\m}{\w}  
\varepsilon^{\alpha\beta} \omega^\m_{ij} \epsilon^j_\beta  \ .
\ee
Next, the $3$-form component of equation (\ref{susy}) reads
\be\label{3form} 
4 \Sigma_{ijkl} \epsilon^l_\alpha - \varepsilon_{\alpha\beta} 
Z^{\Big . }_{[ij} \epsilon^\beta_{k]} = 0 
\ee
where $\Sigma_{ijkl}$ are the scalar charges. Together with (\ref{e1}), 
this equation gives that 
\be\label{3form1} 
\Scal{Ê\Sigma_{ijkl} - \frac{1}{2 \w} Z_{[ij}  Z_{kl]} }Ê\epsilon^l_\alpha = 0 \ .
\ee
This equation is again valid for all $\N$. It is trivially satisfied for 
$\N = 3$; for $\N=4$ and $\N=5$ it implies
\be 
\Sigma_{ijkl} = \frac{1}{2 \w} Z_{[ij}  Z_{kl]} \; , \label{ScalarCont} 
\ee
which is consistent with the $5$-form component of equation (\ref{susy})
\be \Sigma^{[ijkl} \epsilon^{m]}_\alpha = 0 
\ee
for $\N=5$. For these values of $\N$, we have thus made completely explicit
our previous claim that the scalar charges are not independent, but
depend on the electromagnetic charges via Eq.~(\ref{ScalarCont}). 
As we will see latter, (\ref{ScalarCont}) is also the general solution of the equation $\C^3 = c^2 \C$ 
for both $\N=4$ and $5$. Finally, we emphasise that, for $\N>5$, the formula 
(\ref{ScalarCont}) is {\em not} valid in general, unless the BPS degree 
is sufficiently high.

\subsection{$\N=6$ and $\N=8$ supergravity}
We now proceed directly to $\N=8$ because the case $\N=6$ is most
conveniently obtained by consistent truncation of $\N=8$. For maximal 
supergravity, the scalar charge vector is given by
\bea
|\C\rangle &=& \Big( \w + Z_{ij} a^i a^j  + \Sigma_{ijkl} a^i a^j a^k a^l  + 
\frac{1}{6!} \varepsilon_{ijklmnpq} \,  Z^{ij} \, a^k a^l a^m a^n a^p a^q 
\nn\\
&&  \qquad  + \frac{1}{8!} \, \varepsilon_{ijklmnpq} \, \bar \w \,  
a^i a^j a^k a^l a^m a^n a^p a^q \Big) | 0 \rangle \ .
\label{N8C} 
\eea
Its irreducibility as a $Spin^*(16)$ representation, that is, the condition 
$|\C\rangle = \invo |\C\rangle$ (cf. appendix \ref{Spindix}) requires that the scalar 
charges are complex self-dual, \viz
\be\label{csd}
\Sigma_{ijkl} = \frac{1}{4!} \varepsilon_{ijklmnpq} \Sigma^{mnpq} \ .
\ee
By self-duality the $p$-form component of equation (\ref{susy}) is 
equivalent to its $(\N-p)$-form component. The one-form and three-form 
components of this equation were already given in (\ref{e1}) and
(\ref{3form1}), respectively. However, unlike for $\N\leq 5$, we now no longer can 
`peel off' the parameter $\epsilon^l_\alpha$ from 
equation (\ref{3form1}) in general, so formula (\ref{ScalarCont}) 
may fail.

For $\ft18$ BPS solutions, we have $|z_1| = |\w|$, whereas $|z_\m|\neq |\w|$
for $\m =2,3,4$. The non-vanishing components of (\ref{3form1}) are 
then orthogonal, and they determine part of the scalar charges $\Sigma_{ijkl}$.
The remaining components of $\Sigma_{ijkl}$ can then be deduced from
the self-duality constraint (\ref{csd}), in such a way that all scalar
charges are determined as functions of the $Z_{ij}$, but (\ref{ScalarCont})
is not satisfied for all components as $\w^{-1} Z_{[ij}Z_{kl]}$ need not be complex self-dual in general.

For $\ft14$ BPS solutions, $|z_1| = |z_2| = |\w|$ and  $|z_3|, |z_4| \neq |\w|$.
In this case the components corresponding to the two different spinor 
overlap, although the formula (\ref{ScalarCont}) is still not valid for all 
components. The electromagnetic charges must satisfy constraints in order 
to be compatible with the self-duality constraint: inspection shows that 
(\ref{3form1}) now implies
\be\label{1234}
\frac{z_1 z_3}{\w} = \frac{ \bar z_2 \bar z_4}{\bar \w} 
\hspace{10mm}Ê\frac{z_1 z_4}{\w} = \frac{ \bar z_2 \bar z_3}{\bar \w} \ .
\ee
Therefore $|z_3|^2 = |z_4|^2$, and we conclude that {\em there cannot exist
$\ft38$ BPS asymptotically flat stationary solutions of $\N=8$ supergravity} 
(which would require $|z_1|= |z_2|= |z_3|= |\w| \neq |z_4|$). Finally, 
for $\ft12$ BPS solutions, (\ref{3form1}) is valid for any spinor parameter, 
and we at last recover (\ref{ScalarCont}). By self-duality, the 
electromagnetic charges must then satisfy
\be  
\frac{1}{2 \w} Z_{[ij}  Z_{kl]} = 
\frac{1}{4! } \varepsilon_{ijklmnpq}  \ \frac{1}{2 \bar \w} Z^{mn}  Z^{pq} \ .
\ee

The formulas for $\N=6$ can be obtained by truncation of the above results. 
However, $\N=6$ supergravity is somewhat special because its bosonic 
sector, with the coset space $SO^*(12)/U(6)$, is identical to the bosonic 
sector of the magic $\N=2$ supergravity \cite{magic}. 
The two theories differ only in their fermionic sectors, both of which 
can be obtained by truncation of $\N=8$ supergravity. While the bosons are 
truncated in the same way to give the coset $SO^*(12)/U(6)$ for both the 
$\N=2$ and $\N=6$ cases, one retains six gravitinos and 26 spin-$\frac12$ fermions
in the $\N=6$ theory, whereas for the $\N=2$ theory one retains the
complementary set of two gravitinos and 30 spin-$\frac12$ fermions (the latter 
belong to 15 vector multiplets coupled to the $\N=2$ graviton multiplet), 
such that there are altogether 32 fermionic degrees of freedom in each case.
In other words, the bosonic sector by itself `does not know' whether it 
belongs to $\N=6$ supergravity or to the magic $\N=2$ theory. 

These features can be seen directly from the form of the truncated
charge vector which is represented by the state 
\bea
|\C\rangle &=& \Big(\w + \bar Z a^7 a^8\Big) | 0 \rangle 
+ \left( Z_{ij}  + \Sigma_{ij}\,  a^7 a^8 \right) a^i a^j | 0 \rangle
  + \frac{1}{4!} \varepsilon_{ijklmn} \Big( \Sigma^{ij}  +   
 Z^{ij} \, a^7 a^8 \Big) a^k a^l a^m a ^n  | 0 \rangle \nn\\
&& \qquad\quad
 + \frac{1}{6!} \varepsilon_{ijklmn}  \left( Z  +  \bar \w \, a^7 a^8 \right) 
a^i a^j a^k a^l a^m a^n   |0 \rangle  
\label{N6state} 
\eea
and which can be directly obtained from (\ref{N8C}) by truncation. Here
$i,j,\dots= 1,\dots,6$ label the $\N=6$ oscillators while $a^7$ and
$a^8$ correspond to the supercharges of the $\N=2$ theory. When viewed as 
an $\N=2$ theory, equation (\ref{susy}) reduces to its $1$-form component 
which decomposes into (\ref{e1}) for the spinors $\epsilon^7_\alpha$ 
and $\epsilon^8_\alpha$ and a matter component
\be
\Sigma_{ij} \,    \epsilon_\alpha^8  - \varepsilon_{\alpha \beta} \, 
Z_{ij} \, \epsilon^\beta_7 = 0 
\ee
which immediately yields the $\N=2$ $\ft12$ BPS conditions, $|Z|=|\w|$ and
\be
\Sigma_{ij} = \frac{1}{\w} \bar Z Z_{ij} \ . \label{BPSquaternion}
\ee
For $\N=6$, on the other hand, we get the $3$-form and the $5$-form equations
\be 
\epsilon_\alpha^{[k}  \Scal{Ê\Sigma^{ij]} - 
\frac{1}{4 \w } \varepsilon^{ij]mnpq} Z_{mn} Z_{pq} }  = 0 
\hspace{10mm} 
\Scal{Ê\Sigma_{ij} - \frac1{\w} \bar Z \, Z_{ij}}  \epsilon^j_\alpha = 0 
\label{N6Con}
\ee
where we have already substituted the solution (\ref{e1}) for the
supersymmetry generator $\epsilon^i_\alpha$. For $\ft16$ BPS solutions, the non-trivial components 
of these equations are orthogonal, and again suffice to determine the 
scalar charges $\Sigma^{ij}$ as functions of the others. For more 
supersymmetric solutions, the scalar charges are determined by equation 
(\ref{ScalarCont}) to be 
\be
Ê\Sigma^{ij} = \frac{1}{4 \w } \varepsilon^{ijmnpq} Z_{mn} Z_{pq}  \ .
\label{PureDeBPS1}
\ee
Requiring consistency with (\ref{BPSquaternion}) along the $\mathds{C}^4$ 
subspace associated to the unbroken supersymmetries gives 
\be 
\frac{ Z \bar z_1}{\bar \w} =   \frac{ z_2 z_3}{\w} \hspace{10mm} 
\frac{ Z \bar z_2}{\bar \w} =   \frac{ z_1 z_3}{\w} 
\ee
which is just the condition (\ref{1234}) in disguise.
Because $|z_1|^2 = |z_2|^2 = |\w|^2$, both equations reduce to
\be 
Z =  \frac{ z_1 z_2 z_3}{\w^2} \ .
\ee
The charge $Z$ is thus determined to be
\be 
Z =  \frac{1}{6 \w^2 } \varepsilon^{ijklmn} Z_{ij} Z_{kl} Z_{mn}   \label{PureDeBPS2}
\ee
with $|Z|^2 = |z_3|^2$ for $\ft13$ BPS solutions. For $\ft12$ BPS solutions
all the components of  $Ê\Sigma_{ij} - (\bar Z \, Z_{ij})/{\w} $ 
must cancel and we get
\be  
\frac1{\w} \bar Z \, Z_{ij} = \frac{1}{4 \bar \w } 
\varepsilon_{ijmnpq} Z^{mn} Z^{pq}  \ .
\ee
Equivalently, the condition for a solution to preserve some supersymmetry 
in both the $\N=2$ and the $\N=6$ theories requires the remaining eigenvalues 
of $Z_{ij}$ to be equal in modulus, which is consistent with the non-existence
of $\ft38$ BPS solutions in $\N=8$ supergravity. 
 
We conclude this subsection with a few comments on black hole entropy
in $\N=8$ supergravity. In that case, the constraints on the electromagnetic 
charges are related to extremality properties of the $E_{7(7)}$ invariant 
expression of the entropy \cite{PhasesN8}. For static solutions with
$\w = \bar\w = m$ satisfying the $\ft14$ BPS bound condition $|z_1| = |z_2| 
= m$ (and $|z_3|, |z_4|$ possibly different from $m$), equation (\ref{1234}) 
is strictly equivalent to the vanishing of the $E_{7(7)}$ invariant 
expression of the horizon area $A = 4 \pi \sqrt{ \lozenge (Z)  }$, where
\begin{multline}
Ê \lozenge ( Z) \equiv  Z_{ij} Z^{jk} Z_{kl} Z^{li} - 
\frac{1}{4} Z_{ij} Z^{ij} Z_{kl} Z^{kl} \\* + 
\frac{1}{96} \varepsilon_{ijklmnpq}ÊZ^{ij} Z^{kl} Z^{mn} Z^{pq}Ê+  
\frac{1}{96} \varepsilon^{ijklmnpq}ÊZ_{ij} Z_{kl} Z_{mn} Z_{pq} \ .
\end{multline}
This proves the conjecture of \cite{E7entropy} proposing the vanishing of the  
$E_{7(7)}$ invariant expression of the horizon area for $\ft14$ BPS
and $\ft12$ BPS black holes. For asymptotically Taub--NUT solutions, 
$\w$ is complex, and the $\ft14$ BPS condition (\ref{1234}) requires that the 
Ehlers $U(1)$ invariant $\lozenge( \w^{-\frac{1}{2}} Z )$ vanish. This 
leads us to conjecture that the expression for the horizon area of 
asymptotically Taub--NUT BPS black holes in maximal supergravity is 
\be 
A = 4 \pi |\w|  \sqrt{  \lozenge ( \w^{-\frac{1}{2}} Z) } \ . \label{HorizonArea}
\ee
As a matter of fact, this expression is not in 
general invariant with respect to 
the standard action of $E_{7(7)}$ on the electromagnetic charges. This is not in contradiction 
with the $U$-duality invariance of the entropy, however, since the latter cannot 
be identified with the horizon area for asymptotically Taub--NUT spacetimes.

\subsection{Relation to pure spinors} \label{PureSpinorS}
There is an intriguing link between the cubic constraint $\C^3 = c^2 \C$ 
on the charge matrix and pure spinors in pure supergravity theories. Let 
us start with $\N=5$ supergravity, for which the corresponding pure spinor 
equation is more familiar to physicists thanks to the work of N.~Berkovits 
in superstring theory. The duality group of the three-dimensional theory 
is $E_{6(-14)}$ which admits a complex 27-dimensional faithful representation.
With respect to the maximal subgroup $U(1)\times Spin^*(10)$, the ${\bf 27}$ 
decomposes into ${\bf 1} \oplus {\bf 16} \oplus {\bf 10}$ where ${\bf 16}$ 
is the complex chiral spinor representation of $Spin^*(10)$ and ${\bf 10}$ 
the pseudo-real vector representation of $SO^*(10)$. The charge matrix $\C$ 
can be defined in terms of the chiral spinor $|\C\rangle$ as 
\be\label{C} 
\C  \equiv \left( 
\begin{array}{cccc}  \vspace{2mm} \hspace{2mm} 0 \hspace{2mm} &\langle \C | \hspace{2mm} & \hspace{2mm}0  \hspace{2mm}  & \hspace{2mm}0  \hspace{2mm}  \\ \vspace{2mm} \hspace{2mm} |\C\rangle   \hspace{2mm}& \hspace{2mm} 0     \hspace{2mm}&  \hspace{2mm} a_j | \C^\convo \rangle  \hspace{2mm}&  \hspace{2mm} a^j | \C^\convo \rangle \hspace{2mm}   \\  \vspace{2mm} \hspace{2mm}0    \hspace{2mm} & \hspace{2mm} \langle \C^\convo | a^i   \hspace{2mm} &  \hspace{2mm} 0 \hspace{2mm}&  \hspace{2mm} 0 \hspace{2mm} \\  \vspace{2mm} \hspace{2mm}0    \hspace{2mm} & \hspace{2mm} \langle \C^\convo | a_i   \hspace{2mm} &  \hspace{2mm} 0 \hspace{2mm}&  \hspace{2mm} 0 \hspace{2mm}
\end{array}\right) 
\ee
which is understood to act on a complex 27-dimensional vector 
$(\eta,\,| S \rangle , \, V^j,\, V_j )$. $|\C^\convo\rangle$ is the 
antichiral spinor defined from the anti-involution $\invo$ 
\be 
|\C^\convo\rangle \equiv \invo | \C \rangle = \varepsilon_{ijklm} 
\Scal{Ê \Sigma^{jklm} a^i + \frac{1}{3!} Z^{lm}Ê\, a^i a^j a^k + 
\frac{1}{5!}Ê\bar \w \, a^i a^j a^k a^l a^m }Ê| 0 \rangle \ .
\ee
The formula (\ref{C}) makes the claimed relation between the matrix $\C$ and 
the state vector $|\C\rangle$ completely explicit for $\N=5$. Making use of 
the properties 
\begin{gather}  \langle \C | \C \rangle =  \langle \C^\convo | \C^\convo  \rangle \CR
Ê  \langle \C | a_i a_j | \C \rangle =  - \langle \C^\convo |a_i a_j |  \C^\convo  \rangle \hspace{10mm}Ê\langle \C | a^i a_j | \C \rangle =  \langle \C^\convo |a_j a^i  |  \C^\convo  \rangle \end{gather} \ ,
the Fierz identity 
\be a_i | \C^\convo\rangle  \langle \C | a^i + a^i | \C^\convo \rangle \langle \C | a_i  = - \frac{ 1}{2}Ê\langle \C | a^i | \C^\convo \rangle a_i - \frac{ 1}{2}Ê\langle \C | a_i | \C^\convo \rangle a^i \ee
and its conjugate, we compute ${\rm Tr}\,\C^2 = 12\,\langle\C|\C\rangle$ and 
\begin{multline}   
\C^3 -c^2 \C  =  \langle \C | a_k | \C^\convo \rangle \, \left( \begin{array}{cccc}  \vspace{2mm} \hspace{2mm} 0 \hspace{2mm} &\langle \C^\convo  | a^k  \hspace{2mm} & \hspace{2mm}0  \hspace{2mm}  & \hspace{2mm}0  \hspace{2mm}  \\ \vspace{2mm} \hspace{2mm} a^k |\C^\convo \rangle   \hspace{2mm}& \hspace{2mm} 0     \hspace{2mm}&  \hspace{2mm} \scal{Ê\delta^k_j + \sfrac{1}{2} a^k a_j } | \C  \rangle  \hspace{2mm}&  \hspace{2mm} \sfrac{1}{2}Êa^k a^j | \C \rangle \hspace{2mm}   \\  \vspace{2mm} \hspace{2mm}0    \hspace{2mm} & \hspace{2mm} \langle \C | \sfrac{1}{2}Êa^i a^k     \hspace{2mm} &  \hspace{2mm} 0 \hspace{2mm}&  \hspace{2mm} 0 \hspace{2mm} \\  \vspace{2mm} \hspace{2mm}0    \hspace{2mm} & \hspace{2mm} \langle \C | \scal{Ê\delta_i^k + \sfrac{ 1}{2}Êa_i  a^k }   \hspace{2mm} &  \hspace{2mm} 0 \hspace{2mm}&  \hspace{2mm} 0 \hspace{2mm}\end{array}\right)\\*
 + \langle \C | a^k  | \C^\convo \rangle \, \left( \begin{array}{cccc}  \vspace{2mm} \hspace{2mm} 0 \hspace{2mm} &\langle \C^\convo  | a_k  \hspace{2mm} & \hspace{2mm}0  \hspace{2mm}  & \hspace{2mm}0  \hspace{2mm}  \\ \vspace{2mm} \hspace{2mm} a_k |\C^\convo \rangle   \hspace{2mm}& \hspace{2mm} 0     \hspace{2mm}&  \hspace{2mm} \sfrac{1}{2} a_k a_j  | \C  \rangle  \hspace{2mm}&  \hspace{2mm}  \scal{Ê\delta_k^j + \sfrac{1}{2} a_k a^j } | \C \rangle \hspace{2mm}   \\  \vspace{2mm} \hspace{2mm}0    \hspace{2mm} & \hspace{2mm} \langle \C |  \scal{Ê\delta_k^i + \sfrac{1}{2} a^i a_k  }     \hspace{2mm} &  \hspace{2mm} 0 \hspace{2mm}&  \hspace{2mm} 0 \hspace{2mm} \\  \vspace{2mm} \hspace{2mm}0    \hspace{2mm} & \hspace{2mm} \langle \C  |  \sfrac{ 1}{2}Êa_i  a_k    \hspace{2mm} &  \hspace{2mm} 0 \hspace{2mm}&  \hspace{2mm} 0 \hspace{2mm}\end{array}\right) \ .
  \end{multline}
It follows that the constraint $\C^3 = c^2 \C$ is strictly equivalent to the $Spin^*(10)$ pure spinor constraint
\be 
\left< \C \right | a^i |\C^\convo \rangle = 0 \hspace{10mm} 
\left< \C \right | a_i  |Ê\C^\convo \rangle = 0  \ .
\ee
Here, we define a $Spin^*(2\N)$ pure spinor by the direct generalisation 
of the Cartan definition, that is by the requirement that 
$| \C^\convo  \rangle \langle \C |$ lies in the rank $\N$ antisymmetric 
tensor representation of $SO^*(2\N)$. The same computation in $\N=4$ 
pure supergravity shows that the cubic constraint (\ref{cubic}) is 
strictly equivalent to the $Spin^*(8)$ pure spinor constraint 
\be 
\langle  \C | Ê\C^\convo \rangle = 0 
\ee
where 
\be 
| \C^\convo \rangle \equiv \invo \left| \C \right > = \Scal{Ê\varepsilon_{ijkl} \, 
\Sigma^{ijkl}Ê+ \frac{1}{2} \varepsilon_{ijkl}\, ÊZ^{ij} \, a^k a^l + 
\frac{1}{4!} \varepsilon_{ijkl} \, \bar \w \, a^i a^j a^k a^l } 
\left| 0 \right > \ .
\ee
For practical computation it is much easier to consider the coset 
$Spin(2,8)/ ( Spin(2,6) \times U(1) )$ exploiting the isomorphism $Spin(2,6) 
\cong Spin^*(8)$. We postpone the proof of equivalence to the pure spinor 
constraint to Section \ref{N4sec}. For $\N=2$ and $\N=3$, there are no scalar 
charges and the equation $\C^3 = c^2 \C$ is trivially satisfied by any 
element of the coset $\mathfrak{g} \ominus \mathfrak{h}^*$. This is in 
agreement with the fact that any $Spin^*(4)$ or $Spin^*(6)$ chiral spinor 
is pure. 

The general solution of the $Spin^*(2\N)$ pure spinor constraint is
\be 
\left | \C \right > = \w \, 
\exp\left( \frac1{\w} Z_{ij} a^i a^j \right) \left| 0 \right > \ . 
\label{ExpPure} 
\ee
It is well defined only if $\w\ne 0$ but, since $\w= m + in$, it is 
natural to make this requirement. To prove (\ref{ExpPure}), we use the fact 
that for a spinor satisfying $\langle \C | \C \rangle > 0$, there exists
a $U(1) \times Spin^*(2\N)$ transformation that rotates both the electromagnetic 
charges  and the NUT charge to zero, such that in the new `frame'
\be 
\left | \C \right > = c | 0 \rangle \ .
\ee
Then, from the definition of the anti-involution $\invo$ (cf. appendix B), we have
\be 
| \C^\convo  \rangle \langle  \C | = c^2 \invo | 0 \rangle  \langle 0 |  
= \frac{c^2}{\N!}\, \varepsilon_{i_1\cdots i_\N} \, a^{i_1} \cdots  a^{i_\N} 
| 0 \rangle  \langle 0 |
= \frac{c^2}{\N!}\, \varepsilon_{i_1\cdots i_\N} \, a^{i_1} \cdots  a^{i_\N} 
\ee
where we have made use of the fact that we can replace $|0 \rangle\langle 0|$
by the unit operator in this expression because the left state is fully
occupied. To complete the proof, we only need to rotate the spinor back
to its original frame (\ref{ExpPure}); it is then easy to see that the
above result gets replaced by a combination of products of $\N$ fermionic
(creation {\em and} annihilation) oscillators corresponding to the 
$\N$-form representation of $Spin^*(2\N)$. Consequently, $|\C\rangle$ 
is a pure spinor for all $\N$.

Writing out the pure spinor condition for $\N=4$ and $\N=5$, we can easily 
see that it is equivalent to the equation
\be 
\w \Sigma_{ijkl} = \frac{1}{2} Z_{[ij} Z_{kl]} \label{ScalaireFixed}
\ee
which coincides with the equation derived from the requirement for the 
solution to be BPS, cf. (\ref{ScalarCont}). In the preceding section, this 
condition and the BPS bound condition on the eigenvalues of the 
electromagnetic charges were enough for the solution to be BPS. We are 
now going to see that the orders of the zeros of the BPS parameter are 
indeed governed by the number of eigenvalues of the electromagnetic charges 
which satisfy the BPS bound. Inserting the solution (\ref{ScalaireFixed}) 
into the definition of the BPS parameter (\ref{CC}), we get  
\be 
c^2 =  |\w|^2 - 2 Z_{ij} Z^{ij} +  
\frac{2}{ |\w|^2 } \Scal{ \scal{ÊZ_{ij} Z^{ij} }^2 - 
2 Z_{ij} Z^{jk} Z_{kl} Z^{li} } 
\ee
which reduces to
\be 
c^2  =  \frac{ \scal{Ê|\w|^2 -  |z_1|^2 }
\scal{Ê|\w|^2 -  |z_2|^2 }}{Ê |\w|^2 }    \label{BogoDet}
\ee
in terms of the eigenvalues $z_1$ and $z_2$ (the formula is also valid for 
$\N=2,\, 3$ with $z_2 = 0$). Without NUT charge ($|\w|^2=m^2$), $c^2$ to a given power is thus 
proportional to the determinant of the Bogomolny matrix obtained from 
the four-dimensional supersymmetry algebra projected on an asymptotically 
free massive particle state. As we just discussed, once the constraint 
(\ref{cubic}) is solved, the number of preserved supersymmetries can be 
derived from this determinant. It follows also from equation (\ref{BogoDet}) 
that all the extremal solutions admitting a nilpotent charge matrix $\C$ 
are BPS, and thus the moduli space of stationary black holes is 
given by the union of the $U(1) \times Spin^*(2\N)$-orbits of non-extremal 
Kerr--Taub--NUT black holes and the orbits of BPS black holes. 

For $\N = 6$, the $E_{7(-5)}$ constraint $\C^3=c^2 \C$ is equivalent to the 
$SL(2,\mathds{R}) \times Spin^*(12)$ invariant equation
\bea  
\langle \C | a^i a_j | \C \rangle \, a^j | \C \rangle + \langle \C | 
a^i a^j | \C \rangle \, a_j | \C \rangle -  \langle \C^\convo 
| a^i a_j | \C \rangle \, a^j | \C^\convo \rangle - \langle \C^\convo 
| a^i a^j | \C \rangle \, a_j | \C^\convo \rangle &=& 0 \CR
\langle \C^\convo | a^i a_j | \C^\convo \rangle \, a^j | \C^\convo 
\rangle + \langle \C^\convo | a^i a^j | \C^\convo \rangle \, a_j 
| \C^\convo \rangle -  \langle \C | a^i a_j | \C^\convo \rangle \, 
a^j | \C \rangle - \langle \C | a^i a^j | \C^\convo \rangle \, a_j 
| \C \rangle &=& 0 \CR \label{GpureSpinor} 
\eea 
In this case, this equation does not reduce any more to a quadratic constraint on the 
spinor $| \C\rangle $. For $c^2\neq 0$, the scalar charge $\Sigma^{ij}$ 
is generally a non-rational function of $\w$, $Z_{ij}$ and $Z$. For 
instance, the solution of $\C^3=c^2 \C$ for electromagnetic charges that are very 
small compared to the parameter $\w$ defines $\Sigma^{ij}$ as an infinite 
formal series in powers of $\frac{Z_{ij}}{\scriptscriptstyle W} ,\, 
\frac{\bar Z}{\scriptscriptstyle W} $ and their complex conjugates, and 
the resulting expression cannot in general be written in closed form. 
The BPS parameter thus is not simply proportional to the product of 
the determinants of the Bogomolny matrices of the $\N=2$ and $\N=6$ 
supergravities associated to this bosonic theory. Nevertheless, the 
$Spin^*(12)$ pure spinors define solutions of equation (\ref{GpureSpinor}), 
although not all its solutions define 
pure spinors. The $Spin^*(12)$ pure spinor condition reads
\be 
\frac{1}{2}Ê\langle \C | [ a^i , a_j ]Ê | \C^\convo \rangle  = 0 
\hspace{10mm}Ê\langle \C |Êa^i a^j   | \C^\convo \rangle = 0 
\hspace{10mm}Ê\langle \C |Êa_i a_j  | \C^\convo \rangle  = 0 \ .
\ee
Note that although these equations are invariant under the action of $Spin^*(12)$, they are not invariant under the action of $SL(2,\mathds{R})$ in general, and so the general solution of $\C^3 = c^2 \C$ cannot be a pure spinor. The pure spinor condition in components reads
\be 
8 \, \Sigma^{ik}ÊZ_{jk} = \delta^i_j \scal{Ê2 Z_{ij}Ê\Sigma^{ij} - \w Z  }Ê \hspace{7mm}
Ê\w \Sigma^{ij}Ê= \frac{1}{4}Ê\varepsilon^{ijklmn} Z_{kl}ÊZ_{mn}Ê
\hspace{7mm}
ÊZ Z_{ij}Ê= \frac{1}{4}Ê\varepsilon_{ijklmn} \Sigma^{kl}Ê\Sigma^{mn}Ê\ .
\ee
The general solution determines both the scalar charge $\Sigma^{ij}$ and 
the electromagnetic charge $Z$ to be
\be 
\Sigma^{ij} =   \frac{1}{4\w }Ê\varepsilon^{ijklmn} Z_{kl}ÊZ_{mn}Ê
\hspace{10mm} 
Z = \frac{1}{6 \w^2}Ê\varepsilon^{ijklmn}ÊZ_{ij} Z_{kl} Z_{mn} \ .
\ee
Note that according to equation (\ref{PureDeBPS1}) and (\ref{PureDeBPS2}), the $\ft13$ and the $\ft12$ BPS solutions of $\N=6$ supergravity do satisfy these equations, and the charge matrix associated to such a solution defines a pure spinor.  In general, for a charge matrix satisfying the pure spinor equation, one recovers the property that the BPS parameter is proportional to the determinant of the Bogomolny matrix, \viz 
\be 
c^2 =  \frac{ \scal{Ê|\w|^2 -  |z_1|^2 }\scal{Ê|\w|^2 -  |z_2|^2 }
\scal{Ê|\w|^2 -  |z_3|^2 }}{Ê |\w|^4 }  \ .
\ee
Such a restricted solution is $\ft12$ BPS in the quaternionic $\N=2$ magic supergravity if and 
only if it is $\ft12$ BPS in $\N=6$ supergravity. Although the general solution of (\ref{GpureSpinor}) is generically not a pure spinor, it follows from the transitivity property of $SL(2,\mathds{R})\times Spin^*(12)$ on the moduli space  of non-extremal black holes that $\C$ is in the $SL(2,\mathds{R})$-orbit of a pure spinor for $c > 0$. The general solution of (\ref{GpureSpinor}) can thus be parametrised as follows
\bea \w &=& \cosh u \, X + \sinh u \, e^{i\alpha} \, \frac{1}{6 {\bar X}^2} \varepsilon_{ijklmn} X^{ij} X^{kl} X^{mn} \CR
Z &=&  \cosh u \, \frac{1}{6 {X}^2} \varepsilon^{ijklmn} X_{ij} X_{kl} X_{mn}  +  \sinh u \, e^{i\alpha} \, \bar X  \CR
Z_{ij} &=& \cosh u \,  X_{ij} +  \sinh u \, e^{i\alpha} \,  \frac{1}{4 {\bar X}} \varepsilon_{ijklmn} X^{kl} X^{mn}  \CR
\Sigma^{ij} &=&  \cosh u \,   \frac{1}{4 {X}} \varepsilon^{ijklmn} X_{kl} X_{mn} +   \sinh u \, e^{i\alpha} \, X^{ij} \ .
\eea
By taking appropriate limits, one obtains extremal solutions which are BPS either in $\N=6$ supergravity or in the corresponding magic $\N=2$ supergravity associated to the quaternions. Nevertheless, this does not prove that there are no non-BPS extremal solutions with $c=0$.

Although we have not written out explicitly the quintic equation (\ref{polynom}) 
for maximal supergravity, the requirement of $Spin^*(16)$ covariance 
completely fixes the expression for the scalar charges in terms of the 
other charges when $|Z_{ij}|\ll |\w|$. As for $\N=6$, the expression for 
the scalar charges can be expanded into an infinite series in powers of 
$(Z_{ij}/\w)$ in such a way that the solution of (\ref{polynom}) defines 
the scalar charges as non-rational functions of the others. At low orders, 
we have\footnote{Note that the value of $|Z_{ij}/\w|$ is larger than the radius of convergence of the formal series for BPS solutions.} 
\begin{multline}  
\Sigma_{ijkl} = \Scal{ 1 + \frac{1}{24 {\scriptscriptstyle W}^3 
\bar {\scriptscriptstyle W}}  \varepsilon^{mnpqrstu} Z_{mn} Z_{pq} 
Z_{rs} Z_{tu} +  \frac{1}{24 {\scriptscriptstyle W} 
\bar {\scriptscriptstyle W}^3 }  
\varepsilon_{mnpqrstu} Z^{mn} Z^{pq} Z^{rs} Z^{tu} } \\*\hspace{80mm}  
\cdot \Scal{Ê\frac{1}{2  {\scriptscriptstyle W}} Z_{[ij} Z_{kl]} + 
\frac{1}{48 \bar  {\scriptscriptstyle W}} \varepsilon_{ijklvwxy}ÊZ^{vw} 
Z^{xy}Ê} \\ \hspace{-10mm} - \frac{5}{ {\scriptscriptstyle W}^2 \bar  {\scriptscriptstyle W}} Z_{[ij} Z_{kl} Z_{mn]} \Scal{ Z^{mn} - \frac{6}{| {\scriptscriptstyle W}|^2} Z_{pq} Z^{[mn} Z^{pq]} }   \hspace{30mm}  \\ - \frac{5}{ 24 {\scriptscriptstyle W} \bar  {\scriptscriptstyle W}^2} \varepsilon_{ijklmnpq}ÊZ^{[mn} Z^{pq} Z^{rs]} \Scal{ Z_{rs} - \frac{6}{| {\scriptscriptstyle W}|^2} Z^{tu} Z_{[rs} Z_{tu]} } + \mathcal{O} \left(  \frac{ Z^8 }{{\scriptscriptstyle W}^7} \right) \ .
\end{multline} 
It follows that the BPS parameter does not reduce to an expression 
proportional to the determinant of the Bogomolny matrix for asymptotically 
Minkowski solutions. 

The charge matrix transforms as a Majorana--Weyl spinor of $Spin^*(16)$, 
whereas the pure spinor equation is defined for complex spinors. The 
pure spinor equation for a Majorana--Weyl spinor implies that 
$\langle \C | \C \rangle = 0 $, and so there is no non-trivial solution 
in an Euclidean case with the group $Spin(2\N)$. However, since the scalar 
product $\langle \C | \C \rangle$ is indefinite for $Spin^*(16)$, there do 
exist non-trivial solutions in this case. Indeed, if one writes down the constraints 
\be 
\left | \C \right > = \w \, 
\exp\left( \frac1{\w} Z_{ij} a^i a^j \right) \left| 0 \right >  
= \invo \, \w \, 
\exp\left( \frac1{\w} Z_{ij} a^i a^j \right) \left| 0 \right > \ ,
\ee
one gets exactly the $\N=8$ constraints necessary for the corresponding solution 
to be $\ft12$ BPS. As a result, the moduli space of $\ft12$ BPS 
asymptotically flat stationary single-particle solutions of 
$\N=8$ supergravity is isomorphic to the space of $Spin^*(16)$ 
Majorana--Weyl pure spinors.

\section{Isotropy subgroups of BPS solutions}
The formalism developed in the previous sections affords a convenient
tool to investigate, and in fact completely characterise, all the BPS
orbits for different $\N$, thus furnishing a proof for a number of conjectures 
that have been made in the literature.

\subsection{Pure supergravities for $\N\leq 5$}
\label{SpinOrbits}
From the results of the previous section, it follows that the moduli space 
of solutions of the cubic equation (\ref{cubic}) is strictly equivalent to 
the space of pure spinors of $Spin^*(2\N)$ for all $\N\leq5$. Defining 
$\Omega_{ij} \equiv (2/\w) Z_{ij}$, equation (\ref{ExpPure}) tells us
that the general solution can be written as
\be 
\left | \C \right > = \w \, 
\exp\left( \frac{1}{2} \Omega_{ij} a^i a^j \right) \left| 0 \right > \ .
\ee
We emphasise again that for $\N\leq 5$ this form of $|\C\rangle$ is valid 
also for non-BPS solutions: in that case we simply set $\Omega_{ij}=0$
because we can use the duality group to rotate the solution to a `frame'
where it is a pure Kerr--Taub--NUT solution with (complex)
parameter $\w$. We also recall that for $\N\geq 3$, the group $Spin^*(2\N)$ 
is always accompanied by an extra $U(1)$ which must be taken into 
account when analysing the residual symmetries. 

The action of $\mathfrak{u}(1) \oplus \mathfrak{spin}^*(2\N)$ on the 
above spinor can be worked out by means of the formulas given in 
appendix~B to give\footnote{Recall that raising or lowering indices on $\Lambda$ corresponds to 
complex conjugation.}
\be\label{dC} 
\delta   \left | \C \right > = 
\frac{1}{2} \Scal{ \scal{Ê2\,  {\Lambda_i}^k Ê\Omega_{kj}  
+ \Lambda_{ij}Ê+  \Omega_{ik} \, \Lambda^{kl}Ê \Omega_{lj} } a^i a^j 
+  \Omega_{ij} \Lambda^{ij} - Ê {\Lambda_i}^i - 
 i \lambda  } \left | \C \right > \ ,
\ee
where $\lambda$ parametrises the $\mathfrak{u}(1)$ transformation. For a matrix charge $\C$ corresponding to a 
$\ft n\N$ BPS solution,  the matrix $\Omega_{ij}$ can be moved 
via a $Spin^*(2\N)$ rotation to a symplectic form on a subspace 
$\mathds{C}^{2n}\subset \mathds{C}^\N$. In order to analyse the isotropy 
subgroup of $U(1)\times Spin^*(2\N)$ corresponding to such a spinor, it 
is convenient to decompose the $U(\N)$ indices according to the product 
$U(2n)\times U(\N-2n)$ into unbarred ones $A,B,\dots = 1,\dots,2n$ and 
barred ones $\bar A, \bar B, \dots = 1,\dots, \N- 2n$, respectively. 
Splitting the equations (\ref{dC}) in this way and demanding 
$\delta|\C\rangle =0$, we arrive at
\bea\label{dC1}
2{\Lambda_{[A}}^C \Omega_{C|B]} +\Lambda_{AB} +
\Omega_{AC}\Lambda^{CD}\Omega_{DB} 
&=& 0 \qquad\qquad {\Lambda_{\bar A}}^C \Omega_{CB}  + \Lambda_{\bar AB}= 0 
\nn\\
 -i \lambda + \Omega_{AB}\Lambda^{AB} - {\Lambda_A}^A  -  
{\Lambda_{\bar A}}^{\bar A} &=& 0 \qquad\qquad 
\Lambda_{\bar A \bar B } = 0 \ .
\label{Isotropy} 
\eea 
Taking the symplectic trace of the first equation (with 
$\Omega_{AC} \Omega^{CB} = -\delta_A^B$), we get
\be
2\Lambda_A{}^A = \Omega^{AB} \Lambda_{AB} + \Omega_{AB} \Lambda^{AB} \ .
\ee
Let us first consider the subgroup of the isotropy group lying in the
maximal compact subgroup $U(1)\times U(\N)\subset U(1)\times Spin^*(2\N)$.
In this case the constraints on the Lie algebra generators imply 
$\Lambda_{\bar A}{}^B=0$ and $ Ê{\Lambda_{[A}}^C Ê\Omega_{C|B]} =0$,
whence the generators inside $U(2n)$ must leave invariant the symplectic 
form $\Omega_{AB}$, and therefore generate the subgroup $Sp(n)\equiv 
USp(2n)\subset U(2n)$. From the third equation in (\ref{Isotropy}), we deduce that $\lambda$ is 
determined in terms of the other parameters, hence is not independent. The 
maximal compact subgroup of the isotropy subgroup is thus 
$Sp(n) \times U(\N-2n)$. 

To analyse the non-compact generators we define
\be\label{Lpm}
\Lambda^\pm_{AB} := \frac12 \Big( \Lambda_{AB} \pm 
    \Omega_{AC} \Omega_{BD} \Lambda^{\pm\, CD} \Big)\quad \Rightarrow
\qquad \Lambda^\pm_{AB} = \pm \Omega_{AC}\Omega_{BD} \Lambda^{\pm \, CD} \ .
\ee
Then we see that $\Lambda^+_{AB}$ drops out from the first equation in 
(\ref{dC1}), but there is nevertheless still one constraint on it. Namely, from 
(\ref{Lpm}) we get $\Omega^{AB} \Lambda^\pm_{AB} = \pm\Omega_{AB}
\Lambda^{\pm\,AB}$; thus, $\Omega^{AB}\Lambda^+_{AB}$ is real, while 
$\Omega^{AB}\Lambda^-_{AB}$ is imaginary. From the third equation in 
(\ref{dC1}) we then deduce that 
\be
\Omega^{AB}\Lambda^+_{AB} = 0 
\ee
(all other terms being pure imaginary). Together with $Sp(n)$
these parameters combine to give the {\em non-compact} real form 
$SU^*(2n)\subset SL(2n,\mathds{C})$.\footnote{$SU^*(2n) \cong SL(n,\mathds{H})$.} 
In terms of fermionic oscillators, 
a given element of $SU^*(2n)$ is defined via the following generators 
of $SL(2n,\mathds{C})$ 
\bea
{\bf X}^A{}_B &\equiv&  \frac{1}{2} \Scal{Êa^A  a_B - \Omega^{AC}\Omega_{BD} \, a^D a_C } \CR
{\bf X}^{AB} &\equiv&  \frac{1}{2} \Scal{Ê a^A a^B - \Omega^{AC}\Omega^{BD}a_C a_D }Ê  - \frac{1}{4n}Ê\Omega^{AB} \Scal{Ê\Omega_{CD} a^C a^D - \Omega^{CD} a_C a_D }  
\eea
with an anti-Hermitean matrix $\Lambda_A{}^B= - \Lambda^B{}_A$ 
satisfying $\Lambda_A{}^B = - \Omega_{AC} \Omega^{BD} \Lambda_D{}^B $, 
and a traceless element $\Lambda^+_{AB}$ as
\be  
{\bf X}  ( \Lambda ) = \Lambda_A{}^B  {\bf X}^A{}_B + 
  \frac{1}{2} \big(\Lambda^+_{AB}  {\bf X}^{AB} +  \Lambda^{+\, AB} 
   {\bf X}_{AB} \big)
\ee
where ${\bf X}_{AB} \equiv ( {\bf X}^{AB} )^\dagger$. Although the fundamental 
representation of $SU(2n)$ is complex for $n>1$, the fundamental 
representation of  $SU^*(2n)$ is pseudo-real. Indeed, in order to be 
consistent with supersymmetry, the action of $SU^*(2n)$ on a Killing 
spinor $\epsilon^A_\alpha$ must preserve the reality condition (\ref{e1}), 
\ie $\epsilon_\alpha^A = - \Omega^{AB} 
\varepsilon_{\alpha\beta} \epsilon^\beta_B$\ :
\be  
\bigl[Ê\ {\bf X} (\Lambda) \ ,\ 
\epsilon_\alpha^A ( a_A - \Omega_{AB}Êa^B ) \bigr]Ê= 
\scal{ \Lambda^A{}_B  + \Omega^{AC} \Lambda^+_{CB}  }Ê 
\big(\epsilon_\alpha^B ( a_A  - \Omega_{AD}Êa^D )\big) \ . \label{pseudo-vectorR}  
\ee
The part of $U(2n)$ not lying in the $Sp(n)$ subgroup is constrained 
by the condition
\be
{\Lambda_{[A}}^C \Omega_{C|B]} + \Lambda^-_{AB} = 0 
\quad\Rightarrow\qquad \Lambda_A{}^A = \Omega^{AB} \Lambda^-_{AB} \ .
\ee
The explicit computation (using formulas from appendix \ref{Spindix}) 
shows that the associated generators are given by
\be
{\bf N}^{AB} \equiv
\big( a^A + \Omega^{AC} a_C\big) \big( a^B +  \Omega^{BD} a_D \big)\;\; ,\quad 
{\bf N}_{AB} \equiv
\big( a_A - \Omega_{AC} a^C\big) \big( a_B - \Omega_{BD} a^D \big)
\ee
so that $\Omega^{AC} \Omega^{BD}  {\bf N}_{CD} = {\bf N}^{AB}$. Using  
\be 
\big\{ \, a_A -  \Omega_{AC} a^C \ , \ a_B  - \Omega_{BD} a^D \  \big\} = 0 \ ,
\label{AbelianSusy} 
\ee
one easily checks that this particular combination of compact and non-compact 
generators is {\em nilpotent}: 
\be
\big[ {\bf N}_{AB} , {\bf N}_{CD} \big] =
\big[ {\bf N}_{AB} , {\bf N}^{CD} \big] =
\big[ {\bf N}^{AB} , {\bf N}^{CD} \big] = 0
\ee
and that the associated Lie algebra elements transform in the 
$\bf{n}(2\bf n - \bf{1})$ of $SU^*(2n)$, \viz
\begin{multline}
\big[ \, {\bf X}(\Lambda) \,  , \  v^-_{AB} {\bf N}^{AB}  - v^{-\, AB} {\bf N}_{AB}   \big] =   2 \scal{- Ê\Lambda^C{}_A  +  \Omega_{AD} \Lambda^{+\, DC}  }  v^-_{CB}  {\bf N}^{AB}  \\* - 2 \scal{Ê  \Lambda^A{}_C  +  \Omega^{AD} \Lambda^+_{DC}  }  v^{-\, CB}  {\bf N}_{AB} \ .
\end{multline}
The {\em reducibility} of the two-form representation of $SU^*(2n)$ is a 
direct consequence of the pseudo-reality of its fundamental representation. 

The only remaining generators of the $\ft n \N$ BPS isotropy subgroup 
(besides the generators $[a^{\bar A}, a_{\bar B}]$ of $U(\N -2n)$)
correspond to the solutions of 
\be 
Ê{\Lambda_{\bar A}}^C Ê\Omega_{CB}  + \Lambda_{\bar AB}Ê= 0 
\ee
in the complex ${\bf 2n} \otimes ({\bf \N- 2n})$ representation of 
$SU^*(2n) \times U(\N-2n)$. In terms of fermionic oscillators,
the associated generators are
\be
{\bf N}^{\bar A B} = a^{\bar A} \big(a^B + \Omega^{BC} a_C \big)
\;\; ,
\qquad 
{\bf N}_{\bar A B} = a_{\bar A} \big(a_B - \Omega_{BC} a^C \big)\ . \ee
This is once again a combination of compact and non-compact generators, which 
commutes to give the nilpotent generators in the ${ \bf n (2n -1)}$ of 
$SU^*(2n)$ given above, \viz
\begin{gather}  
\bigl[\,  {\bf N}_{\bar A B}\,  , \, {\bf N}^{\bar C D }\, Ê\bigr]Ê
=  \delta^{\bar C}_{\bar A} \, \Omega_{BE}\,  {\bf N}^{ED} \hspace{20mm} 
 \bigl[\,  {\bf N}_{\bar A B}\,  , \, {\bf N}_{\bar C D }\, Ê\bigr]Ê=  0 \CR 
 \bigl[\,  {\bf N}_{\bar A B}\,  , \, {\bf N}^{C D }\, Ê\bigr]Ê= 0 
\hspace{10mm} 
\bigl[\,  {\bf N}^{\bar A B}\,  , \, {\bf N}^{C D }\, Ê\bigr]Ê= 0 
\label{NilCom} \ .
\end{gather}
We thus arrive at the conclusion that {\em the isotropy subgroups $\J_n(\N)$
are non-reductive subgroups of} $U(1)\times Spin^*(2\N)$ for $\N\leq 5$, that is, 
`Poincar\'e-like' groups with the product $SU^*(2n)\times U(\N-2n)$
as the semi-simple `Lorentz-like' subgroups; schematically, we have
\be 
\J_n(\N) = \scal{ÊSU^*(2n) \times 
U(\N-2n) } \ltimes \scal{\ (Ê{{ \Yboxdim8pt  {\yng(1)}}
\otimes \yngd )\oplus 
{{ \Yboxdim8pt  {\yng(1,1)}}_- \otimes {\bf 1}}}\ } \ ,
\ee
where the Young tableaux of $SU^*(2n)$ and $U(\N-2n)$ are to be built
with undotted and dotted boxes, respectively.
The $\ft n\N$ isotropy subgroup of $U(1)\times Spin^*(2\N)$ (for $n\ge 1$) 
is thus of dimension $\N^2  + (2n+1) (n - 1)$. As we will see below,
similar statements hold for $\N=6$ and $\N=8$.

From equation (\ref{AbelianSusy}) it follows that the `Heisenberg-like' 
subgroup of the isotropy subgroup leaves invariant the Killing spinor 
$\epsilon_\alpha^A = - \Omega^{AB} \varepsilon_{\alpha\beta} \epsilon^\beta_B$.
Therefore the isotropy subgroup acts on the Killing spinors in the fundamental
representation of $SU^*(2n)$. The isotropy subgroups $\J_n(\N) \subset\H^*$ 
for $\N\leq5$ are given in the following table; we omit the extra-index  
on $\J_n(\N)$ as there is only one such group for each pair $(n,\N)$ for
$\N\leq 5$.

\begin{gather}
\begin{array}{|c|c|c|c|c|}
\hline
  & \,\, \N=2  \,\, & \,\,\N=3 \,\,  & \,\,\N=4 \,\, & \,\,\N=5 \,\,\\*
\hline
& & & & \vspace{-4mm} \\*
\, \, \H_4Ê \hspace{2mm} & \hspace{4mm} U(1) \hspace{4mm} & \hspace{2mm} U(3) \hspace{2mm}  & \hspace{2mm}  U(4)  \hspace{2mm}  & \hspace{2mm} U(5) \hspace{2mm}  \\*
& & & & \vspace{-4mm} \\*
 \hline
& & & & \vspace{-4mm} \\*
\, \, \J_{1}(\N)Ê \hspace{2mm}&\hspace{4mm} \mathds{R}  \hspace{4mm} & \hspace{2mm}  \Ic U(2) \hspace{2mm}  & \hspace{2mm}  \Ic ( SO(2)\times SO(4) )  \hspace{2mm}  & \hspace{2mm} \Ic (U(1) \times SU(2) \times SU(3)) \hspace{2mm}  \\*
& & & & \vspace{-4mm} \\*
 \hline
& & & & \vspace{-4mm} \\*
\, \, \J_2(\N) Ê \hspace{2mm} &\hspace{4mm}   \hspace{4mm} & \hspace{2mm}   \hspace{2mm}  & \hspace{2mm}  ISO(5,1) \hspace{2mm}  & \hspace{2mm} \scal{Ê U(1) \times Spin(5,1)}  \ltimes (S_+ \oplus V ) \hspace{2mm}  \vspace{-4mm} \\*
& & & &  \\*
 \hline
\end{array} \nonumber 
\end{gather}
\begin{centerline} {\small Table IV : Isotropy subgroups $\J_n(\N)\subset\H^*$ 
for pure $\N\leq 5$ supergravities  }\end{centerline}

\vspace{0.4cm}\noindent
Here $IG$ is defined to be the semidirect product of the group $G$ with 
the abelian translation group in the fundamental representation of $G$, 
and $\Ic G$ is defined to be the semidirect product of the group $G$ 
with the Heisenberg group defined as the translation group in the 
fundamental representation of $G$ with a central charge. 

\subsection{$\N=6$ supergravity}

This description is valid for $\N=6$ supergravity if one restricts to the $U(1)\times Spin^*(12)$-orbits of solutions for which the charge matrix satisfies the pure spinor condition. However the decomposition into $SL(2,\mathds{R}) \times Spin^*(12)$-orbits of the solutions is more involved and requires one to consider BPS degrees with respect to both $\N=6$ supergravity and the quaternionic $\N=2$ magic supergravity, as well as the vanishing of the horizon area. Indeed the invariance of the extremality parameter $\varkappa \equiv \sqrt{ c^2 - a^2 }$ with respect to  $SL(2,\mathds{R}) \times Spin^*(12)$ implies that the condition for the horizon area to vanish is left invariant by $SL(2,\mathds{R}) \times Spin^*(12)$. 

The representation under $SL(2,\mathds{R}) \times Spin^*(12)$ of the $\N=6$ charge matrix can be conveniently described by the state (\ref{N6state})
\begin{multline} \left|\C\right> = \Scal{Ê1 +  a^7 a^8  \, \invo }  \biggl(Ê\w + Z_{ij} a^i a^j + \frac{1}{4!} \varepsilon_{ijklmn}\,  \Sigma^{mn}Êa^i a^j a^k a^l \biggr .  \\*\biggl .  + \frac{1}{6!} \varepsilon_{ijklmn}Ê\, Z a^i a^j a^k a^l a^m a^n \biggr) \left| 0 \right > \end{multline}
where $\invo$ is the $ Spin^*(12)$ anti-involution defined on chiral spinors. The action of $\mathfrak{spin}^*(12)$ on $\C$ is defined as for lower $\N$, and the action of $\mathfrak{sl}(2,\mathds{R})$ is defined as follows
\be \delta \left|\C\right> = \frac{1}{2}  \Scal{Êi \lambda \scal{ a^7 a_7  + a^8 a_8 - 1} + \xi  a^7 a^8 - \bar \xi a_7 a_8  }  \left|\C\right> \ .\ee
Using the explicit form of the state, one gets 
\begin{multline} \delta \left|\C\right> = \frac{1}{2} \biggl(  \Scal{Ê1 + a^7 a^8  \, \invo } (-i \lambda) Ê+ \Scal{Ê     \invo  +  a^7 a^8 Ê} \xi \biggr) \biggl(Ê\w + Z_{ij} a^i a^j \biggr .  \\*\biggl . + \frac{1}{4!} \varepsilon_{ijklmn}\,  \Sigma^{mn}Êa^i a^j a^k a^l  + \frac{1}{6!} \varepsilon_{ijklmn}Ê\, Z a^i a^j a^k a^l a^m a^n \biggr) \left| 0 \right > \label{SL2trans} \end{multline}
where we used the fact that $\invo$ is an {\em anti}-involution to exhibit the fact that the $U(1)$ factor $\lambda$ acts as in the lower $\N$ cases. Let us consider first the non-BPS solutions with a non-vanishing horizon area that would be $\ft12$ BPS in $\N=2$ magic supergravity. In this case, the state $\left|\C\right>$ can be moved to a basis in which 
\be  \left|\C\right> = \Scal{Ê1 + a^{7} a^{8}   }  \Bigl(1 + \frac{1}{6!} \varepsilon_{ijklmn}Ê a^i a^j a^k a^l a^m a^n \Bigr) \left| 0 \right > \ .\ee
There is then no way that the generators of $\mathfrak{sl}(2,\mathds{R})$ and $ \mathfrak{spin}^*(12)$ can cancel against each other. The only solution for generators of $\mathfrak{sl}(2,\mathds{R})$ is given by
\be  \xi = - i \lambda \ee
and by the traceless ${\Lambda_i}^j$ for $ \mathfrak{spin}^*(12)$. The isotropy subgroup is thus given in this case by 
\be \J_\gra{0}{1} (6) \cong \mathds{R} \times SU(6) \ .\ee
For solutions with a non-vanishing horizon area which are $\ft16$ BPS with $\N=6$ supergravity, but not BPS in $\N=2$ magic supergravity, the  state $\left|\C\right>$ can be rotated to a basis in which 
\be \left|\C\right> = \Scal{Ê1 + a^7 a^8 \invo  }  \Bigl(1 + \frac{1}{2} \Omega_{AB} a^A a^B \Bigr) \left| 0 \right > \ee
where $\Omega_{AB}$ defines a symplectic form on a subspace $\mathds{C}^2$ of $\mathds{C}^6$. In this case, the non-compact generators of $\mathfrak{sl}(2,\mathds{R})$ must be zero in order to leave the state invariant. The computation of the isotropy subgroup is in fact identical to the case of lower $\N$ and one obtains 
\bea \J_\gra{1}{0} (6)  &\cong& \scal{ÊSU(2) \times U(4) } \ltimes \scal{\ Ê{{ \Yboxdim8pt  {\yng(1)}}} \otimes \yngd\oplus \mathds{R} \ } \CR
&\equiv&   I\hspace{-0.6mm}c\scal{ SU(2) \times U(4) } \ .\eea
For solutions which are $\ft12$ BPS in both $\N=2$ and $\N=6$, and for which the horizon area thus necessarily vanishes, the state takes the form
\bea \left|\C\right> &=& \Scal{Ê1 +  a^7 a^8 } \scal{ 1 + \invo}  \Bigl(1 + \frac{1}{2} \Omega_{AB} a^A a^B \Bigr)  \left| 0 \right > \CR
&=& \Scal{Ê1 + a^7 a^8   }  \Bigl(1 + \frac{1}{2} \Omega_{AB} a^A a^B \Bigr) \Bigl(1 + \frac{1}{4!} \varepsilon_{\bar A\bar B\bar C\bar D} a^{\bar A} a^{\bar B} a^{\bar C} a^{\bar D}  \Bigr)  \left| 0 \right > \ .\eea
Using the same arguments as in the case of lower $\N$, one derives the isotropy subgroup associated to $\N=2$, \ie $\mathds{R}$, and the isotropy subgroup associated to $\N=6$, \ie $  I\hspace{-0.6mm}c\scal{ SU(2) \times SU(4) }$. However, the self-duality property of the state implies that the constraint on  $\Lambda_{\bar A\bar B}$ reduces to 
\be  \Lambda_{\bar A\bar B} - \frac{1}{2} \varepsilon_{\bar A \bar B \bar C \bar D}Ê\Lambda^{\bar C \bar D} = 0 \ee
in such a way that the $SU(4)\cong Spin(6)$ factor is enlarged to $Spin(6,1)$. The product representation of the fundamental of $SU(2)$ and $SU(4)$ is promoted to the $SU(2)$-Majorana representation of $Spin(6,1)$.  The remaining generators of the isotropy subgroup correspond to mixed non-compact transformations of $\mathfrak{sl}(2,\mathds{R})$ and $\mathfrak{spin}^*(12)$. Indeed, one can compute that
\bea  \scal{Êa^7 a^8 -  a_7 a_8 }  \Scal{Ê1 +  a^7 a^8  } | 0 \rangle &=& \Scal{Ê1 + a^7 a^8  } |0 \rangle \CR
\frac{1}{2}  \scal{Ê\Omega_{AB} a^A a^B - \Omega^{AB} a_A a_B }  \Scal{Ê1 + \frac{1}{2} \Omega_{AB} a^A a^B } | 0 \rangle &=& \Scal{1 + \frac{1}{2} \Omega_{AB} a^A a^B } |0 \rangle
\eea
in such a way that 
\be  \Scal{Ê a^7 a^8 - a_7 a_8 -  \frac{1}{2} \Omega_{AB} a^A a^B + \frac{1}{2} \Omega^{AB} a_A a_B}  |\C \rangle = 0 \ .\ee
The commutation relation of this generator with the nilpotent $\mathds{R}$ generator gives 
\begin{multline} \Bigl[Êa^7 a^8 - a_7 a_8  \, , \,  i a^7 a_7 + i   a^8 a_8  - i - i a^7 a^8 - i a_7 a_8  \Bigr] 
\\* = 2 \Scal{Ê i a^7 a_7 + i   a^8 a_8  - i - i a^7 a^8 - i a_7 a_8 } \end{multline}
which defines the Lie algebra of the maximal parabolic subgroup $IGL_+(\mathds{R})$ of $SL(2,\mathds{R})$. This generator commutes with all the other generators, in such a way that the $\ft16$ BPS orbit of solutions that are $\ft12$ BPS in $\N=2$ magic supergravity is
\be \J_\gra{1}{1}(6)  \cong  IGL_+(\mathds{R}) \ltimes I\hspace{-0.6mm}c\scal{ SU(2) \times Spin(6,1) } \ .\ee
As discussed in the preceding section, the charge $Z$ is constrained to be a function of the others for solutions of  $\N=6$ supergravity that preserve at least one third of the supersymmetry charges, and $|\C\rangle$ is then a $Spin^*(12)$ pure spinor. Therefore, we have only one orbit to consider for the $\ft13$ and $\ft12$ BPS solutions. A $\ft13$ BPS charge matrix can be transformed to
\be \C =  \Scal{Ê1 + a^7 a^8  \, \invo }  \, e^{\frac{1}{2} \Omega_{AB} a^A a^B} \, | 0 \rangle \ , \ee
where $\Omega_{AB}$ defines a symplectic form over $\mathds{C}^4$. The same analysis as for lower $\N$ gives that the subgroup 
\be \scal{ SU^*(4) \times SU(2) } \ltimes \scal{\ Ê{{ \Yboxdim8pt  {\yng(1)}}} \otimes \yngd\oplus  {{ \Yboxdim8pt  {\yng(1,1)}}}_- \otimes {\bf 1}Ê\ } \subset Spin^*(12) \ee
is in the isotropy subgroup. None of the generators of $\mathfrak{sl}(2,\mathds{R})$ leave the charge matrix invariant on their own. However, let us consider the generators of $\mathfrak{sl}(2,\mathds{R})$ of the $Spin^*(4)\cong SL(2,\mathds{R}) \times SU(2)$ subgroup of $Spin^*(12)$ acting on the $\mathds{C}^2$ subspace orthogonal to the symplectic form $\Omega_{AB}$. Let $\Omega_{\bar A \bar B}$ be the $SU(2)$ symplectic form on this subspace; since
\be \frac{1}{2}Ê\Omega_{\bar A \bar B} \  \frac{1}{8} \Omega_{[AB}Ê\Omega_{CD]} = \frac{1}{6!} \varepsilon_{\bar A \bar B ABCD} \ , \ee
we have that 
\be  \frac{1}{2} \Omega_{\bar A \bar B} a^{\bar A} a^{\bar B} \,  e^{ \frac{1}{2}Ê\Omega_{AB} a^A a^B } \, | 0\rangle  = \invo \,  e^{ \frac{1}{2}Ê\Omega_{AB} a^A a^B } \, | 0\rangle\ee
and the generators of $\mathfrak{sl}(2,\mathds{R}) \subset Spin^*(4)$ act on $|\C\rangle$ as follows
\begin{multline}  \frac{1}{2} \Scal{Ê i b \scal{Êa^{\bar A} a_{\bar A} - 1} + \frac{1}{2}Ê\zeta \Omega_{\bar A \bar B} a^{\bar A} a^{\bar B} -   \frac{1}{2}Ê\bar \zeta \Omega^{\bar A \bar B} a_{\bar A} a_{\bar B} } \, e^{ \frac{1}{2}Ê\Omega_{AB} a^A a^B } \, | 0\rangle \\*
=  \frac{1}{2} \Scal{Ê- i b + \zeta \invo } \, e^{ \frac{1}{2}Ê\Omega_{AB} a^A a^B } \, | 0\rangle 
\end{multline}
\ie in the same way as do the generators of $\mathfrak{sl}(2,\mathds{R})$ (\ref{SL2trans}). The isotropy subgroup thus also contains the diagonal subgroup of these two $SL(2,\mathds{R})$ subgroups. Since one of these $SL(2,\mathds{R})$ groups lies in $Spin^*(4)$, the nilpotent generators that transform in the fundamental of $SU(2)$ also transform in the fundamental of $SL(2,\mathds{R})$. The isotropy subgroup of $\ft13$ BPS solutions of $\N=6$ supergravity is thus
  \be  \J_\gra{2}{0}(6)  \cong \scal{ SU^*(4) \times SO^*(4) } \ltimes \scal{\ Ê{{ \Yboxdim8pt  {\yng(1)}}} \otimes \yngd\oplus  {{ \Yboxdim8pt  {\yng(1,1)}}}_- \otimes {\bf 1}Ê\ } \ .\ee
For the $\ft12$ BPS solutions of $\N=6$ supergravity, the charge matrix can be transformed to
\be \C =  \Scal{Ê1 + a^7 a^8   }  \, e^{\frac{1}{2} \Omega_{AB} a^A a^B} \, | 0 \rangle  \ee
where $\Omega_{AB}$ is a symplectic form on $\mathds{C}^6$. Again, the same analysis as for lower $\N$ gives the subgroup
\be   SU^*(6) \ltimes  {{ \Yboxdim8pt  {\yng(1,1)}}}_- \subset Spin^*(12) \ee
and the subgroup $\mathds{R} \subset SL(2,\mathds{R})$. Then, following the same argument as for the $\ft16$ BPS solutions, one obtains the mixed solution
\be  \Scal{Êa^7 a^8 - a_7 a_8 - \frac{1}{2}\Omega_{AB} a^A a^B +\frac{1}{2} \Omega^{AB} a_A a_B}  |\C \rangle = 0 \ ,\ee
which defines together with the translation generator the non-semi-simple group $IGL_+(\mathds{R})$. The isotropy subgroup of the $\ft12$ BPS solutions is
\be \J_\gra{3}{1}(6) \cong IGL_+(\mathds{R}) \ltimes  SU^*(6) \ltimes  {{ \Yboxdim8pt  {\yng(1,1)}}}_-  \ .\ee
So far, we have only discussed the isotropy subgroups associated to the various BPS solutions represented by simple charges for which all the central charges that are not saturated vanish (either $|z_\m| =|\w|$ or $z_\m = 0$). However, some solutions define different orbits. This is the case, for instance, for the solutions that are either $\ft16$ BPS in the $\N=6$ theory or $\ft12$ BPS in the corresponding $\N=2$ theory and which have, moreover, a vanishing horizon area. The horizon area can be computed for such BPS solutions by embedding them into maximal supergravity and then using the conjectured formula for the horizon area of BPS black holes (\ref{HorizonArea}). In these cases, the computations shows that the corresponding isotropy subgroups  $\J_{\gra{0}{1}^\circ}(6)$ and  $\J_{\gra{1}{0}^\circ}(6)$ contain an extra $\mathds{R}_+^*$ factor with respect to the generic ones $\J_\gra{0}{1}(6)$ and  $\J_\gra{1}{0}(6)$, and that some compact generators become nilpotent. We do not consider solutions for which the $E_{7(7)}$ invariant is negative valued, since the energy is negative in this case and all the solutions of the corresponding orbit have naked singularities \cite{N8Moduli}.

The $\N=6$ isotropy subgroups are displayed in the following Table.
\begin{gather}
\begin{array}{|c|c|c|}
\hline
\,  \mbox{dim} \,  & \,\, \mbox{$0$-BPS / $\N=2$ \& $\ft n6$-BPS / $\N=6$  }\,\, & \,\,\mbox{$\ft12$-BPS / $\N=2$ \& $\sfrac{n-1}{6}$-BPS / $\N=6$} \,\, \\*
\hline
& &  \vspace{-4mm} \\*
\, 36\,  & \hspace{2mm} U(6) \hspace{4mm} & \hspace{2mm}\hspace{2mm}   \\*
& & \vspace{-4mm} \\*
 \hline
& &  \vspace{-4mm} \\*
\,  36 \,  & \hspace{2mm} \Ic(SU(2)\times U(4)) \hspace{2mm} & \hspace{2mm} \mathds{R} \times SU(6) \hspace{2mm}   \\*
& & \vspace{-4mm} \\*
 \hline
& &  \vspace{-4mm} \\*
\, 37 \,  & \hspace{2mm}IGL_+(1,\mathds{R}) \ltimes \Ic(SU(2)\times Sp(2) \ltimes {{ \Yboxdim6pt  {\yng(1,1)}}}_+)   \hspace{2mm} & \hspace{2mm} IGL_+(1,\mathds{R})  \ltimes Sp(3) \ltimes {{ \Yboxdim6pt  {\yng(1,1)}}}_+  \hspace{2mm}   \\*
& & \vspace{-4mm} \\*
 \hline
& &  \vspace{-4mm} \\*
\, 43 \,& \hspace{2mm} \scal{ SU^*(4) \times SO^*(4) } \ltimes \scal{\ Ê{{ \Yboxdim8pt  {\yng(1)}}} \otimes \yngd\oplus  {{ \Yboxdim6pt  {\yng(1,1)}}}_- \otimes {\bf 1}Ê\ }  \hspace{2mm} & \hspace{2mm}IGL_+(\mathds{R}) \ltimes \Ic(SU(2)\times Spin(6,1)) \hspace{2mm}   \\*
& & \vspace{-4mm} \\*
 \hline
& &  \vspace{-4mm} \\*
\, 52\, & \hspace{4mm} \hspace{4mm} & \hspace{2mm} IGL_+(\mathds{R}) \ltimes  SU^*(6) \ltimes  {{ \Yboxdim6pt  {\yng(1,1)}}}_-  \hspace{2mm}    \vspace{-4mm} \\*
& &  \\*
 \hline
\end{array} \nonumber 
\end{gather}
\begin{centerline} {\small Table V : Isotropy subgroups 
$\J_\gra{n-i}{i}(6)\subset SL(2,\mathds{R}) \times Spin^*(12)$ in 
$\N=6$ supergravity  }\end{centerline}

\subsection{$\N=8$ supergravity} 

The arguments  work the same way in the case of maximal supergravity. Let 
us first discuss the $\ft 1 2$ BPS solutions. Using a $U(8)\subset Spin^*(16)$ transformation, 
one can always reach a charge matrix such that $\w$ and $Z_{ij}$ are 
real and such that $Z_{ij} = \frac{\scriptscriptstyle W}{2} \Omega_{ij}$, 
where $\Omega_{ij}$ defines a symplectic matrix of $\mathds{C}^8$. Using 
a non-compact element of $Spin^*(16)$ one can then fix $\w$ to $1$. As 
for the $\N=1$ to $5$ cases, the $0$-form, the $2$-form and the $4$-form components 
of $\left | \C\right> $ then match with $e^{ \frac12 \Omega_{ij} a^i a^j} 
|0 \rangle$. Moreover, because $\Omega_{ij}$ defines a real symplectic 
form of $\mathds{C}^8$, $e^{ \frac{1}{2} \Omega_{ij} a^i a^j} 
| 0 \rangle$ is real with respect to the anti-involution $\invo$ and matches 
with $\left | \C\right> $ for all form-degree components. 
\be 
|\C\rangle = \exp \left( \frac12 \Omega_{ij} a^i a^j \right) | 0 \rangle \ .
\ee
The computation of the isotropy subgroup works as for lower $\N$, except that 
there is no extra $U(1)$ generator. The $\ft 12$ BPS isotropy subgroup is 
\be 
\J_{4}(8) = SU^*(8) \ltimes   {\Yboxdim8pt   {\yng(1,1)}}_- \ ,
\ee 
which is again non-reductive, with the Lorentz-like subgroup $SU^*(8)$ acting 
on 28 translations $\mathds{R}^{28}$; the latter antisymmetric rank-two 
tensor representation is again real for $SU^*(8)$ (but not for
$SU(8)$).

As discussed in the preceding section, there is no $\ft 3 8$ BPS stationary solution in $\N=8$ supergravity. For both the $\ft 1 8$ and $\ft1 4$ solutions, one can reach a basis such that $\w = 1$ and $2 Z_{ij}$ defines a symplectic form on a $\mathds{C}^2$, respectively $\mathds{C}^4$, subspace of $\mathds{C}^8$. In these cases, $\invo e^{\frac{1}{2} \Omega_{AB} a^A a^B } \left | 0 \right >$ only involves the creation operators $a^{\bar A}$ in such a way that it is orthogonal to $e^{\frac{1}{2} \Omega_{AB} a^A a^B } \left | 0 \right >$. Thus 
\be  \left | \C \right > = ( 1 + \invo )\,   e^{ \frac{1}{2} \Omega_{AB} a^A a^B} \left| 0 \right > \ .\ee
By definition, the generators of  $\mathfrak{spin}^*(16)$ commute with the involution $\invo$, and one gets that the variation of the Majorana--Weyl spinor $\left | \C \right > $ is given by
\begin{multline} \delta   \left | \C \right > = ( 1 + \invo ) \,  \Bigl(  \scal{2\, Ê{\Lambda_A}^C Ê\Omega_{CB}  + \Lambda_{AB}Ê+  \Omega_{AC} \, \Lambda^{CD}Ê \Omega_{DB}  } a^A a^B + \Lambda_{\bar A \bar B} a^{\bar A} a^{\bar B} \Bigr . \\* \Bigl .  + 2 \scal{Ê{\Lambda_{\bar A}}^C \Omega_{CB} +  \Lambda_{\bar A B} } a^{\bar A} a^B +  \Omega_{AB} \Lambda^{AB} - Ê {\Lambda_A}^A - {\Lambda_{\bar A}}^{\bar A}   \Bigr) e^{ \frac{1}{2} \Omega_{EF} a^E a^F }    \left | 0 \right > \ .\end{multline}
In the case of the $\ft 14$ BPS orbit,
\be \Omega_{[AB} \Omega_{CD]} = \frac{1}{3} \varepsilon_{ABCD} \ , \ee
where $\varepsilon_{ABCD}$ defines the $SL(4,\mathds{C})$ invariant epsilon tensor. By counting the degree of the various components with respect to the decomposition under $U(4)\times U(4) \subset U(8)$ one obtains that the only components for which the operator $(1+\invo)$ introduces a further mixing are 
\be (1 + \invo ) \,  \Lambda_{\bar A \bar B} a^{\bar A} a^{\bar B}  \, e^{ \frac{1}{2} \Omega_{EF} a^E a^F }    \left | 0 \right > \ . \ee
Then using the fact that 
\be \invo\  \varepsilon_{ABCD} a^A a^B a^C a^D \left| 0 \right> = \varepsilon_{\bar A \bar B \bar C \bar D } a^{\bar A}Êa^{\bar B}Êa^{\bar C} a^{\bar C}ÊÊ \left| 0 \right> \ , \ee
it follows that the condition $\delta \left | \C \right > = 0 $ gives the equations 
\be\begin{split}
 Ê2\, {\Lambda_{[A}}^C Ê\Omega_{C|B]}  + \Lambda_{AB}Ê+  \Omega_{AC} \, \Lambda^{CD}Ê \Omega_{DB} &= 0 \\
Ê{\Lambda_{\bar A}}^C Ê\Omega_{CB}  + \Lambda_{\bar AB}Ê&= 0 
  \end{split}\hspace{10mm} \begin{split} 
  \Lambda_{\bar A \bar B }  - \frac{1}{2} \varepsilon_{\bar A \bar B \bar C \bar D }Ê\Lambda^{\bar C \bar D} &= 0 \\
\Omega_{AB} \Lambda^{AB} -  {\Lambda_A}^A  -  {\Lambda_{\bar A}}^{\bar A} &= 0 \ .
  \end{split} \label{IsotropyN8} \ee 
The traceless condition and the condition for the $\mathfrak{su}(4)$ generators of the first $U(4)$ factor to leave invariant the symplectic form imply that the maximal compact subgroup of the $\ft 14$ BPS isotropy subgroup is $Sp(2)\times SU(4) \cong Spin(5) \times Spin(6)$. The conditions on the non-compact generators 
\be  \Lambda_{AB}Ê+  \Omega_{AC} \, \Lambda^{CD}Ê \Omega_{DB} = 0 \hspace{10mm}Ê\Omega_{AB} \Lambda^{AB} = 0  \hspace{10mm}    \Lambda_{\bar A \bar B }  - \frac{1}{2} \varepsilon_{\bar A \bar B \bar C \bar D }Ê\Lambda^{\bar C \bar D} = 0 \ee
restrict the parameters to lie in the vector representation of $SO(5)$ and $SO(6)$ respectively. The maximal semi-simple  subgroup of the $\ft 14$ BPS isotropy subgroup is thus $Spin(5,1)\times Spin(6,1)$. As for the lower $\N$ case, the nilpotent generators of the isotropy subgroup lie in the ${\bf 4} \otimes {\bf 4} $ complex representation of $SU^*(4)\times SU(4)$ and in the ${\bf 6}$ representation of $SU^*(4)$. They transform in the ${\bf 32}$ Majorana--Weyl spinor representation of $Spin(5,1)\times Spin(6,1)$ and the vector representation of $SO(5,1)$, respectively. Both the ${\bf 4}$ Weyl representation of $Spin(5,1)$ and the ${\bf 8}$ representation of $Spin(6,1)$ are pseudo-real, but their respective pseudo-anti-involutions permit one to define a real ${\bf 32}$ spinor representation of $Spin(5,1)\times Spin(6,1)$. The $\ft 14$ BPS 
isotropy subgroup is
\be  
\J_{2}(8) =  \scal{ÊSpin(5,1) \times Spin(6,1) } \ltimes 
\scal{\ ({\bf 4} \otimes {\bf 8})_{\mathds{R}}Ê\oplus {\bf 6}Ê\otimes {\bf 1} \ } \ .
\ee
In the case of the $\ft18$ BPS orbit of non-vanishing horizon area, the actions on the two components $e^{\frac{1}{2} \Omega_{AB} a^A a^B } \left | 0 \right >$  and $\invo e^{\frac{1}{2} \Omega_{AB} a^A a^B } \left | 0 \right >$  do not mix, and the equations defining the isotropy subgroup of $Spin^*(16)$ reduce to equations (\ref{Isotropy}), with $\lambda = 0$, since there is no $U(1)$ factor in this case. This slight modification of the equation implies that the $U(\N-2)$ factor of the isotropy subgroup reduces to $SU(\N-2)$ for $\N=8$. As a result, one gets that 
\be 
\J_{1^+}(8) =  \scal{ÊSU(2) \times SU(6) } \ltimes 
\scal{\ Ê{{ \Yboxdim8pt  {\yng(1)}} \otimes \yngd  \oplus  {\bf 1}}\ } \ .
\ee
A representative of a $\ft18$ BPS solution with vanishing horizon area can be parametrised by three positive real numbers $0 < \rho_\un < \rho_\deux < \rho_\trois < 1$ which satisfy $ 1+\rho_\un - \rho_\deux - \rho_\trois = 0$, as follows:
\be  \left | \C \right > = ( 1 + \invo )\,   \Scal{Ê1 +  a^1 a^2 } \Scal{Ê1 +  \rho_\un  \, a^3 a^4  + \rho_\deux  \, a^5 a^6  +  \rho_\trois  \, a^7 a^8 } \ \left| 0 \right > \ .\ee
The generic $\ft18$ BPS isotropy subgroup $\J_{1^+}(8)$ is not modified by the deformation associated to the parameters $\rho_\un,\, \rho_\deux$ and $\rho_\trois$ as long as they satisfy $0 < \rho_\un < \rho_\deux < \rho_\trois < 1$ and $1+\rho_\un - \rho_\deux - \rho_\trois > 0$. The subgroup $SU(2) \times Sp(3) \subset SU(2) \times SU(6)$ remains unchanged for any value of  $1+\rho_\un - \rho_\deux - \rho_\trois$, but the signature of the remaining generators with respect with the Cartan form depends on the sign of $1+\rho_\un - \rho_\deux - \rho_\trois$, in such a way that when the latter is negative, the isotropy subgroup is 
\be 
\J_{1^-}(8) =  \scal{ÊSU(2) \times SU^*(6) } \ltimes 
\scal{\ Ê{{ \Yboxdim8pt  {\yng(1)}} \otimes \yngd  \oplus  {\bf 1}}\ } \ .
\ee
This corresponds to $\ft18$ BPS solutions for which $\lozenge(\w^{-\frac{1}{2}} Z) < 0$. Such solutions  carry a naked singularity and will be disregarded \cite{N8Moduli}.  For $ 1+\rho_\un - \rho_\deux - \rho_\trois = 0$ most of the generators are nilpotent and there is an extra $\mathds{R}_+^*$ invariance of the charge matrix which decreases the dimension of the corresponding orbit by one. The isotropy subgroup of the $\ft18$ BPS solutions of vanishing horizon area is 
 \bea 
\J_{1^\circ} (8) &=&  \scal{Ê\mathds{R}_+^* \times SU(2) \times Sp(3) } \ltimes 
\Scal{\ Ê {\scal{Ê(\yngd \otimes { \Yboxdim8pt  {\yng(1)}} )_+ \oplus  {{ \Yboxdim6pt  {\yng(1,1)}}}_+ }^\ord{1}  \oplus (\yngd \otimes { \Yboxdim8pt  {\yng(1)}} )^\ord{2}_+ \oplus   {\bf 1}^\ord{3}}\ } \CR
&=& \Ic \scal{Ê SU(2) \times ( \mathds{R}_+^* \times Sp(3)) \ltimes {{ \Yboxdim6pt  {\yng(1,1)}}}_+} \ .
\eea
\section{Orbits of stationary single-particle solutions}
Under the action of an element $g\in \G$, the coset representative $\V$ 
transforms as 
\be 
\V \quad \rightarrow \quad \V(g) = g \V h(g,\V) \ ,
\ee
where $h(g,\V)$ is the element of $\H^*$ that permits one to reach the specific 
representative of the class $[ g \V ]$ in the chosen parametrization of the 
coset space $\G / \H^*$. The subgroup of $\G$ preserving the asymptotic 
flatness condition $\V\rightarrow  \mathds{1}$ is thus $\H^*$. As we explained 
in the first section, all the non-extremal asymptotically flat axisymmetric 
stationary single-particle solutions which are regular outside the horizon 
are in the $\H^*$-orbit of some Kerr solution.\footnote{ We recall that these solutions do not exhaust the full set of stationary solutions to the equations of motion. However, all non-extremal solutions lying off $\H^*$ orbits passing through regular Kerr solutions are comprised entirely of singular solutions without horizons. } In the 
following, we will discuss these orbits in detail for all pure supergravity theories. 

In general, both horizon area and surface gravity (hence also the 
associated thermodynamic quantities, \ie entropy and temperature) 
are invariant with respect to the four-dimensional duality group $\G_4$. 
However, neither of them is invariant under the action of the 
three-dimensional group $\H^*$ since the relevant expressions depend 
explicitly on the mass and the NUT charge. Nevertheless it has been 
observed that {\em the product of the horizon area and the surface 
gravity} is equal to the deviation from extremality $4\pi \varkappa$ \cite{OrtinN4}, 
which {\em is} invariant under the action of $\H^*$. This statement is 
still valid for non-extremal multi-black-hole solutions. It turns out that both the horizon area and the surface gravity 
are modified by the presence of other black holes, but their product 
remains equal to $4\pi \varkappa$. 
We should mention that the statistical interpretation of the horizon 
area and the surface gravity in the case of an asymptotically Taub--NUT 
solution is not clear \cite{Gibbons}. One important fact that follows from this invariance is that the horizon area transforms by a non-linear rescaling with respect to the action $\H^*$. Therefore, although the horizon area $A$ is generally not invariant with respect to the action of $\H^*$, the condition $A=0$ is. 

\subsection{Stratified structure of the moduli spaces of charges}

The $\H^*$-orbits of single-particle solutions can be characterised in
terms of the $\H^*$-orbits of the charge matrix $\C$ in $\mathfrak{g} 
\ominus \mathfrak{h}^*$. The decomposition of the set of asymptotically 
flat axisymmetric stationary single-particle solutions, including the 
extremal solutions, which can be obtained as special limits of non-extremal 
ones, can be derived from the decomposition into $\H^*$-orbits of charge
matrices $\C$ satisfying the cubic equation $\C^3 = c^2 \C$ or its 
quintic analogue (\ref{polynom}). The set of such charge matrices 
({\it alias} the {\em moduli space of solutions} of (\ref{cubic}) or
(\ref{polynom})) is a stratified space $\mathcal{M}$, that is, a partially  
ordered union of manifolds 
\be
\mathcal{M} = \bigcup_{n\in I} \mathcal{M}_n \ ,\ee 
where the submanifolds $\mathcal{M}_n$, are such that all their intersections are empty, that is, 
$\mathcal{M}_n \cap \mathcal{M}_m = \emptyset$, and the intersection of the closure of a 
given stratum $\overline{\mathcal{M}_n}$ with another stratum $\mathcal{M}_m$ is either empty or $\mathcal{M}_m$ itself
\be \overline{\mathcal{M}_n}\cap \mathcal{M}_m \ne \emptyset \quad  \Rightarrow \quad \mathcal{M}_m \subset \overline{\mathcal{M}_n} \ .\ee
There is a main stratum $\mathcal{M}_0$, whose closure is $\mathcal{M}$ itself. The stratification is said to be ordered if for any $m$ and $n$ in $I$, either $\mathcal{M}_m \subset \overline{\mathcal{M}_n} $ or $\mathcal{M}_n \subset \overline{\mathcal{M}_m}$.  For an ordered stratification, we label the strata by integers, such that $m>n$ means that  $\mathcal{M}_m \subset \overline{\mathcal{M}_n} $. 

The main stratum $\mathcal{M}_0$ corresponds 
to solutions with $c^2\neq 0$, hence to non-BPS solutions; it has the structure 
\be\label{M0}
\mathcal{M}_0 \, = \, \mathds{R}^*_+\times \H^* / \H_4 \ ,
\ee 
where the coset $\H^*/\H_4$ encodes the gravitational and electromagnetic 
charges for fixed $c^2$, and the extra factor $\mathds{R}^*_+$ corresponds 
to the non-zero values of the BPS parameter $c^2$. Clearly, re-scalings 
of $c$ are {\em not} part of the group $\H^*$; however, as we will show in 
the following section, they are associated to the so-called `trombone 
symmetry' \cite{Active}. Modulo certain conformal diffeomorphisms, the latter can be 
incorporated into the full three-dimensional duality group $\G$, as we 
will show below.

The other strata $\mathcal{M}_n$ with $n \ne 0$ parametrise solutions with 
$c^2 =0$. The charge matrix $\C$ of such strata parametrises stationary non-rotating extremal solutions, like spherically symmetric extremal black holes or multi-black-hole solutions. These strata are $\H^*$-orbits with
\be\label{Mn}
\mathcal{M}_n \, \cong \,  \H^* / \J_n  \, , 
\ee
where the $\J_n =\J_n(\C)$ are the isotropy groups that leave invariant 
the given charge matrix $\C$, and which were analysed in the previous section 
for pure supergravity. We note that the space of single-particle-like stationary 
solutions is likewise a stratified space. It differs from the above moduli space 
of charges only by the extra information not captured by $\C$, namely
the value of the angular momentum parameter $a$, which is restricted 
to lie in the interval $-c\le a \le c$ because we are excluding hyper-extremal 
solutions (the values $a=\pm c$ give extremal Kerr solutions). 

We next show that {\em each $\H^*$-orbit in $\mathcal{M}$ is a Lagrangian
submanifold} of a $\G$-orbit space. For this purpose, we define a larger
isotropy group $\J'_n\equiv\J'_n(\C)\subset\G$ consisting of all transformations
$g\in\G$ leaving invariant the given charge matrix $\C$; clearly $\J_n\subset\J'_n$.
To see that the inclusion
\be
\H^*/\J_n \subset \G/\J'_n
\ee
embeds $\H^*/\J_n$ as a Lagrangian submanifold we introduce the
symplectic form
\be\label{sympl} 
\omega( x, y) \big|_\C  \equiv \trace \C [ \x , \y ]Ê 
\ee
on $\G/\J'_n$. Here, $x$ and $y$ are invariant vector fields $\in T\big(\G/\J'_n\big)$
which coincide with the class of Lie algebra elements 
$[\x]  , \, [\y] \in \mathfrak{g} \, / \,   \j'_n \cong T_\C\big(\G/\J'_n\big)$ at 
$\C\in\G/\J'_n$;\footnote{Here $ \mathfrak{g} \, / \,   \j'_n $ is the class of elements of $\mathfrak{g} $ that become identified when their difference lies in $ \j'_n $.} observe that the r.h.s.\ of (\ref{sympl}) vanishes when $\x$ or $\y$ or both
are in $\j'_n$ and thus it is well-defined on $\mathfrak{g} \, / \,  \j'_n$. On a point $\C\in \H^*/\J_n \subset \G/\J'_n$, since $\C\in\mathfrak{g}\ominus\mathfrak{h}^*$ it follows 
that, if $[\x]$ admits a representative $\x  \in\mathfrak{h}^*$, the 
symplectic form $\omega(x,y)\big|_\C$  is non-zero only if
$[\y]$ admits a non-trivial representative $\y \in  \mathfrak{g}\ominus\mathfrak{h}^*$, which proves that $T_\C\big(\H^*/\J_n\big) \subset T_\C\big(\G/\J'_n\big)$ is isotropic with respect with $\omega |_\C$. Moreover, for any non-trivial representative $\y\in \mathfrak{g}\ominus\mathfrak{h}^*$, $[ \C , \y ]$ is a non-zero element of $\mathfrak{h}^*$ such that there exits $\x \in\mathfrak{h}^*$ for which $\trace \C [Ê\x , \y]  \ne 0$ ( the existence being ensured by the non-degeneracy of the symplectic form $\omega$). Therefore $T_\C\big(\H^*/\J_n\big) \subset T_\C\big(\G/\J'_n\big)$ is Lagrangian with respect with $\omega |_\C$. We conclude that $\H^*/\J_n $ is a Lagrangian submanifold of $\G/\J'_n$ 
with respect to the symplectic form $\omega$.

It is important to emphasise the link between the moduli spaces 
$\mathcal{M}_n$ (for $n\geq 1$) and the nilpotent adjoint orbits of the 
corresponding group, which have been extensively studied by mathematicians \cite{coadjoint}.\footnote{We are grateful to B.~Pioline for having drawn our attention to \cite{coadjoint}.} 
This link was already emphasised in \cite{QuantumAttractors}, and we can now 
state it in a precise way. Although we are interested in real simple Lie algebras $\mathfrak{g}$, the characterisation of their nilpotent orbits requires one to consider the complexification $\mathfrak{g}_\mathds{C}$ of  $\mathfrak{g}$. Define $\mathfrak{N}_{\G_\mathds{C}}$ as the variety of nilpotent elements of $\mathfrak{g}_\mathds{C}$. $\mathfrak{N}_{\G_\mathds{C}}$ is a stratified space and each stratum is a $\G_\mathds{C}$-orbit, where $\G_\mathds{C}$ is the complexification of the 
simple Lie group $\G$,
\be 
\mathfrak{N}_{\G_\mathds{C}} \cong \bigcup_{n \in I_\G} \,  \frac{\G_\mathds{C}}{\I_{\G_\mathds{C}}^\ord{n}} \ .
\ee
where the index-set $I_{\G}$ labels the different isotropy subgroups and thus
the inequivalent orbits. The subspaces
\be \mathfrak{N}_{\G} \equiv \mathfrak{N}_{\mathfrak{G}_\mathds{C}} \cap \mathfrak{g} \hspace{10mm} \mathfrak{N}_{\H_\mathds{C}} \equiv \mathfrak{N}_{\G_\mathds{C}} \cap ( \mathfrak{g}_\mathds{C} \ominus \mathfrak{h}_\mathds{C} ) \ee
are also stratified spaces which decompose into (real)  $\G$-orbits and $\H_\mathds{C}$-orbits respectively. The Kostant--Sekiguchi correspondence \cite{Sekiguchi} states that their 
stratifications are identical since there exists a homeomorphism  \cite{homeoKS}
\be \frac{\mathfrak{N}_\G}{\G} \cong \frac{\mathfrak{N}_{\H_\mathds{C}}}{\H_\mathds{C}}  \label{KS} \ . \ee
Thanks to this homeomorphism, the problem of determining the stratification of the real algebraic variety $\mathfrak{N}_\G$ reduces to the much easier problem of determining the stratification of the complex algebraic variety $\mathfrak{N}_{\H_\mathds{C}}$. 

In supergravity, the charge matrix lies in $\mathfrak{g} \ominus \mathfrak{h}^*$, and we are thus interested in the subvariety $\mathfrak{N}_{\H^*} \subset  \mathfrak{N}_\G$
\be \mathfrak{N}_{\H^*} \equiv  \mathfrak{N}_\G \cap ( \mathfrak{g} \ominus \mathfrak{h}^* ) \ee
which defines the moduli space of charge matrices of (possibly singular) extremal spherically symmetric black hole solutions. As we have just proved, $\mathfrak{N}_{\H^*} $ is in fact a Lagrangian subvariety of $  \mathfrak{N}_\G$ in the sense that each $\H^*$-orbit inside $\mathfrak{N}_{\H^*} $ is a Lagrangian submanifold of a $\G$-orbit inside $ \mathfrak{N}_\G$.  Nevertheless, some $\G$-orbits of  $\mathfrak{N}_\G$ do not contain any $\H^*$-orbit  inside $\mathfrak{N}_{\H^*} $. The $\H^*$-orbits inside $\mathfrak{N}_{\H^*} $ can be classified by a determination of the inequivalent embeddings of $\mathfrak{h}^* \subset \mathfrak{g}$ such that a given representative of the corresponding nilpotent orbit in $\mathfrak{N}_\G$ lies inside $ \mathfrak{g} \ominus \mathfrak{h}^*$. In this way, one can compute the isotropy subgroups of $\H^*$-orbits without knowing explicitly the charge matrix $\C$ of any of its representatives as a function of the conserved charges $\w$ and $Z$. As we shall see, this permits one to show the existence of an $\H^*$-orbit of non-BPS extremal solutions inside $\mathcal{M}$ in both $\N=8$ and $\N=6$ supergravities. 

Among the $\G$ nilpotent orbits, there is a minimal non-trivial nilpotent orbit which is at the boundary of any orbit inside $\mathfrak{N}_\G$. In pure supergravity theories, the minimal $\G$-orbit (\ie  $\G / \scal{Ê\G_4 \ltimes ( \mathfrak{l}_4 \oplus \mathds{R} )}$ in these cases), generically does not contain any $\H^*$-orbit in $\mathfrak{g} \ominus \mathfrak{h}^*$. Only in $\N=6$ and $\N=8$ supergravities do the respective minimal orbits contain $\H^*$-orbits of $\ft12$ BPS charge matrices. The minimal nilpotent orbits seem to be associated to maximally supersymmetric black holes in general. 

Since there is no uniqueness theorem for extremal solutions which would generalise Mazur's theorem for non-extremal solutions, it is natural to enquire whether higher-order orbits of $\mathfrak{N}_{\H^*}$, which do not lie on the boundary of $\mathcal{M}_0$, can correspond to regular extremal solutions of supergravity. There is no such orbit when the theory contains no scalar fields, but there can be many otherwise. 

\subsubsection*{Pure supergravity}

As we have shown in Section \ref{PureSpinorS}, for all supergravity theories with $\N\le5$, all solutions with a vanishing BPS parameter $c=0$ are BPS and the stratification is ordered with respect to the BPS degree. Indeed, $\mathcal{M}$ is then the space of $Spin^*(2\N)$ pure spinors, which admits the following stratification by BPS degree
\be 
\mathcal{M}_0 \cong
\mathds{C}^{\scriptscriptstyle \times } \times \frac{ Spin^*(2\N)}{U(\N)} 
\,  , \ 
\mathcal{M}_n \cong 
\frac{ U(1) \times Spin^*(2\N)}{\scal{ÊSU^*(2n) \times 
U(\N-2n) } \ltimes \scal{\ (Ê{{ \Yboxdim8pt  {\yng(1)}}
\otimes \yngd )\oplus 
{{ \Yboxdim6pt  {\yng(1,1)}}_- \otimes {\bf 1}}}\ }  } 
\ee
such that the last stratum is just a single point $\{0\}$ (the trivial solution). The orbits of  $\ft n\N$ BPS stationary solutions are of dimension $\N^2 -\N+1 - (2n+1) (n - 1) $.

The stratification is more involved in the case of  $\N=6$ supergravity. In this case $\mathcal{M}_\gra{p}{q}$ corresponds to solutions which are $\ft p6$ BPS in $\N=6$ supergravity and $\ft q2$ BPS in the corresponding magic supergravity associated to the quaternions. $\mathcal{M}_\gra{p}{q} \subset\overline{\mathcal{M}}_\gra{r}{s}$ if and only if both $p> r$ and $q>s$, and $\partial \mathcal{M}_\gra{p}{q} = \overline{\mathcal{M}}_{\gra{p}{q}^\circ}$.
\begin{gather}
\mathcal{M}_\gra{0}{0} \cong 
\mathds{R}_+^*\times \frac{ SL(2,\mathds{R})\times  Spin^*(12)}{U(6)}  \CR
\mathcal{M}_\gra{1}{0}  \cong  \frac{ SL(2,\mathds{R})\times  Spin^*(12)}{\Ic(SU(2)\times U(4))} \hspace{20mm} \mathcal{M}_\gra{0}{1} \cong  \frac{ SL(2,\mathds{R})\times  Spin^*(12)}{ \mathds{R} \times SU(6)} \\* \nn
\mathcal{M}_{\gra{1}{0}^\circ}   \cong  \frac{ SL(2,\mathds{R})\times  Spin^*(12)}{ IGL_+(1,\mathds{R}) \ltimes \Ic(SU(2)\times Sp(2) \ltimes {{ \Yboxdim6pt  {\yng(1,1)}}}_+) } \hspace{5mm} \mathcal{M}_{\gra{0}{1}^\circ}  \cong  \frac{ SL(2,\mathds{R})\times  Spin^*(12)}{ IGL_+(1,\mathds{R}) \ltimes Sp(3) \ltimes {{ \Yboxdim6pt  {\yng(1,1)}}}_+ } \CR
\hspace{60mm} \mathcal{M}_\gra{1}{1} \cong  \frac{ SL(2,\mathds{R})\times  Spin^*(12)}{IGL_+(\mathds{R}) \times \Ic(SU(2)\times Spin(6,1))} \CR
\mathcal{M}_\gra{2}{0} \cong \frac{ SL(2,\mathds{R})\times  Spin^*(12)}{\scal{ SU^*(4) \times SO^*(4) } \ltimes \scal{\ Ê{{ \Yboxdim8pt  {\yng(1)}}} \otimes \yngd\oplus  {{ \Yboxdim6pt  {\yng(1,1)}}}_- \otimes {\bf 1}Ê\ }} \hspace{60mm} \CR
\mathcal{M}_\gra{3}{1} \cong  \frac{ SL(2,\mathds{R})\times  Spin^*(12)}{ IGL_+(\mathds{R}) \times  SU^*(6) \ltimes  {{ \Yboxdim6pt  {\yng(1,1)}}}_- } \ .
\end{gather}
This stratification is in agreement with the stratification of $\mathfrak{N}_{E_{7(-5)}}$ \cite{E7strat}, although the latter suggests that there is an additional stratum $\mathcal{M}_{\gra{0}{0}^\circ} $ of dimension $33$ in the boundary of the main stratum $\mathcal{M}_\gra{0}{0}$, 
\be \mathcal{M}_{\gra{0}{0}^\circ} \cong  \frac{ SL(2,\mathds{R})\times  Spin^*(12)}{ Sp(3)  \ltimes {{ \Yboxdim6pt  {\yng(1,1)}}}_+ \times \mathds{R} } \ee
whose boundary is
\be \partial \mathcal{M}_{\gra{0}{0}^\circ}  = \mathcal{M}_{\gra{1} {0}^\circ}  \cup \mathcal{M}_{\gra{0} {1}^\circ} \cup  \mathcal{M}_\gra{1}{1} \cup \mathcal{M}_\gra{2}{0} \cup \mathcal{M}_\gra{3}{1} \ .\ee
 This stratum does indeed exist, and corresponds to non-BPS extremal solutions, such as for example the ones discovered in \cite{nonBPSextrem} within the $STU$ model. We note also that the first strata (corresponding to elements satisfying ${\ad_\C}^5 = 0$) of the nilpotent orbits of $F_{4(4)},\, E_{6(2)}$ and $E_{8(-24)}$ all have the same stratification ordering as those of $E_{7(-5)}$ \cite{MagicStrat}. This suggests that the moduli spaces of all four magic $\N=2$ supergravity theories might have the same stratification, \ie that the quotients $\mathcal{M}/\H^*$ associated to these theories might all be homeomorphic.

The moduli space of solutions to the quintic $\N=8$ characteristic equation decomposes into the strata
\bea 
\mathcal{M}_0 &\cong& \mathds{R}_+^* \times \frac{Spin^*(16)}{SU(8)} 
\quad , \qquad
\mathcal{M}_1 \cong \frac{Spin^*(16)}{  \Ic(ÊSU(2) \times SU(6) ) } \\
&&\mathcal{M}_{1^\circ}  \cong \frac{Spin^*(16)}{ \Ic \scal{ SU(2) \times Ê(\mathds{R}_+^* \times Sp(3) ) \ltimes {{ \Yboxdim6pt  {\yng(1,1)}}}_+}  } \CR
\mathcal{M}_2 &\cong& \frac{Spin^*(16)}{  \scal{ÊSpin(5,1) \times Spin(6,1) }
\ltimes \scal{\ {\bf 4} \otimes {\bf 8}Ê\oplus {\bf 6}Ê\otimes {\bf 1} \ } } 
\quad , \quad
\mathcal{M}_4 \cong \frac{Spin^*(16)}{ ÊSU^*(8) \ltimes  
{ \Yboxdim6pt {\yng(1,1)}}_-}  \nn
\eea
together with the trivial solution $\{0\}$. The ordering $0,\, 1,\, 1^\circ ,\, 2,\, 4$ is in agreement with the stratification of $\mathfrak{N}_{E_{8(8)}}$ \cite{E8strat},  although the latter suggests that there is an additional stratum $\mathcal{M}_{0^\circ} $ of dimension $57$ in the boundary of $\mathcal{M}_0$,
\be \mathcal{M}_{0^\circ} \cong \frac{Spin^*(16)}{ ÊSp(4) \ltimes  
{ \Yboxdim6pt {\yng(1,1)}}_-} \ee
which has the same boundary as $\mathcal{M}_1$.   This stratum does indeed exist, and corresponds to non-BPS extremal solutions. None of the central charges of the solutions lying in this orbit is saturated (\ie $|z_{\mathpzc{m}} |^2 < |\w|^2$), and they all satisfy $\lozenge(\w^{-\frac{1}{2}}Z) < 0$.

Let us compare these moduli spaces with the moduli spaces of  $\ft12$ and $\ft14$ BPS static black holes \cite{N8Moduli} (\ie with vanishing NUT charge)
\bea \mathcal{M}_4^{\scriptscriptstyle \rm static} &\cong& \frac{E_{7(7)}}{E_{6(6)} \ltimes \boldsymbol{27}}  \cong \mathds{R}_+^* \times  \frac{SU(8)}{Sp(4)}  \CR
 \mathcal{M}_2^{\scriptscriptstyle \rm static} &\cong& \frac{E_{7(7)}}{ Pin(5,6) \ltimes ( \boldsymbol{32} \oplus \mathds{R} )}   \cong \mathds{R}_+^*  \times \mathds{R}_+ \times  \frac{SU(8)}{Sp(2)\times Sp(2)} \eea 
where the $E_{7(7)}$ coset spaces correspond to orbits of the active duality group \cite{ActiveModuli}. Note that the active $E_{7(7)}$ transformations on solutions with non-vanishing NUT charge do not preserve the BPS degree in general, so that there is no well-defined action of the active duality group $E_{7(7)}$ on the strata $\mathcal{M}_2$ and $\mathcal{M}_4$. The fact that the action does preserve the BPS degree for static solutions is related to the fact that the $\ft14$ BPS condition is associated to the vanishing of the quartic $E_{7(7)}$ invariant $\lozenge(Z)$ for asymptotically Minkowskian solutions, whereas it is related to the vanishing of $\lozenge(\w^{-\frac{1}{2}} Z)$ in general. These strata are therefore non-trivial fibre bundles with respect to 
the Ehlers $U(1)$:
\be
\begin{array}{ccc} U(1) & \rightarrow &  \frac{Spin^*(16)}{ ÊSU^*(8) \ltimes  
{ \Yboxdim4pt {\yng(1,1)}}_-}  \\*
  & & \downarrow \, \, \\* & & \frac{E_{7(7)}}{E_{6(6)} \ltimes \boldsymbol{27}} \end{array}\hspace{15mm}
  \begin{array}{ccc} U(1) & \rightarrow & \frac{Spin^*(16)}{  (ÊSpin(5,1) \times Spin(6,1) )
\ltimes (\, {\bf 4} \otimes {\bf 8}Ê\oplus {\bf 6}Ê\otimes {\bf 1} \, ) }  \\*
  & & \downarrow \, \, \\* & &  \frac{E_{7(7)}}{ Pin(5,6) \ltimes ( \boldsymbol{32} \oplus \mathds{R} ) }  \end{array} \ .                          
  \ee 
It follows that there is no action of $E_{7(7)}$ on $\mathcal{M}_2$ and $\mathcal{M}_4$  that would  agree on a fixed $SU(8)$ subgroup, with the action of  $Spin^*(16)$. In fact, this would be inconsistent since their closure would then generate a well-defined action of $E_{8(8)}$ on the $29$ (respectively $46$) dimensional  strata whereas the minimal representation of $E_{8(8)}$ is $57$-dimensional \cite{PseudoConform}. Although there is no $29$-dimensional representation of $E_{8(8)}$, the minimal unitary representation of $E_{8(8)}$ acts on the space of functions defined on a $29$-dimensional Lagrangian submanifold of the $56$-dimensional minimal adjoint orbit \cite{minimalEisenstein,GKN}, which we have just proved to be diffeomorphic to $\mathcal{M}_4$. We will come back to this observation when we discuss the nilpotency degree of the charge matrix on each stratum. 

The dimensions of the various strata of pure supergravity theories are summarised in the following Table:
 
\begin{gather}
\begin{array}{|c|c|c|c|c|c|c|}
\hline
  & \,\, \N=2  \,\, & \,\,\N=3 \,\,  & \,\,\N=4 \,\, & \,\,\N=5 \,\, & \, \, \N=6 \,\, &\, \, \N=8 \,\, \\*
\hline
& & & & & & \vspace{-4mm} \\*
{\rm dim} ( \mathcal{M}_0) Ê \hspace{2mm} &\hspace{4mm} 4  \hspace{4mm} & \hspace{2mm} 8  \hspace{2mm}  & \hspace{2mm}  14  \hspace{2mm}  & \hspace{2mm} 22  \hspace{2mm}  & \hspace{2mm} 34  \hspace{2mm}   & \hspace{2mm} 58  \hspace{2mm}   \\*
& & & & & & \vspace{-4mm} \\*
 \hline
& & & & & &\vspace{-4mm} \\*
{\rm dim} ( \mathcal{M}_1) Ê \hspace{2mm} &\hspace{4mm} 3  \hspace{4mm} & \hspace{2mm} 7  \hspace{2mm}  & \hspace{2mm}  13  \hspace{2mm}  & \hspace{2mm} 21  \hspace{2mm} & \hspace{2mm} 33  \hspace{2mm}  & \hspace{2mm} 57  \hspace{2mm}   \\*
& & & & & & \vspace{-4mm} \\*
 \hline
& & & &  & &\vspace{-4mm} \\*
{\rm dim} ( \mathcal{M}_{1^\circ} ) Ê \hspace{2mm} &\hspace{4mm}   \hspace{4mm} & \hspace{2mm}   \hspace{2mm}  & \hspace{2mm}    \hspace{2mm}  & \hspace{2mm}   \hspace{2mm} & \hspace{2mm} 32  \hspace{2mm}  & \hspace{2mm} 56  \hspace{2mm}   \\*
& & & & & & \vspace{-4mm} \\*
 \hline
& & & &  & &\vspace{-4mm} \\*
{\rm dim} ( \mathcal{M}_2) Ê \hspace{2mm} &\hspace{4mm}      \hspace{4mm} & \hspace{2mm}    \hspace{2mm}  & \hspace{2mm}  8  \hspace{2mm}  & \hspace{2mm} 16   \hspace{2mm} & \hspace{2mm} 26  \hspace{2mm} & \hspace{2mm} 46   \hspace{2mm} \\*
& & & & & & \vspace{-4mm} \\*
 \hline
& & & &  & &\vspace{-4mm} \\*
{\rm dim} ( \mathcal{M}_4) Ê \hspace{2mm} &\hspace{4mm}      \hspace{4mm} & \hspace{2mm}    \hspace{2mm}  & \hspace{2mm}     \hspace{2mm}  & \hspace{2mm}     \hspace{2mm} & \hspace{2mm} 17  \hspace{2mm} & \hspace{2mm} 29   \hspace{2mm} \vspace{-4mm} \\*
& & & &  & & \\*
 \hline
\end{array} \nonumber 
\end{gather}
\begin{centerline} {\small Table VI : Dimensions of strata in pure supergravity} \end{centerline}

It follows from the cubic equation (or its quintic analogue) that a charge matrix of $\mathcal{M}_1$ satisfies $\C^3 = 0$ (or $\C^5= 0$ for $E_8$). It turns out that the order of the stratum $n$ is related to the nilpotency degree of the charge matrix in general and thus that for pure supergravity theories, the BPS degree of the solutions is characterised in a $\G$ invariant way by the nilpotency degree of the charge matrix. For $\N=2,\, 3$ the condition $\C^2 = 0 $ implies that the charge matrix vanishes and that $\mathcal{M}_1$ is the last non-trivial stratum. As we will see in Section \ref{N4sec}, for $\N=4$ supergravity, $\C^2 = 0$ on $\mathcal{M}_2$.  To summarise briefly, we have for low values of $\N$ that
\begin{gather}
\begin{array}{|c|c|c|c|}
\hline
  & \,\, \N=2  \,\, & \,\,\N=3 \,\,  & \,\,\N=4 \,\, \\*
\hline
& & & \vspace{-4mm} \\*
\  \mathcal{M}_1Ê \hspace{2mm} & \hspace{2mm} \C^3 = 0   \hspace{2mm}  & \hspace{2mm}  \C^3 = 0  \hspace{2mm}  & \hspace{2mm} \C^3 = 0   \hspace{2mm}  \\*
& & & \vspace{-4mm} \\*
\hline
 & & & \vspace{-4mm} \\*
\  \mathcal{M}_2 Ê \hspace{2mm} & \hspace{2mm}  \hspace{2mm}  & \hspace{2mm}   \hspace{2mm}  & \hspace{2mm} \C^2 = 0  \hspace{2mm}  \vspace{-4mm} \\*
 & & &  \\*
 \hline
\end{array} \nonumber
\end{gather}
\begin{centerline} {\small Table VII : Nilpotency degree of charge matrices 
for $\N= 2,\, 3 ,\, 4$} \end{centerline}

\noindent
For $\N\ge 5$ supergravity, the nilpotency degree in the fundamental 
representation of $\mathfrak{e}_{6(-14)},\, \mathfrak{e}_{7(-5)} $ or 
$\mathfrak{e}_{8(8)}$ is not enough to characterise the degree of the 
strata. It is then useful to consider $\N=4$ supergravity as a consistent 
truncation of $\N=5$ supergravity,  both of them as consistent truncations 
of $\N=6$ supergravity, and all three of them as consistent truncations 
of $\N=8$ supergravity. These truncations can be understood from the 
decompositions of $\mathfrak{e}_{8(8)}$ 
\bea 
\mathfrak{e}_{8(8)} &\cong& \mathfrak{su}(2) \oplus \mathfrak{e}_{7(-5)} \oplus \scal{Ê\boldsymbol{2} \otimes \boldsymbol{56}}_{\mathds{R}} \\*
&& \hspace{-7mm} \cong  \mathfrak{su}(2) \oplus  \scal{Ê\mathfrak{u}(1) \oplus \mathfrak{e}_{6(-14)} \oplus \boldsymbol{27} } \oplus \scal{Ê\boldsymbol{2} \oplus Ê\boldsymbol{2} \otimes \boldsymbol{27} } \CR
&& \hspace{-10mm} \cong \mathfrak{su}(2) \oplus \mathfrak{u}(1) \oplus \scal{Ê\mathfrak{u}(1) \oplus \mathfrak{so}(2,8) \oplus {\bf 16}_+ }Ê\oplus \scal{Ê{\bf 10} \oplus  {\bf 16}_- \oplus {\bf 1} }Ê \oplus \boldsymbol{2} \oplus Ê\boldsymbol{2} \otimes \scal{Ê {\bf 10} \oplus  {\bf 16}_- \oplus {\bf 1} } \nn
\eea
where the representations are complex when unspecified. It follows that 
a solution of $\N=5,\, 6$ supergravity, corresponding upon embedding 
into $\N=8$ supergravity to a solution with an $\mathfrak{e}_{8(8)} $ 
charge matrix satisfying $\C^n = 0$, has an $\mathfrak{e}_{6(-14)}$
or $\mathfrak{e}_{7(-5)} $ charge matrix that satisfies both 
 \be 
\C^n = 0 \quad \mathrm{and}\quad {\ad_\C}^n = 0 \ .
\ee
The condition $\C^3= 0 $ on $\mathcal{M}_1$ implies ${\ad_\C}^5 = 0 $. For $\ft14$ BPS solutions in $\N=8$ supergravity, it is convenient to consider the case for which they can be understood as $\ft12$ BPS solutions in $\N=4$ supergravity. The $\mathfrak{spin}(2,8)$  charge matrix then satisfies $\C^2= 0$ in the spinor representations, which implies ${\ad_\C}^3 = 0$. However, one checks that the charge matrix is not nilpotent in the vector representation $[Ê\C , [ \C , \Upgamma^{\mathpzc{M}} ] ] \ne 0 $.  Since the fundamental representation of $E_{6(-14)}$ decomposes into the direct sum of the antichiral spinor representation, the vector, and the trivial representation with respect to $\mathfrak{spin}(2,8)$, it follows that the charge matrix of $\ft25$ BPS solutions of $\N=5$ supergravity satisfy both $\C^3= 0$ and ${\ad_\C}^3 = 0$, but $\C^2 \ne 0$. The same property holds then for charge matrices of the $\ft14$ BPS solutions of $\N=8$ supergravity and for the elements of $\mathcal{M}_2 = \mathcal{M}_\gra{2}{0} \cup \mathcal{M}_\gra{1}{1}$ in $\N=6$ supergravity. 

One computes that the $\ft12$ BPS solutions of $\N=6$ supergravity have charge matrices which satisfy $\C^2= 0$, from which it follows that ${\ad_\C}^3 = 0$, and so the  $\ft12$ BPS solutions of $\N=8$ supergravity have charge matrices which satisfy $\C^3= 0$. Note finally that $\C^2 =0$ implies $\C=0$ for $\mathfrak{e}_{8(8)} \ominus \mathfrak{spin}^*(16)$, and therefore the nilpotency degree of the charge matrix in the adjoint representation does not disentangle the $\ft12$ BPS solutions from the $\ft14$ BPS ones. It is useful then to consider the embedding of $\N=4$ supergravity coupled to six vector multiplets inside maximal supergravity. The latter can be understood from the decomposition 
\be 
\mathfrak{e}_{8(8)} \cong \mathfrak{spin}(8,8) \oplus S_+ \ .
\ee
Both the $\ft14$ and the $\ft12$ BPS solutions of $\N=8$ supergravity that are 
also $\ft12$ BPS solutions of $\N=4$ supergravity coupled to six vector 
multiplets have charge matrices which are nilpotent in the spinor 
representation ${\C_{S_-}}^2 = 0$. As it will be explained in the final section, 
the difference between $\ft14$ and $\ft12$ BPS solutions of $\N=8$ 
supergravity is characterised in $\N=4$ supergravity by the fact 
that the $\mathfrak{spin}(8,8)$ charge matrix corresponding to the 
latter are also nilpotent in the vector representation $[Ê\C_{S_-} ,[ \C_{S_-} ,\Upgamma^{\mathpzc{M}} ] ] = 0 $, whereas the $\ft14$ BPS ones are not. 

In order to characterise this $\ft14 / \ft12$ difference in maximal supergravity, one has to consider (for example) the charge matrix in the $\bf 3875$ representation of $\mathfrak{e}_{8(8)}$ that arises in the decomposition of the rank two symmetric tensor of the adjoint representation. As well as the adjoint representation, the $\bf 3875$ is five-graded with respect to the subgroup $SL(2,\mathds{R}) \times E_{7(7)}$ (see appendix \ref{groups}), therefore the quintic characteristic equation is also valid in the $\bf 3875$ representation. It follows that the BPS charge matrix satisfies  ${\C_{\scriptscriptstyle \bf 3875}}^5 = 0$. The $\bf 3875$ of $E_{8(8)}$ decomposes into the following representations of $Spin(8,8)$ \cite{3875}
\be { \bf 3875} \cong  ( V \otimes V )_{\Yboxdim4pt  {\yng(2)}}  \oplus (S_- \otimes S_- )_{\Yboxdim3pt  {\yng(1,1,1,1)}} \oplus ( V \otimes S_-)_{\scriptscriptstyle \bf 1920} \ee
The action of $\C$ in the tensor product representation
\be \C_{\scriptscriptstyle S_- \otimes S_-} \equiv \mathds{1} \otimes \C_{S_-} + \C_{S_-} \otimes \mathds{1} \ee
to the third power
\be  { \C_{\scriptscriptstyle S_- \otimes S_-}}^3 = \mathds{1} \otimes {\C_{S_-}}^3 + 3 \C_{S_-} \otimes {\C_{S_-}}^2 + 3 {\C_{S_-}}^2 \otimes \C_{S_-} + {\C_{S_-}}^3 \otimes \mathds{1} \ee
vanishes if ${\C_{S_-}}^2 =0$. Then if both ${\C_{S_-}}^2 = 0$ and ${\C_{\scriptscriptstyle V}}^2 = 0$, it follows in the same way that
\be {\C_{\Yboxdim4pt  {\yng(2)}}}^3 = {\C_{\Yboxdim3pt  {\yng(1,1,1,1)}}}^3 = {\C_{\scriptscriptstyle  \bf 1920}}^3 = 0 \ .\ee
The charge matrices associated to $\ft12$ BPS solutions of $\N=8$ supergravity thus satisfy that 
\be {\C_{\scriptscriptstyle \bf 3875}}^3 = 3 \scal{Ê\C \otimes \C^2 + \C^2 \otimes \C}_{\scriptscriptstyle {\bf 3875}\otimes {\bf 3875}} =  0 \ .\ee
However, if  ${\C_{S_-}}^2= 0$ but ${\C_{\scriptscriptstyle V}}^2 \ne 0 $, 
\be  {\C_{\Yboxdim4pt  {\yng(2)}}}^4 = 6 \scal{Ê{\C_{\scriptscriptstyle V}}^2 \otimes {\C_{\scriptscriptstyle V}}^2 }_{\Yboxdim4pt  {\yng(2)} \otimes \Yboxdim4pt  {\yng(2)}} \ne 0 \ .\ee
Therefore, the charge matrices associated to $\ft14$ BPS solutions of $\N=8$ supergravity are such that ${\C_{\scriptscriptstyle \bf 3875}}^4 \ne 0$.

To summarise, we have that
\begin{gather}
\begin{array}{|c|c|c|c|}
\hline
 & \,\, \N=5  \,\,  & \,\, \N=6  \,\, & \,\,\N=8 \,\,  \\*
\hline
& & &  \vspace{-4mm} \\*
\  \mathcal{M}_1Ê \hspace{2mm} &\hspace{4mm} \C^3 = 0  \hspace{5mm} {\ad_\C}^5 = 0   \hspace{4mm} & \hspace{2mm} \C^3 = 0  \hspace{5mm} {\ad_\C}^5 = 0   \hspace{2mm}  & \hspace{2mm}  \C^5 = 0   \hspace{5mm} {\C_{\scriptscriptstyle \bf 3875}}^5 = 0  \hspace{5mm}   \lozenge > 0 \hspace{2mm}    \\*
& & &  \vspace{-4mm} \\*
\hline
& & &  \vspace{-4mm} \\*
\  \mathcal{M}_{1^\circ} Ê \hspace{2mm} &\hspace{4mm}    \hspace{5mm}   \hspace{4mm} & \hspace{2mm} \C^3 = 0  \hspace{5mm} {\ad_\C}^4 = 0   \hspace{2mm}  & \hspace{2mm}  \C^4 = 0   \hspace{5mm} {\C_{\scriptscriptstyle \bf 3875}}^5 = 0    \hspace{5mm}   \lozenge = 0 \hspace{2mm}    \\*
& & &  \vspace{-4mm} \\*
\hline
& & &  \vspace{-4mm} \\*
\  \mathcal{M}_2 Ê \hspace{2mm} &\hspace{4mm}  \C^3 = 0  \hspace{5mm} {\ad_\C}^3 = 0   \hspace{4mm} & \hspace{2mm}  \C^3 = 0  \hspace{5mm} {\ad_\C}^3 = 0 \hspace{2mm}  & \hspace{2mm}  \C^3 = 0     \hspace{5mm} {\C_{\scriptscriptstyle \bf 3875}}^5 = 0  \hspace{5mm}   \lozenge =0 \hspace{2mm}    \\*
& & &  \vspace{-4mm} \\*
 \hline
& & &  \vspace{-4mm} \\*
\  \mathcal{M}_4 \hspace{2mm} &\hspace{4mm}  \hspace{4mm} & \hspace{2mm}  \C^2 = 0  \hspace{5mm} {\ad_\C}^3 = 0 \hspace{2mm}  & \hspace{2mm} \C^3 = 0  \hspace{5mm} {\C_{\scriptscriptstyle \bf 3875}}^3 = 0  \hspace{5mm}  \lozenge =0   \hspace{2mm}  \vspace{-4mm} \\*
& & &   \\*
 \hline
\end{array} \nonumber 
\end{gather}
\begin{centerline} {\small Table VIII : Nilpotency degree of the charge matrices for $\N= 5,\, 6,\, 8$} \end{centerline}
The conjectured additional stratum $\mathcal{M}_{0^\circ}$ is in the same complex orbit of $E_8$ as $\mathcal{M}_1$, and the corresponding charge matrix thus satisfy the same nilpotency condition. However, the $E_{7(7)}$ invariant is strictly negative in this case. Such non-BPS solutions would correspond to particular values of the conserved charges for which the purely gravitational contribution to horizon area cancels exactly the one associated to central charges of negative $E_{7(7)}$ invariant. 

Of course these nilpotency conditions also define the corresponding nilpotent orbits in $\mathfrak{N}_\G$. As we have explained in this section, the moduli spaces $\mathcal{M}_n$ are Lagrangian submanifolds of the corresponding orbits in $\mathfrak{N}_\G$, with respect to the symplectic structure associated to the Lie algebra. The link between extremal black hole solutions of maximal supergravity and these nilpotent orbits was already noticed in \cite{QuantumAttractors}. It turns out that the representations of $E_{8(8)}$ on the nilpotent orbits of $\mathfrak{N}_{E_{8(8)}} $ lead to unitary representations of $E_{8(8)}$ on the space of functions supported on Lagrangian submanifolds  (see \cite{E82} for the case of $E_{8(-24)}$). There have been speculations that such ``quantised'' representations of $E_{8(8)}$ would play a role in the quantisation of black holes \cite{QuantumAttractorsF}. It is rather natural to conjecture that there exist unitary representations of the group $\G$ on the moduli spaces $\mathcal{M}_n$ which are induced by the adjoint action of $\G$ on the corresponding nilpotent orbits of $\mathfrak{N}_\G$ in which $\mathcal{M}_n$ can be embedded as Lagrangian submanifolds. The associated $\G$ symmetric quantum mechanics on the moduli spaces of extremal spherically symmetric black holes might permit one to compute non-perturbative corrections to the action defining the stationary equations of motion of supergravity theories.

\subsection{Active duality transformations and parabolic cosets}
\label{Active1section}
Unlike the elements of the divisor subgroup $\H^*\subset \G$, a general element  $g \in \G$ does not in general preserve asymptotic conditions through the standard non-linear action. Nevertheless, for $d\ge 4$, it is possible to 
define an action of the whole duality group, different from the 
standard non-linear action, which preserves asymptotic conditions in such 
a way that the action on electromagnetic charges is the same as the standard non-linear action \cite{Active}. 
\subsubsection*{Action of the four-dimensional duality group $\G_4$}
In four dimensions the electromagnetic 
charges transform in a representation $ \mathfrak{l}_4$  
of the duality group $\G_4$. Given any $g\in\G_4$ and any particular set 
$Z\in\mathfrak{l}_4$ of such charges, there exists a Borel subgroup 
$\mathfrak{B}_Z\subset\G_4$ that leaves $Z$ conformally invariant  
(that is, invariant up to a factor), 
\be\label{gZ} 
g \, Z =\lambda_{(g,Z)} \, Z  \hspace{10mm}Ê\lambda_{(g,Z)}  
\in \mathds{R}_+^* 
\ee
and which is big enough to act transitively on the symmetric space 
$\G_4 / \H_4$. Furthermore, there is a distinguished generator 
$\boldsymbol{z}\in\mathfrak{B}_Z$ such that any element of $\mathfrak{B}_Z$ 
decomposes as the product of an element $\exp(\ln\lambda \,\boldsymbol{z} )$ 
and an element that leaves invariant the charge $Z$, such that 
$\mathfrak{B}_Z \cong \mathds{R}_+^* \ltimes \mathfrak{B}_{0 \, Z}$.
By the Iwasawa theorem, we can represent $g$ in the form
\be\label{g1} 
g  = u_{(g,Z)}  \, \exp\Big( \ln \lambda_{(g,Z)} \, \boldsymbol{z}  \Big) \, 
b_{(g, Z)}  
\ee
with $u_{(g,Z)}\in\H_4$ and $b_{(g,Z)}\in\mathfrak{B}_{0\,Z}$. Of the 
three factors in (\ref{g1}), only the first leaves invariant the asymptotics 
of the scalar fields. However, due to the invariance of $Z$ under the last 
factor, we need only worry about implementing the action of the middle 
(scaling) operator in a way compatible with the asymptotics. This is what 
the so-called `trombone symmetry' is needed for.

As originally defined in \cite{Active}, the trombone symmetry is a symmetry 
of the equations of motion of any pure supergravity in any dimension, but 
it is not a symmetry of the action. It  acts on the fields as a rescaling 
of the various tensor fields with a weight given by their rank; on the 
metric, the vectors and the scalars it thus acts as
\be\label{Trombone}
g_{\mu\nu}(x) \rightarrow \lambda^2 \, g_{\mu\nu}(x) 
\hspace{10mm}ÊA_\mu(x) \rightarrow \lambda A_\mu (x)
\hspace{10mm}Ê\phi(x) \rightarrow \phi (x) \ .
\ee
In other words, this symmetry acts like a Weyl transformation with 
a constant parameter $\lambda$. By the diffeomorphism invariance of the
theory, the above action is equivalent to a coordinate rescaling $\varphi(x)
\rightarrow \varphi(\lambda^{-1}x)$ on all fields
{\em without} rescaling the various tensor fields according  
to their rank. By definition, this compensated trombone transformation 
preserves the asymptotic behaviour of the solution, and acts on the charge 
$Z$ by a rescaling, precisely as in (\ref{gZ}). Consequently, the action
of an element $g\in \G_4$ of the active duality group on a solution with
charge $Z$, is defined, via the Iwasawa decomposition (\ref{g1}), as the 
successive action of the compensated trombone symmetry of parameter 
$\lambda_{(g,Z)}$ and the standard non-linear action of the element 
$u_{(g,Z)}\in\H_4 $. By construction, the action of the active duality 
group preserves the asymptotic behaviour of the solution and acts on 
the charge $Z$ as in (\ref{gZ}). However, it does not preserve the number 
of preserved supersymmetry charges in general. Nevertheless, non-supersymmetric solutions 
remain non-supersymmetric under the action of the active duality group 
$\G_4$. Although (\ref{gZ}) would seem to suggest that one can take  
$\lambda\rightarrow 0$, this limit is {\em not} in the orbit space: 
the Iwasawa decomposition (\ref{g1}) holds for {\em any} element $g\in\G_4$ 
with {\em non-zero} $\lambda >0$. In other words, {\em the group $\G_4$  does not mix BPS and non-BPS solutions}. As we will see below this is a 
crucial difference with respect to the action of the three-dimensional duality 
group $\G$ whose maximal subgroup $\H^*$ is non-compact.

From the above discussion, it follows that the $\G_4$-orbits are of the form
\be\label{G4B}
\frac{\G_4 }{\mathfrak{B}_{0\,Z}} \cong \mathds{R}_+^* \times 
\frac{\H_4}{\mathfrak{B}_{0\,Z} \cap \H_4} \ .
\ee
The fact that these orbits take the form of {\em parabolic cosets} over
the group $\G_4$ explains why we have a proper group action of the full 
group $\G_4$ on them. Since the active transformations act on the 
charges linearly, one can furthermore restrict the action of $\G_4$ to 
an arithmetic subgroup that preserves the Dirac quantisation condition 
and acts linearly on the lattice of quantised charges \cite{Active}. 
For maximal $\N=8$ supergravity, the parabolic stability groups 
$\mathfrak{B}_{0\,Z} \subset E_{7(7)}$ of the 56 electromagnetic charges
and their $E_{7(7)}$-orbits were analysed and classified in \cite{N8Moduli}.

\subsubsection*{Action of three-dimensional duality group $\G$}
We now wish to generalise this construction to three dimensions in
such a way that an action of the full duality group $\G$ can be
implemented on the orbits. In three dimensions, the charges are associated 
to the scalar fields themselves, and they transform in the adjoint 
representation of $\G$. In the adjoint representation, the subgroup of 
$\G$ that leaves a given element of its Lie algebra $\mathfrak{g}$ 
conformally invariant (the would-be analogue of $\mathfrak{B}_0$) is 
not big enough to act transitively on $\G / \H^*$. However, as we are 
going to see, one can nevertheless generalise the concept of active 
transformations to three dimensions. There are several new features 
and subtleties here, which we will now explain in turn. 

From the five-graded decomposition of $\mathfrak{g}$, one can define a 
maximal parabolic subgroup $\P\subset\G$, whose Lie algebra $\mathfrak{p}$ 
consists of all generators with non-negative gradation, \ie 
\be 
\mathfrak{p} \cong {\bf 1}^\ord{0} \oplus \mathfrak{g}_4 ^\ord{0}  
\oplus \mathfrak{l}_4^\ord{1} \oplus {\bf 1}^\ord{2} \ .
\ee
The gradation is defined with respect to the generator $\boldsymbol{h}\in 
\mathfrak{g}$, and $\P \cong \mathds{R}_+^* \ltimes \P_0$ where 
$\P_0\subset\P$ is the subgroup generated by 
\be\label{p0} 
\mathfrak{p}_0 \cong  \mathfrak{g}_4 ^\ord{0}  \oplus 
\mathfrak{l}_4^\ord{1} \oplus {\bf 1}^\ord{2} 
\ee
from which the generator $\boldsymbol{h}$ has been omitted.  
The maximal parabolic subgroup $\P$ can be associated to the 
charge matrix $\C = c\boldsymbol{h}$ similarly to the way that the Borel subgroup 
$\mathfrak{B}_Z\subset\G_4$ can be associated to a given charge $Z$ in higher 
dimensions (we assume $c>0$ for the moment). By contrast, the adjoint 
action of  $\P_0$ does not leave the generator $\boldsymbol{h}$ 
invariant, but only its subgroup $\G_4$ does: from the four-dimensional 
point of view, a solution associated to the charge matrix $\C = c 
\boldsymbol{h}$ is purely gravitational, while the action of the $\G_4$ 
subgroup only shifts the scalar fields by constants.

We use the common convention that the $\G/\H^*$ coset representative 
$\mathcal{V}$ is defined as a function on the parabolic subgroup $\P$, 
for which the $\G_4$ component is defined to be a given representative 
of a coset element $\G_4 / \H_4$. Then the action of an element $p\in\P$ 
on $\mathcal{V}$ only requires a right compensating transformation 
$h_4\in\H_4\subset\H^*$
\be 
 \V(p) = p\,  \V\,  h_4(p,\V) \ ,
\ee
needed to compensate for the component of $p$ lying in $\G_4$.
It follows that the generators of $ \mathfrak{l}_4^\ord{1} $ act on the 
electromagnetic scalars by constant shifts. The latter decompose into two 
subsets. Half of them act on the scalars arising from the time components of the 
Maxwell one-forms as global gauge transformations\footnote{For a non-zero 
  NUT charge, the timelike isometry orbits are compact and there is a 
  topological quantisation condition on the parameter $\alpha$. However we 
  will interpret the action of these generators as large gauge transformations 
  when acting on a solution with charge matrix $\C = c \h$ for which the 
  timelike isometry orbits are non-compact and $\alpha$ can then take arbitrary 
values.}  
\be 
A  + i \alpha d t = e^{-i\alpha \, t } \scal{Êd + A}  e^{i \alpha \, t } \ .
\ee 
The other half correspond to shifts of the integration constants appearing in the definitions of the scalar fields dual to three-dimensional one-forms associated to the dimensionally reduced Maxwell fields. We conclude 
that the action of the generators of $ \mathfrak{l}_4^\ord{1} $ on a solution 
can be interpreted as large gauge transformations. Likewise, the action of 
the generator  $\e \in {\bf 1}^\ord{2}$ on a solution amounts to a shift 
of the integration constant appearing in the definition of the axion 
field obtained from the four-dimensional metric by dualisation in three 
dimensions. Therefore, the action of the group $\P_0$ on a solution 
with a charge matrix $\C = c \h$ amounts to a reparametrisation of the solution.
In other words: although the map $\V\rightarrow p_0\V$ for $p_0\in\P_0$ 
changes the asymptotics of $\V$, in which case the scalar field configurations $\V$ 
and $p_0\V$, for $p_0\in\P_0$, would be regarded as inequivalent from the
point of view of the three-dimensional theory, they are in fact physically
equivalent from the point of view of the four-dimensional theory because the constant 
shifts induced by $p_0$ all drop out in the relevant charges as computed in
four dimensions. The present construction thus retains a `memory' of the
four-dimensional origin of the three-dimensional theory.

The remaining generator of the maximal parabolic subgroup $\P$ is the 
generator $\boldsymbol{h}$ itself. It follows from the five-graded  
decomposition (\ref{five}) of $\mathfrak{g}$ that its action on a given 
solution is again a trombone-like symmetry. The latter is a modified version of
(\ref{Trombone}) which scales spacelike and timelike indices differently,
and which {\em only exists for stationary solutions}. More specifically, 
we have
\begin{gather}  
g_{00}(x) \rightarrow \lambda^2 g_{00}(x) 
\hspace{10mm}Êg_{0\mu}(x) \rightarrow  g_{0\mu}(x) \hspace{10mm} 
g_{\mu\nu} (x)\rightarrow \lambda^{-2} g_{\mu\nu} (x) \CR
A_0(x)  \rightarrow \lambda A_0 (x)\hspace{10mm} 
A_\mu(x) \rightarrow \lambda^{-1} A_\mu(x) \label{TromT}
\end{gather}
where $x^\mu$ now denotes the {\em spatial} coordinates, and Greek indices 
are understood to run from $1$ to $3$. By diffeomorphism covariance,
this action on stationary solutions is equivalent, to the `compensated
trombone' transformation
\be 
t \rightarrow \lambda t  \hspace{10mm}  
x^\mu \rightarrow \lambda^{-1}  x^\mu  \ ,
\ee
{\it i.e.}, to a `weighted' rescaling of the four-dimensional coordinates
$(t,x^\mu)$ without rescaling the tensor fields with respect to their 
rank.\footnote{That (\ref{TromT}) is indeed correct is most easily 
 seen for pure gravity in four dimensions: using $g_{00}= - H,\; g_{0\mu} = 
 - H\hat{B}_\mu$, the duality relation $H^2 d\hat{B} = \star  dB$ and the 
 standard Kaluza Klein formula
 $$
 g_{\mu\nu} = H^{-1} \gamma_{\mu\nu} - H\hat{B}_\mu\hat{B}_\nu \ ,
 $$
 we see that the three-dimensional fields scale as $H\rightarrow \lambda^2 H$
 and $B\rightarrow \lambda^2 B$ (as it must be, since $(H,B)$ coordinatise 
 the $\sigma$-model manifold $SL(2,\mathds{R})/SO(1,1)$), while 
 $\gamma_{\mu\nu}$ is invariant. This corresponds precisely to the action 
 of $\boldsymbol{h}$ in the five-graded decomposition (\ref{five}).}

For any other charge matrix $\C$ in the $\H^*$-orbit of $c\boldsymbol{h}$ 
we have $\C = U_\C (c\boldsymbol{h})U_\C^{-1}$ for some $U_\C\in\H^*$. 
Consequently we can define the associated maximal parabolic subgroup  
$\P_\C = U_\C\P U_\C^{-1}\subset\G$ whose Lie algebra $\mathfrak{p}_{\C}
\subset\mathfrak{g}$ is generated by the eigenvectors of the adjoint 
action of $\C$ with positive eigenvalues. As for $\P$, any element 
of $\P_{\C}$ can be written as the product of an element of the form 
$\exp(c^{-1} \ln \lambda \C)$ and an element of the subgroup 
$\P_{0\,{\C}} \subset \P_{\C}$. 

Inspired by the definition of the active duality group transformations 
in \cite{Active}, we now define the active transformations in  
the three-dimensional theory in such a way that the action of an element 
of the maximal parabolic subgroup $\P_\C$ on a solution of charge matrix 
$\C$ is given by the compensated trombone transformation with parameter 
given by the component of the $\P_\C$ element associated to the 
generator $\C$. However, there is another 
subtlety which distinguishes the three-dimensional theory from the 
four-dimensional one, and which is related to the fact that the maximal 
subgroup $\H^*$ is not compact, unlike the group $\H_4$ in (\ref{g1}). 
If we were dealing with the compact form $\H$ instead (as would be the case
for Lorentzian solutions corresponding to the reduction with a spacelike
Killing vector), the Iwasawa theorem would entail the isomorphism
\be\label{GP}
\frac{\G}{\P} \cong \frac{\H}{\H\cap\P} = \frac{\H}{\H_4}
\ee
such that the moduli space of charges could be identified with the 
parabolic coset
\be\label{GP1}
\mathcal{M}^{\rm \scriptscriptstyle Lorentz}  \cong \frac{\G}{\P_0} \ ,
\ee
in complete analogy with (\ref{G4B}). The formula (\ref{GP}) would 
furthermore ensure that a proper action of the full group $\G$ can be 
implemented on the full orbit space. Here, by contrast, the maximal 
subgroup $\H^*\subset\G$ is {\em non-compact}.
Because the Iwasawa decomposition does not generally hold with maximal 
non-compact subgroups, the isomorphism (\ref{GP}) is no longer valid if 
we replace $\H$ by $\H^*$, so stationary solutions cannot fully be described in 
terms of parabolic coset spaces. Rather, the breakdown of the 
Iwasawa theorem is precisely linked to the existence of BPS orbits,
whereas the isomorphism (\ref{GP1}) is possible for spacelike reductions 
because of the absence of BPS colliding plane wave solutions. Indeed, the following analysis 
will trace out in detail the link between different types of BPS orbits 
and the subsets of $\G$ for which the Iwasawa decomposition fails, and will
relate them to the strata $\mathcal{M}_n$ discussed in the foregoing section.

For a non-compact maximal subgroup $\H^*$, the Iwasawa theorem only 
holds on a {\em dense subset $\mathring{\G}\subset\G$}. Every element 
$g\in\mathring{\G}_{\C} \equiv  U_\C \mathring{\G}  U_\C^{-1}$ in this 
dense subset can be decomposed into a 
product of an element $u_{(g,\C)}\in\H^*$, a `diagonal' element
$\exp(c^{-1}\ln \lambda \C)$ (with $\lambda_{g,\C} >0$) and 
an element $p \in {\P_{0\,{\C}}}$ as follows
\be 
g = u_{(g,\C)}  \, \exp\Big( c^{-1}\ln \lambda_{(g,\C)} \C \Big) \, 
p_{(g, \C)} \ .
\label{Iwasawa} 
\ee
The singular elements $g\in\G\setminus\mathring{\G}_{\C}$ (where the 
Iwasawa decomposition breaks down) correspond to limits of regular elements 
$g_k\in \mathring{\G}_{\C}$ for which $\lambda_{(g_k,\C)}\rightarrow 0 $, 
while simultaneously the element $u_{(g_k,\C)}$ goes to the boundary of the 
non-compact group $\H^*$, in such a way that the limit $g= \lim g_k\in\G$ 
is well-defined. This is one main difference with (\ref{g1}) for which 
no such limit can be taken because $\H_4$ is compact.

The active duality group transformation corresponding to an element 
$g\in \mathring{\G}_\C$ on a solution $\V(x)$ with a charge matrix $\C$ with 
$c>0$ is now defined as the successive action of the compensated trombone 
transformation with parameter $\lambda_{(g,\C)}$, followed by the standard 
non-linear action of the group element $u_{(g,\C)}\in \H^*$ [as computed 
from (\ref{Iwasawa}); note that this decomposition depends on the initial 
solution $\V$ via its associated charge $\C\,$], \ie
\be\label{gV} 
g \, : \, \mathcal{V}(x) \, \rightarrow \, \V'(x) := 
u_{(g,\C)} \cdot  \mathcal{V}\big(\lambda^{-1}_{(g,\C)}  x \big) \cdot 
h\big(u_{(g,\C)} , 
\mathcal{V}(\lambda^{-1}_{(g,\C)}  x ) \big) \ ,
\ee
where the matrix $\V(x)$ is triangular (\ie $\V\in\P_\C$), and the 
compensator $h\in\H^*$ restores the triangular gauge, but now with
respect to $\P_{\C(g)}$, where the transformed charge matrix is 
computed from (\ref{Charge}) as
\be\label{gC}
\C(g) = \lambda_{(g,\C)} \,  u_{(g,\C)} \, \C \,  u^{-1}_{(g,\C)}  
\ee
while the BPS parameter transforms as 
\be\label{gc}
c(g) = \lambda_{(g,\C)} \, c \ .
\ee
The remarkable fact is now that these transformations define regular (and 
non-trivial!) solutions even when $u_{(g,\C)}$ and $\lambda_{(g,\C)}$ become 
singular separately. For $\lambda_{(g_k,\C)}\rightarrow 0$ we have $\lim c(g_k) =0$,
and therefore the initial non-BPS solution is mapped to a BPS solution.
From (\ref{Iwasawa}), we see that the limiting matrix $g= \lim_k g_k$
no longer admits an Iwasawa decomposition with respect to $\C$.
Consequently, {\em the elements $g\in\G$ for which the Iwasawa 
decomposition fails are precisely the ones that map non-BPS to BPS 
solutions.} However, as we already indicated, this procedure fails to 
define a proper Lie group action in general owing to the existence of 
non-trivial solutions with $c=0$. As defined above the action of the active 
duality group cannot be `inverted' in the sense that the above procedure
cannot be applied to solutions with vanishing BPS parameter, because 
there are generators in the Lie algebra $\mathfrak{g}$ whose action 
diverges in the limit $c \rightarrow 0$. In other words, the group 
$\G$ cannot act properly on all solutions.

One can understand the `almost action' of the active duality group from a 
more geometrical point of view. The `almost Iwasawa decomposition' 
(\ref{Iwasawa}) permits one to define\footnote{Note that the action of $\P_0$ is well-defined on the submanifold $\mathring{\G}\subset \G$, since, by definition, its action on $\G$ preserves the property of admitting an Iwasawa decomposition.} a homeomorphism between $\mathcal{M}_0 \cong 
\mathds{R}_+^* \times \H^* / \H_4 $ and $\mathring{\G} / \P_0$.  The  `almost action' 
of $\mathring{\G}$ on $\mathcal{M}_0$ can then be derived from the 
action of $\G$ on $\mathring{\G} / \P_0$ using this homeomorphism. One cannot 
extend this `almost action' to a Lie group action on $\mathcal{M}$ 
because the codimension of $\mathcal{M}_1$ in $\mathcal{M}_0$ (which
equals $1$) does not match the codimension of the subset $\G \setminus \mathring{\G}$ 
on which the Iwasawa decomposition fails. More specifically, the homeomorphism between  $\mathcal{M}_0$ and $\mathring{\G} / \P_0$  does not extend to a 
homeomorphism between $\mathcal{M}$ and $\G / \P_0$. Basically, the 
dimension of the complement of $\mathcal{M}_0$ inside $\G / \P_0$ is of lower
dimension than the next stratum $\mathcal{M}_1 \cong \H^* / \J_1$,
in such a way that the moduli space of charges  $\mathcal{M}$ cannot be 
homeomorphic to the coset space $\G / \P_0$. When $\H^*$ admits a $U(1)$ 
factor, 
\be 
\mathcal{M}_0 \cong \mathds{R}_+^* \times \H^* / \H_4 \cong 
\frac{\mathds{C}^{\scriptscriptstyle \times} \times \H_4 
\ltimes \mathfrak{l}_4}{\H_4} 
\ee
and $\mathcal{M}_0$ is locally isomorphic to $\mathds{C}^{\scriptscriptstyle 
\times} \times \mathfrak{l}_4$. The complement of the image of the embedding 
of $\mathcal{M}_0 \cong \mathds{R}_+^* \times \H^* / \H_4 $ into $\G / \P_0$ 
inside $\G / \P_0$ corresponds to limit points of 
$\mathds{C}^{\scriptscriptstyle \times} \times \mathfrak{l}_4$ for which 
the complex parameter goes to zero as the vector of $ \mathfrak{l}_4$ 
diverges. It follows that this subspace has same the dimension as $\mathfrak{l}_4$,
whereas the stratum $Ê\mathcal{M}_1$ is of dimension 
${\rm dim} [ \mathfrak{l}_4 ] + 1$.
\be 
{\rm dim}Ê\bigl[Ê\mathcal{M}_1 \bigr] = 
{\rm dim} \left[ \frac{ \G}{\P_0} \setminus  \mathcal{M}_0 \right] 
+1 Ê \quad \Rightarrow \quad   \frac{ \G}{\P_0}  \ncong 
\mathcal{M} \cong \mathcal{M}_0 \cup \mathcal{M}_1 \cup \cdots \ .
\ee
Note, however, that the above argument works only for $\N\leq 5$; for
$\N =8$ one would need to better understand how to characterise the
subsets of $E_{8(8)}$ on which the Iwasawa decomposition fails.
These conclusions can also be stated differently as follows: while there exists
an `almost action' of $\G$ on the main stratum $\mathcal{M}_0$, no proper action
of $\G$ can be implemented on the various BPS strata: these being Lagrangian
submanifolds, they have only half the dimension that would be required for a 
non-linear realisation of $\G$.

In this discussion, we have not really been able to precisely generalise the notion of active 
duality-group transformations to the three-dimensional theories. In this connection, one 
can identify two noteworthy differences with respect to the higher-dimensional cases which 
seem to be unavoidable. First, the action of the active duality group on the relevant charges 
is no longer equivalent to the standard non-linear action
of the group. Second, this action is highly non-linear, which 
follows from the fact that the charge matrix involves gravity degrees 
of freedom as well. We conclude that the common idea that the 
three-dimensional duality group $\G$ is broken at the quantum level to 
an arithmetic subgroup, with the relevant representation simply defined over 
the integers, might be too na\"{i}ve. 

Nevertheless, the difficulties that appear in trying to define a non-linear realisation of 
an arithmetic group, could as well give a solution to the singular behaviour of the 
`almost representation' on the BPS strata. Our expectation is that even if there is no well-defined action of $\G$ on the moduli space $\mathcal{M}$, the space of functions on $\mathcal{M}$ could admit a non-linear action of the duality group $\G$. We have already seen in the last section that the strata of  $\mathcal{M} \setminus \mathcal{M}_0$ are Lagrangian submanifolds of the corresponding nilpotent orbits in $\mathfrak{N}_\G$. $\H^*/\H_4$ is itself a Lagrangian submanifold of the $\G / \G_4$-orbit of the generator $\h \in \mathfrak{g}$. It seems possible that the action of $\G$ on a solution in $\mathfrak{g}$ of the characteristic equation (\ref{polynom},\, \ref{cubic}) induces an action of $\G$ on the space of functions defined on $\mathcal{M}$. For instance, the stratum $\mathcal{M}_4$ of $\ft12$ BPS solutions of $\N=8$ supergravity is a $29$ dimensional Lagrangian submanifold of the minimal adjoint orbit $\G/ \P_0$, and the minimal unitary realisation of $E_{8(8)}$ \cite{minimalEisenstein,GKN}  might be defined on the functions supported on $\mathcal{M}_4$. The non-perturbative corrections to the three-dimensional Euclidean theory describing stationary solutions of $\N=8$ supergravity should be invariant under the action of an arithmetic subgroup $E_{8(8)} (\Z)$ of  $E_{8(8)}$. The corresponding automorphic forms can be written \cite{minimalEisenstein}
\be \mathscr{E}^{E_{8(8)}} (\Psi) =  \bigl<  \Psi_{E_{8(8)} (\Z)} , \, \rho( \V ) \, \Psi_{Spin^*(16)} \bigr> \ ,\ee 
where $\Psi_{Spin^*(16)} $ is the so-called spherical vector, which would in this case be a $Spin^*(16)$ invariant function over $\mathcal{M}_4$. $\rho( \V )$ is  the coset element $\V$ in the minimal unitary representation, and $\Psi_{E_{8(8)} (\Z)} $ is an $E_{8(8)} (\Z)$ invariant distribution defined over $\mathcal{M}_4$. A spherical vector $\Psi_{Spin^*(16)} $ and its p-adic equivalent defining   $\Psi_{E_{8(8)} (\Z)} $ have been computed in \cite{minimalEisenstein} and \cite{adelic} respectively. This formula suggests that non-perturbative corrections can be identified as observables of the quantum mechanics of a particle living on $\mathcal{M}_4$ associated to the operator $\rho( \V )$.

We are next going to illustrate the definitions of this section with the two simplest examples,
namely pure gravity and Maxwell--Einstein theory.

\subsection{The $SL(2,\mathds{R})$-orbit of Taub--NUT solutions}
\label{MEAD}
The simplest example is pure gravity in four dimensions, for which we 
can define an `active' realisation of the Ehlers group $SL(2,\mathds{R})$ 
on stationary solutions following the steps described in the foregoing 
section. The $\mathfrak{sl}(2,\mathds{R})$ generators of the Ehlers group 
decompose as 
\be 
h \h + e \e + \beta \boldsymbol{\beta}  =  
\left( \begin{array}{cc} \  h \  & \, e + \beta \, \\ - \beta & -h \end{array} \ .
\right) 
\ee  
Here, the $SO(2)$ generator $\boldsymbol{\beta} \equiv \boldsymbol{e}-
\boldsymbol{f}$ preserves the asymptotics, while $\e$ is the nilpotent 
generator of the subgroup $\P_0 \cong \mathds{R}$. The Iwasawa 
decomposition of an $SL(2,\mathds{R})$ matrix\footnote{Because
 $SO^*(2)=SO(2)$ is compact, a breakdown of the Iwasawa decomposition 
 is not an issue here, which is consistent with the fact that
 pure gravity in four dimensions does not admit BPS solutions.} implies that 
an element of the coset $SL(2,\mathds{R}) / \mathds{R}$ decomposes as the 
product of an $SO(2)$ element and an element of the parabolic subgroup 
$\P\cong IGL_+(1,\mathds{R})$ as follows
\be\label{SL2}
\left( \begin{array}{cc} \, \mu \,  & \, 0 \, \\ \, \mu b 
\, &\,  \mu^{-1} \, \end{array} \right)  = \frac{1}{\sqrt{1 + b^2}}   
\left( \begin{array}{cc} \, 1 \,  & \, -b \, \\ b & 1  \end{array} \right)   
\left( \begin{array}{cc} \, { \scriptstyle \sqrt{1 + b^2} \mu } 
\,  & \, 0 \, \\ 0 & \frac{1}{\sqrt{1 + b^2} \mu} \end{array} \right)   
\left( \begin{array}{cc} \,1 \,  & \, \frac{b}{ (1 + b^2) \mu^2} 
\, \\ 0 & 1  \end{array} \right) \ .
\ee
The charge matrix is
\be\label{CSL2} 
\C \equiv \left( \begin{array}{cc}  \vspace{2mm} \hspace{2mm} m 
\hspace{2mm} &  \hspace{2mm}  n \hspace{2mm} \\ \vspace{2mm} 
\hspace{2mm}  n   \hspace{2mm}& \hspace{2mm}-m    
\hspace{2mm}\end{array}\right) \;\; \in \;\;
 \mathfrak{sl}(2,\mathds{R}) \ominus \mathfrak{so}(2) \ .
\ee
Following the steps of the preceding section (in particular
formulas (\ref{gC}) and (\ref{gc})), the active action of 
an element of $SL(2,\mathds{R})$ on the Schwarzschild solution of unit 
mass and vanishing NUT parameter is such that the upper triangular matrix 
on the right in (\ref{SL2}) can be disregarded, while the diagonal 
element raises the BPS parameter from $c=m=1$ to $c=\sqrt{m^2 + n^2} =
\sqrt{1 + b^2} \mu$ through the action of the trombone transformation. Finally, 
the $SO(2)$ element determines the value of the mass and the NUT charge 
in such a way that the new solution has mass $m = \frac{1- b^2}{\sqrt{1+b^2}} 
\mu$ and NUT charge $n = \frac{2 b }{\sqrt{ 1 + b^2}} \mu$. 
This defines the isomorphism
\be 
\mathcal{M}_0 \cong \frac{SL(2,\mathds{R})}{\mathds{R}} 
\ee
between the moduli space $(m,n) \neq (0,0)$ of Taub--NUT solutions, 
and the parabolic coset  $SL(2,\mathds{R}) / \mathds{R}$. The 
triangular form of the coset element defines only local coordinates on this 
space. The coset space is a (trivial) line bundle of fibre $\mathds{R_+^*}$ 
over the parabolic coset $SL(2,\mathds{R}) / IGL_+ (1,\mathds{R}) \cong S^1$, 
and thus is diffeomorphic to a cylinder. This cylinder is covered by the 
coordinates $(\mu , b ) \in \mathds{R}_+^* \times \mathds{R}$ plus 
an $\mathds{R}_+^*$ half-line defined by the limit  $b \rightarrow \pm 
\infty $ and $\mu \rightarrow  0 $ in such a way that $|b| \mu$ 
is a finite positive number. The map $\mu^\prime = \mu |b| , \, 
b^\prime = - \frac{1}{b} $ defines a complementary open set of coordinates 
for which the limit point coordinates are now regular. This limit point corresponds 
to the Schwarzschild solution with negative mass $- \mu |b|$. The cylinder
is closed at one end by adding the trivial stratum $\mathcal{M}_1$
consisting only of the point $(m,n)= (0,0)$.

In this way one obtains an action of $SL(2,\mathds{R})$ on the Taub--NUT 
solutions defined from the left action on $SL(2,\mathds{R}) / \mathds{R}$ 
through the map $(m, n)  = \left(\frac{1 - b^2}{\sqrt{1 + b^2}} \mu, 
\frac{2 b }{\sqrt{ 1 + b^2}} \mu \right) $. This map has as inverse
\be 
\mu = \frac{1}{2} \sqrt{ ( c + m)^2 + n^2 } 
\hspace{10mm}Êb= \frac{ 2 n c}{ (c + m)^2 + n^2} \ .
\ee
For a general element of $SL(2,\mathds{R})$, $ g\equiv \scriptscriptstyle  
\left( \begin{array}{cc} \, \alpha\,  & \, \beta \, \\ \, \gamma \, &\,  
\delta \, \end{array} \right) $, with $\alpha \delta - \beta \gamma = 1$, 
one obtains the following transformation of the solution's charges:
\bea\label{mn'} 
m^\prime &=& \frac{ ( \alpha^2 - \gamma^2 + \beta^2 - \delta^2 ) c + ( \alpha^2 - \gamma^2 - \beta^2 + \delta^2 ) m + 2 ( \alpha \beta - \gamma \delta ) n }{\sqrt{ 2 ( \alpha^2 + \gamma^2 + \beta^2 + \delta^2 )  + 2 ( \alpha^2 + \gamma^2 - \beta^2 - \delta^2 ) \frac{m}{c} + 4 ( \alpha \beta + \gamma \delta ) \frac{n}{c} }} \CR
n^\prime &=& \frac{ 2 ( \alpha \gamma + \beta \delta ) c + 2 ( \alpha \gamma - \beta \delta ) m + 2 ( \alpha \delta + \beta \gamma ) n }{\sqrt{ 2 ( \alpha^2 + \gamma^2 + \beta^2 + \delta^2 )  + 2 ( \alpha^2 + \gamma^2 - \beta^2 - \delta^2 ) \frac{m}{c} + 4 ( \alpha \beta + \gamma \delta ) \frac{n}{c} }} \ .\label{nonlinear}  
\eea
To derive these formulas, one first expresses $(\mu,b)$ via $(m,n)$, then works out
the non-linear action of $SL(2,\mathds{R})$ in order to obtain
\be
\mu' = (\alpha + \beta b) \mu \quad , \qquad
b' = \frac{\gamma + \delta b}{\alpha + \beta b} 
\ee
and finally expresses $(m',n')$ in terms of the new parameters $(\mu',b')$ 
as functions of $(m,n)$. This construction extends trivially to non-vanishing 
angular momentum by taking a Kerr solution as the reference solution. 
The action is the same with the value of $(a/c)$ kept fixed.

In order to see explicitly that the `active action' (\ref{mn'}) is actually the same as the abstract
formula (\ref{gC}), we must perform an Iwasawa decomposition of the
general $SL(2,\mathds{R})$ element $g$, but with $\C$ from (\ref{CSL2})
rather than $\h$ as the diagonal element, as in (\ref{Iwasawa}). After 
some algebra we arrive at 
\begin{multline} 
\left( \begin{array}{cc} \, \alpha\,  & \, \beta \, \\ \, \gamma \, &\,  
\delta \, \end{array} \right)  =  \frac{1}{\sqrt{1 + b^2}}   
\left( \begin{array}{cc} \, 1 \,  & \, -b \, \\ b & 1  \end{array} \right)   
\frac{1}{2c} \left( \begin{array}{cc} \, ( c + m ) \lambda  + (c-m) \lambda^{-1}  & \, n ( \lambda  -\lambda^{-1} ) \, \\ \, n ( \lambda  -\lambda^{-1} ) \, &\,  
( c - m ) \lambda  + (c+m) \lambda^{-1}\, \end{array} \right)  \\
\!\!\!\!\!\!\!\!\!\!\!\!\!\!\!\!\!\!\!\!\!\!\!\!\!\!\!\!
\times \frac{1}{2c} \left( \begin{array}{cc} \, 2 c - n e \,  & \, (c+m) e \, \\ \, (-c+m) e \, &\,  
2 c + n e  \, \end{array} \right) 
\end{multline}
where the matrix in the middle is just $\exp\big[c^{-1}(\ln \lambda) \C\big]$,
the matrix on the left is the $SO(2)$ rotation $u_{(g,\C)}$, and the matrix
to the right is the parabolic element $p_{(g,\C)}$ that leaves invariant the 
charge matrix $\C$ from (\ref{CSL2}) through the active action (\ref{gC}). The parameters in these matrices
are given by
\bea b &=& \frac{ ( \gamma - \beta ) c + ( \gamma + \beta ) m + ( \delta - \alpha ) n}{( \alpha + \delta ) c + ( \alpha - \delta ) m + ( \beta + \gamma) n }  \hspace{10mm}Ê\frac{1}{\sqrt{ 1 + b^2}}Ê= \frac{ ( \alpha + \delta ) c + ( \alpha - \delta ) m + ( \beta + \gamma) n  }{ 2 c \lambda }Ê \CR
\lambda &=& \frac{1}{\sqrt{2c}}Ê \sqrt{  ( \alpha^2 + \gamma^2 + \beta^2 + \delta^2 ) c  +  ( \alpha^2 + \gamma^2 - \beta^2 - \delta^2 )  m + 2 ( \alpha \beta + \gamma \delta )  n  } \CR
e &=& \frac{  n \beta - ( c-m) \alpha + \lambda^{-2} \scal{Ên \gamma + ( c - m) \delta }}{ ( c-m) \beta + n \alpha} \ .
 \eea

Finally, we would like to point out that it is by no means evident from 
(\ref{mn'}) whether and how the continuous duality group $SL(2,\mathds{R})$
can be broken to an arithmetic subgroup such as $SL(2,\Z)$ upon
quantisation. Although one can of course restrict the action (\ref{mn'})
to elements $g$ in such an arithmetic subgroup, the resulting (discrete) set 
of admissible charges $(m,n)$ does not appear to have a nice structure  satisfying a Dirac quantisation condition. As this is the simplest example involving gravitational degrees of freedom,
similar comments apply to the larger duality groups of all supergravity
theories with $\N\geq 1$.

\subsection{Maxwell--Einstein theory [$\G =SU(2,1)$]}

The simplest example including nontrivial BPS solutions (which 
do not exist for pure gravity)~\footnote{Related results for Maxwell--Einstein theory 
have been obtained by L.~Houart, A.~Kleinschmidt, 
  N.~Tabti and J.~Lindman-H\"ornlund (A.~Kleinschmidt, priv. comm.).}
in which one can make completely explicit the failure of
the construction to define a group action for a non-compact divisor group 
$\H^*$ is Maxwell--Einstein theory, for which the coset space is 
$SU(2,1)/U(1,1)$. Defining a group action of the duality group 
$\G$ on the space of solutions requires the vector fields defining the 
Lie algebra $\mathfrak{g}$ to be regular. If the divisor group $\H^*$
is non-compact, these vector fields are regular only on a dense subspace 
of the space of solutions, but diverge like $\frac{1}{c}$ as one approaches the subspace 
of BPS solutions. For this reason, the action of the duality group will 
be ill defined on this subspace so that some of the directions 
in the group degenerate and do not define transformations. Nevertheless, the vector fields 
do allow for transformations that allow one to move from any non-BPS solution to 
any other solution with the same angular momentum ratio $(c/a)$, 
including the BPS solutions (with this fixed ratio).

The Lie algebra $\mathfrak{su}(2,1)$ decomposes into a direct sum of 
$\mathfrak{su}(1,1)$ and the parabolic subalgebra $\mathfrak{p}$ generated 
by $\h,\,\boldsymbol{\beta} ,\,  \e,\, \x$ and $\y$ (with the corresponding
parameters $h,\beta,e,x$ and $y$, respectively). Hence, any
element $\boldsymbol{u}\in\mathfrak{su}(2,1)$ has the form
\be 
\boldsymbol{u} =  \left(\begin{array}{ccc} \hspace{2mm} i \alpha   \hspace{2mm} &  \hspace{2mm}  \alpha  \hspace{2mm}  & \hspace{2mm} \xi + i \zeta  \hspace{2mm}  \\  \hspace{2mm}- \alpha  \hspace{2mm}& \hspace{2mm} i \alpha    \hspace{2mm}&  \hspace{2mm}- \zeta  + i \xi \hspace{2mm} \\  \hspace{2mm}\xi -  i \zeta  \hspace{2mm} & \hspace{2mm} - \zeta -i \xi   \hspace{2mm} &  \hspace{2mm}- 2 i \alpha  \hspace{2mm}\end{array}\right) + \left(\begin{array}{ccc} \hspace{2mm}  h  + i\beta  \hspace{2mm} &  \hspace{2mm}  e   \hspace{2mm}  & \hspace{2mm} x  + iy  \hspace{2mm}  \\  \hspace{2mm} 0   \hspace{2mm}&   \hspace{2mm}  - h + i\beta   \hspace{2mm} & \hspace{2mm} 0   \hspace{2mm}\\   \hspace{2mm}0     \hspace{2mm}&  \hspace{2mm} - y -i x  \hspace{2mm}  &  \hspace{2mm}  - 2i\beta   \end{array}\right) \label{u}
\ee 
where the left summand is in $\mathfrak{su}(1,1)$. The charge matrix is
\be\label{CSU21} 
\C \equiv \left( \begin{array}{ccc}  \vspace{2mm} \hspace{2mm} m \hspace{2mm} &  \hspace{2mm}  n \hspace{2mm}  & \hspace{2mm}- \frac{z}{\sqrt{2}} \hspace{2mm}  \\ \vspace{2mm} \hspace{2mm}   n   \hspace{2mm}& \hspace{2mm}-m    \hspace{2mm}&  \hspace{2mm}  i\frac{z}{\sqrt{2}}  \\  \vspace{2mm} \hspace{2mm}\frac{\bar z }{\sqrt{2}}    \hspace{2mm} & \hspace{2mm}i \frac{\bar z }{\sqrt{2}}   \hspace{2mm} &  \hspace{2mm} 0 \hspace{2mm}\end{array}\right) 
\;\in \; \mathfrak{su}(2,1) \ominus \mathfrak{u}(1,1) \ . 
\ee
As explained in Section \ref{Active1section}, $\h$ acts like the 
trombone transformation, up to a pseudo-conformal diffeomorphism. The action of $\e$ 
amounts to the addition of an irrelevant constant to the axion field, and 
$\y$ defines a shift of the magnetic scalar 
in a similar way. The final generator $\x$ acts as a global gauge 
transformation. We thus define the `active' $SU(2,1)$ on the space of 
solutions in such a way that the generators $\e,\, \x$ and $\y$  
act trivially on the Kerr solution, and the generator $\h$ is defined to act 
as the compensated trombone transformation on it. 

The generator $\boldsymbol{\beta}$ of the four-dimensional duality group 
leaves the Kerr solution invariant as it does any pure gravity solution. The 
isotropy subgroup of $SU(2,1)$ of the Kerr solution under the active 
transformations is thus the group $\P_0 \cong \Ic U(1)$ generated by the 
Lie algebra elements $\boldsymbol{\beta},\, \e,\, \x$ and $\y$, which 
together with the $\h$ generator define a maximal parabolic subgroup 
$\P \cong \mathds{R}_+^* \ltimes  \Ic U(1)\subset SU(2,1)$. The 
na\"{\i}ve model for the full space of solutions at fixed angular momentum 
is thus the coset space $SU(2,1) / \Ic U(1)$. However, the map from the 
space of solutions into this coset space fails to be an isomorphism on 
the subspace of BPS solutions.

The subsequent analysis proceeds along the same lines as for pure gravity. 
Let us first look to the coset space itself. It is a trivial fibre bundle 
over $SU(2,1) / \P \cong S^3$ with fibre $\mathds{R}_+^* $. Its (lower)
triangular matrix form is
\be 
\left( \begin{array}{ccc}  \vspace{2mm} \hspace{2mm} \mu  \hspace{2mm} &  
\hspace{2mm}  0\hspace{2mm}  & \hspace{2mm} 0 \hspace{2mm}  \\ 
\vspace{2mm} \hspace{2mm} \mu ( b + i |q|^2 )   \hspace{2mm}& \hspace{2mm}  
\mu^{-1}     \hspace{2mm}&  \hspace{2mm} \sqrt{2} i q   \hspace{2mm}   \\  
\vspace{2mm} \hspace{2mm} \sqrt{2} \mu q^* \hspace{2mm} & \hspace{2mm} 0 
\hspace{2mm} &  \hspace{2mm} 1 \hspace{2mm}\end{array}\right) 
\label{coset matrix}
\ee
with local coordinates $\mu>0\,,\, b\in\mathds{R}$ and $q\in\mathds{C}$. This 
coordinate system does not cover the whole coset space. $b$ and $q$ can 
be regarded as stereographic coordinates on the three-sphere $S^3$, 
such that the map 
\be 
\mu^\prime = \sqrt{b^2 + |q|^4} \mu \hspace{10mm} b^\prime 
= - \frac{b}{b^2 + |q|^4} \hspace{10mm}Êq^\prime  = \frac{q}{b - i |q|^2} 
\ee
gives the coordinates on the other hemisphere. It remains for one to add the points 
at infinite $b$ and $q$ with a finite strictly positive value of 
$\sqrt{b^2 + |q|^4} \, \mu $. As for the pure gravity case, such 
points correspond to Kerr solutions with a negative mass 
$ - \sqrt{b^2 + |q|^4} \, \mu$.

The coset matrix (\ref{coset matrix}) admits a `singular Iwasawa 
decomposition' as a product of an element of $U(1,1)$, an element 
generated by $\h$ and an element of $\Ic U(1)$, \viz
\begin{multline}\label{Iwa} 
\left( \begin{array}{ccc}  \vspace{2mm} \hspace{2mm} \frac{1}{\sqrt{ (1 - |q|^2)^2 + b^2}} \hspace{2mm} &  \hspace{2mm}  \frac{ - b - i |q|^2}{\sqrt{ (1 - |q|^2)^2 + b^2}} \hspace{2mm}  & \hspace{2mm} \frac{\sqrt{2} q}{1 - |q|^2 + i b } \hspace{2mm}  \\ \vspace{2mm} \hspace{2mm} \frac{ b + i |q|^2 }{\sqrt{ (1 - |q|^2)^2 + b^2}}   \hspace{2mm}& \hspace{2mm} \frac{1}{\sqrt{ (1 - |q|^2)^2 + b^2}}     \hspace{2mm}&  \hspace{2mm} \frac{\sqrt{2} i q }{1 - |q|^2 + i b}    \hspace{2mm}  \\  
\vspace{2mm} \hspace{2mm} \frac{\sqrt{2}  q^*     }{\sqrt{ (1 - |q|^2)^2 + b^2}}\hspace{2mm} & \hspace{2mm} 
\frac{-i\sqrt{2} q^*}{\sqrt{ (1 - |q|^2)^2 + b^2}}    \hspace{2mm} &  
\hspace{2mm} \frac{ 1 + |q|^2 + i b}{1 - |q|^2 + i b} \hspace{2mm}\end{array}
\right) \times \\*
  \left( \begin{array}{ccc}  \vspace{2mm} \hspace{2mm} {\scriptstyle \sqrt{ (1 - |q|^2)^2 + b^2} \mu }\hspace{2mm} &  \hspace{2mm}  0 \hspace{2mm}  & \hspace{2mm}0 \hspace{2mm}  \\ \vspace{2mm} \hspace{2mm} 0  \hspace{2mm}& \hspace{2mm} \frac{1}{\sqrt{ (1 - |q|^2)^2 + b^2} \mu }     \hspace{2mm}&  \hspace{2mm}0   \hspace{2mm}  \\  \vspace{2mm} \hspace{2mm}0 \hspace{2mm} & \hspace{2mm} 0    \hspace{2mm} &  \hspace{2mm} 1 \hspace{2mm}\end{array}\right)  
 \left( \begin{array}{ccc}  \vspace{2mm} \hspace{2mm}1 \hspace{2mm} &  \hspace{2mm}  \frac{ b - i |q|^2}{((1 - |q|^2)^2 + b^2) \mu^2 } \hspace{2mm}  & \hspace{2mm} \frac{- \sqrt{2} q}{(1 - |q|^2 + i b ) \mu} \hspace{2mm}  \\ 
\vspace{2mm} \hspace{2mm}0  \hspace{2mm}& \hspace{2mm}1   \hspace{2mm}&  
\hspace{2mm} 0  \hspace{2mm}  \\  \vspace{2mm} \hspace{2mm} 0 \hspace{2mm} & 
\hspace{2mm} \frac{i\sqrt{2} q^*}{(1 - |q|^2 - ib ) \mu}    
\hspace{2mm} &  \hspace{2mm} 1 \hspace{2mm}\end{array}\right)  \ .  
\end{multline} 
We see that the decomposition becomes singular for the subspace of unit 
modulus $|q|^2 = 1$ and zero $b$; that is, the subset of $SU(2,1)$
on which the Iwasawa decomposition fails is always conjugate to a 
cylinder  $\mathds{R}_+^* \times S^1$ in $SU(2,1)$.\footnote{Which also
  corroborates our previous claim that the set on which the Iwasawa
  decomposition fails is of codimension 2 in $\G=SU(2,1)$.}
Nevertheless, one can associate a solution of the 
Maxwell--Einstein equations to any generic point. If we apply this coset 
element to the Schwarzschild solution of unit mass (\ie $m\!=\!1, n\!=\!z\!
=\!0$ in (\ref{CSU21})), the subgroup $\P_0$ does not act, while the diagonal 
element changes the BPS parameter from $c \equiv \sqrt{ m^2 + n^2 - |z|^2}=1$ 
to $\lambda\equiv\sqrt{ (1 - |q|^2)^2 + b^2} \,\mu $, and the 
$U(1,1)$ element yields the transformed charges as 
\begin{gather} 
m = \frac{ 1  - |q|^4 - b^2 }{ \sqrt{ (1 - |q|^2)^2 + b^2} }\,\mu\hspace{10mm}
n = \frac{ 2b }{ \sqrt{ (1 - |q|^2)^2 + b^2} } \,\mu \CR
z = \frac{ 1 - |q|^2 + i b}{\sqrt{ (1 - |q|^2)^2 + b^2} } \, 2 q \mu \ .
\end{gather}
Inverting this map we obtain
\begin{gather} \mu = \frac{1}{2} \sqrt{ ( c + m)^2 + n^2 } \CR
b = \frac{ 2n c }{ (c + m)^2 + n^2 } \hspace{10mm}Êq = 
\frac{ z }{ c + m + i n } \ .
\end{gather}
In the BPS limit, the map projects out the overall phase of $z$ and 
$m + in$, which corresponds to the action of the $U(1)$ center of 
$U(1,1)$ on these solutions (this $U(1)$ rotates $m+in$ and $z$ in
the same way, but is not `seen' by the coordinates $(\mu,b,q)$).
With these formulas at hand, we can explicitly verify our previous 
claim that the combined `active action' of the two left matrices
in (\ref{Iwa}) according to (\ref{gC}) and (\ref{gc}) remains well-defined
even though the matrices separately become singular.

The action of a Lie algebra element of $\mathfrak{su}(2,1)$
\be 
\left(\begin{array}{ccc} \hspace{2mm} i \alpha+ h \hspace{2mm} &  \hspace{2mm}  b + \beta    \hspace{2mm}& \hspace{2mm} \sqrt{2} ( x + iy + r + i s )  \hspace{2mm} \\  \hspace{2mm} b - \beta    \hspace{2mm}&   \hspace{2mm} i\alpha- h  \hspace{2mm} & \hspace{2mm}  \sqrt{2} (y - i x - s + i r)  \hspace{2mm}\\   \hspace{2mm} \sqrt{2} ( - x + i y + r - is ) \hspace{2mm}&  \hspace{2mm} - \sqrt{2} ( y + i x + s + ir )  \hspace{2mm}  &  \hspace{2mm}- 2 i \alpha\end{array}\right) 
\ee
is obtained in complete analogy with (\ref{mn'}). A slightly tedious
calculation yields the following infinitesimal action on the elements 
of the charge matrix:
\bea 
\delta m &=& h \scal{Êc + \frac{n^2}{c}Ê} + x \frac{ pn}{c} + b \frac{mn}{c} + y \frac{qn}{c} + r q + s p + \beta  n \CR
\delta n &=& - h \frac{ mn}{c} - x \frac{ pm }{c} - b \scal{ c + \frac{m^2}{c}} - y \frac{qm}{c} - \beta  m + r p - s q \CR
\delta q &=& h \frac{np}{c} - x \scal{ c - \frac{p^2}{c} } + b \frac{mp}{c} - y \frac{qp}{c} - \alpha p + r m - s n \CR
\delta p &=& - h \frac{nq}{c} - x \frac{qp}{c} - b \frac{mq}{c} + y \scal{ c - \frac{q^2}{c}} + \alpha q + s  m + qn  \nonumber 
\eea
\vspace{-6mm} \be
\delta c = h m + x q - bn - yp \hspace{57mm}
\ee
with $z \equiv q + i p$. This transformation is singular for $c = 0$. 
Specialising to BPS solutions with $n = p = 0$ and $q,m \neq 0$, we get the action of the four generators of $\mathfrak{su}(2,1) \ominus \mathfrak{u}(1,1)$
\bea 
\delta m &=& 0 \CR
\delta n &=& - (b + y ) \frac{m^2}{c}   \CR
\delta q &=& 0  \CR
\delta p &=& - (b +y) \frac{m^2}{c}  \CR
\delta c &=& (h + x) m \ .
\eea
The two generators corresponding to $b+y=h+x=0$ leave the charges invariant, the one
corresponding to $h+x\neq 0$ breaks the BPS condition, and the one 
corresponding to $b+y\neq 0$ is singular. In fact, this is not the only 
pathology of the construction. Indeed, if one can reach the BPS 
Reissner--Nordstr\"{o}m solution from any non-BPS one through an action 
generated by the $\h$ generator, one can also reach it from any Kerr--Newman 
solution with an arbitrary value of the angular momentum per unit of mass. 
The generalisation to arbitrary angular momentum is trivially obtained 
by substituting the Kerr solution for the Schwarzschild solution as 
the starting reference solution on which the maximal parabolic subgroup 
is defined to act as the trombone symmetry. The orbits are then exactly 
the same, each with its own value of $(a/c)$. When one reaches a BPS solution, 
we have $a,c\rightarrow 0$ in such a way that this ratio is kept fixed.
However, it is not possible to invert this transformation in the sense 
that there is no preferred value $(a/c)$ from which to start when $a=c=0$.
We conclude that the action of the generators of $\mathfrak{su}(2,1)\ominus 
\mathfrak{u}(1,1)$ on the BPS solutions is either trivial or ill-defined. 

Let us see, anyway, how one can reach BPS solutions from non-BPS solutions through the active action of $SU(2,1)$. For a global transformation $\exp ( \ln \lambda \h )$, one gets 
\bea m(\lambda) &=& \lambda \frac{ \sqrt{(c  + m )^2 + n^2} \, m  +( 1 - \lambda^{-4}) \frac{ \frac{|z|^2}{4c } + c n^2 }{\sqrt{ (c  + m )^2 + n^2}}} { \sqrt{ \scal{Ê c + m + (1 - \lambda^{-2}) \frac{|z|^2}{2c}}^2 + \lambda^{-4} n^2}}  \CR
n(\lambda) &=& \lambda^{-1} n\,  \sqrt{ \frac{ (c  + m )^2 + n^2 }{\scal{Êc + m + (1 - \lambda^{-2}) \frac{|z|^2}{2c}}^2 + \lambda^{-4} n^2}} \CR
z(\lambda) &=& z \,   \sqrt{ \frac{ (c  + m )^2 + n^2 }{\scal{Êc + m + (1 - \lambda^{-2}) \frac{|z|^2}{2c}}^2 + \lambda^{-4} n^2}} \Scal{ 1 + (1 - \lambda^{-2}) \frac{ \frac{ |z|^2}{2c} - i n }{c + m + i n}} 
\eea
The BPS parameter is given by 
\be c(\lambda) = \lambda c \,  \sqrt{ \frac{  \scal{Êc + m + (1 - \lambda^{-2}) \frac{|z|^2}{2c}}^2 + \lambda^{-4} n^2 }{  (c  + m )^2 + n^2}} \ .\ee
The discriminant for the equation $c(\lambda)=0$ is strictly negative for non-zero NUT charge $n$, $\Delta = - \frac{(m+c)^4 n^2}{c^2} + \mathcal{O}(n^4)$. One thus obtains that $\lambda$ can be chosen in such a way that $c(\lambda)=0$ if and only if $n=0$. In the latter case, both the NUT charge and the electromagnetic charges are left invariant, and the mass transforms as follows 
\be m(\lambda) = \frac{ \lambda + \lambda^{-1}}{2} m + \frac{\lambda - \lambda^{-1}}{2} c \ .\ee
For $\lambda = \sqrt{\frac{m-c}{m+c}}$ one gets the BPS 
Reissner--Nordstr\"{o}m solution.

\section{$\N=4$ supergravity as an example}
\label{N4sec}
Our final example comprises the cases of pure and matter-coupled $\N=4$ supergravity. We discuss these models in finer detail here mainly in order to illustrate the 
efficiency of our methods. From Table~III we see that the relevant
duality groups are $\G_4= SO(6,n)$, which is enlarged to $\G= SO(8,2+n)$
in the reduction to three dimensions, and where $n$ denotes the number 
of vector multiplets in four dimensions. In particular, we will analyse
the charge matrix $\C$ directly in terms of $Spin(8,2)$ for pure
$\N=4$ supergravity.

\subsection{The non-linear sigma model formulation}
Since we will only consider stationary axisymmetric solutions, it is convenient to use the so-called Weyl coordinates
\be ds^2 = H^{-1} e^{2\sigma} \delta_{\alpha\beta} dx^\alpha dx^\beta + \rho^2 H^{-1} d \varphi^2 - H ( dt + \hat{B} d \varphi ) ^2 \label{Weyl} \ .\ee
The bosonic sector of $\N=4$ supergravity includes six vector fields $U^a d t + \hat{A}^a d \varphi $ which transform in the vector representation of $SO(6)$. They are coupled to scalar fields lying in the coset $SL(2,\mathds{R}) / SO(2)$  \cite{N4}. We will write $X$ for the dilaton and $Y$ for the axion field. The two-dimensional action leading to the equations of motion of stationary axisymmetric fields configurations is given by
\begin{multline} 
\int dx^2 \Bigl( - 2 \partial^\alpha \sigma \partial_\alpha \rho + \frac{1}{2} \rho H^{-2} \partial^{\alpha} H \partial_\alpha H - \frac{1}{2} \rho^{-1} H^{2} \partial^\alpha \hat{B} \partial_\alpha \hat{B}  \\*
-  \rho H^{-1}  X \partial^\alpha U_a \partial_\alpha U^a +  \rho^{-1} H X  \scal{ \partial^\alpha \hat{A}_a + U_a \partial^\alpha \hat{B} } \scal{ \partial_\alpha \hat{A}^a + U^a  \partial_\alpha \hat{B} } \\*
 + \frac{1}{2} \rho X^{-2} \scal{ \partial^\alpha X \partial_\alpha X + \partial^\alpha Y \partial_\alpha Y } + 2 \varepsilon^{ij} Y \partial_\alpha U_a \scal{ \partial_\beta \hat{A}^a + U^a  \partial_\beta \hat{B} } \Bigr)\ .
\end{multline}
This action is invariant with respect to a non-linear representation of $SL(2,\mathds{R}) \times  SL(2,\mathds{R}) \times SU(4) \cong Spin(2,2) \times Spin(6)$, where $SU(4)$ is linearly represented on the vector fields as the vector representation of $SO(6)$, and the $SL(2,\mathds{R})$'s correspond to an $SL(2,\mathds{R}) / SO(1,1) \times  SL(2,\mathds{R})  /  SO(2) $ non-linear sigma model. 
After dualising the fields $\hat{A}^a$ and $\hat{B}$ through the definitions
\bea \rho^{-1} H X \scal{ \partial_\alpha \hat{A}^a + U^a \partial_\alpha \hat{B} } &=& \varepsilon_{\alpha\beta} \scal{ \partial^\beta A^{a} + Y \partial^\beta U^a }  \CR
\rho^{-1} H^2 \partial_\alpha \hat{B} &=& \varepsilon_{\alpha\beta} \scal{ \partial^\beta B +  U_a \partial^\beta A^{a} -   A_a \partial^\beta U^a } \ ,
\eea
the equations of motion of the dual fields follow from the action 
\begin{multline}  \label{n4action}
\int dx^2 \biggl( - 2 \partial^\alpha \sigma \partial_\alpha \rho + \frac{1}{2} \rho \Bigl( X^{-2} \partial^\alpha X \partial_\alpha X + X^{-2} \partial^\alpha Y \partial_\alpha Y + H^{-2} \partial^{\alpha} H \partial_\alpha H \\* +   H^{-2} \scal{ \partial^\alpha B +  U_a \partial^\alpha A^{a} -   A_a \partial^\alpha U^a } \scal{ \partial_\alpha B +U_a \partial_\alpha A^{a} -  A_a \partial_\alpha U^a }  \\*
-  2 H^{-1}  X \partial^\alpha U_a \partial_\alpha U^a - 2 H^{-1} X^{-1}   \scal{ \partial^\alpha A_a + Y \partial^\alpha U_a  } \scal{ \partial_\alpha A^{a} + Y \partial_\alpha U^a } \Bigr) \biggr)\ .
\end{multline}
This action is itself invariant with respect to non-linear transformations of $Spin(2,8)$, and can be identified as a non-linear sigma model over the coset $SO(2,8) /  SO(2,6) \times SO(2)$. 

In order to make explicit the four-dimensional character of the solutions 
in which we are interested, we use a representation of $Spin(2,8)$ that 
makes the four-dimensional duality group $SL(2,\mathds{R})\times SU(4)$ 
explicit, as well as the   $SL(2,\mathds{R})$ duality group of pure gravity 
in three dimensions. We thus choose a representation for which the subgroup 
$Spin(2,2)\times Spin(6)$ is block diagonal. This representation is given 
by matrices valued in the Clifford algebra associated to $\mathds{R}^6$, 
which is defined as follows
\begin{gather} 
\{ \gamma^a , \gamma^b \} = 2 \delta^{ab} \hspace{10mm} 
\gamma^{ab} \equiv \frac{1}{4} [ \gamma^{a},  \gamma^b ]    \CR
C^2 = 1   \hspace{10mm}    C \gamma^a C = - {\gamma^a}^t  \hspace{10mm}   C \gamma^{ab} C = - {\gamma^{ab}}^t \ .
\end{gather}
We thus define the generators of $\mathfrak{spin}(2,8)$ in terms of the 
six numbers $h,\, e,\, f, \, h^\prime , \, e^\prime $ and $f^\prime $, 
as well as the four six-dimensional vectors contracted with $\gamma^a$, $\ba q_1,\, \ba p_1 ,\, \ba q_2,\, 
\ba p_2$, and the generators of $\mathfrak{spin}(6)$, $\ba \upsilon$. We use the familiar `slash' 
notation in order to make clear which objects are Clifford-algebra valued, \ie  $\ba q_1 \equiv q_{1\, a} \gamma^a$, $\ba \upsilon \equiv \frac{1}{2} \upsilon_{ab} \gamma^{ab}$, {\it etc.}  An element $\mathfrak{u}$ of  $\mathfrak{spin}(2,8)$ is parametrised by these submatrices as follows 
\be 
\mathfrak{u} = \left(\begin{array}{cccc} \hspace{2mm}h + \ba \upsilon \hspace{2mm}& \hspace{2mm}e \hspace{2mm}& \hspace{2mm}  \ba q_1 \hspace{2mm}& \hspace{2mm}\ba p_1  \hspace{2mm}  \\
 \hspace{2mm}f\hspace{2mm} &\hspace{2mm} - h + \ba \upsilon\hspace{2mm} & \hspace{2mm}- \ba p_2\hspace{2mm} & \hspace{2mm}\ba q_2\hspace{2mm} \\ \ba q_2 & - \ba p_1 & h^\prime  + \ba \upsilon & e^\prime  \\ \ba p_2 & \ba q_1 & f^\prime  & -h^\prime  + \ba \upsilon \end{array} \right) 
\ee 
where objects without a slash are to be multiplied by the unit matrix; thus,
$\mathfrak{u}$ can be viewed as a complex 16-by-16 or as a real 32-by-32 matrix. We will generally identify the elements of the Clifford algebra proportional to the unit matrix with the real numbers. 
The subalgebra $  \mathfrak{spin}(2,6) \oplus  \mathfrak{so}(2) $ is 
defined by the elements $\boldsymbol{\alpha} $ of $\mathfrak{spin}(2,8)$ 
satisfying $ C \boldsymbol{\alpha}^t  C = - \boldsymbol{\alpha} $ 
(where $C$ is considered as the diagonal four by four matrix with 
all diagonal entries equal to the Clifford element $C$), and can be written
\be 
\boldsymbol{\alpha} = \left(\begin{array}{cccc} \hspace{2mm}  \ba \upsilon \hspace{2mm}& \hspace{2mm}b \hspace{2mm}& \hspace{2mm}  \ba q  \hspace{2mm}& \hspace{2mm} \ba p   \hspace{2mm}  \\
 \hspace{2mm}- b \hspace{2mm} &\hspace{2mm}   \ba \upsilon\hspace{2mm} & \hspace{2mm}- \ba p \hspace{2mm} & \hspace{2mm}\ba q \hspace{2mm} \\ \ba q  & - \ba p  &  \ba \upsilon & - b \\ \ba p  & \ba q  &  b &  \ba \upsilon \end{array} \right)  +  \left(\begin{array}{cccc} \hspace{2mm} 0 \hspace{2mm}& \hspace{2mm} a \hspace{2mm}&\hspace{2mm} 0\hspace{2mm}&0 \hspace{2mm} \\\hspace{2mm}-a \hspace{2mm}&0 &0&0\\ 0 &0  & 0 &a\\ 0  & 0  & -a & 0 \end{array} \right) \ .
\ee 
We define the coset representative $\V$ with generators $\mathbf{h}$ and $\mathbf{e}$ for the gravity fields $H$ and $B$, $\mathbf{ h^\prime }$ and $\mathbf{e^\prime  }$ for the dilaton $X$ and the axion $Y$, and $\mathbf{q}^a$ for the six electric fields $U^a$ and $\mathbf{p}^a$ for the six magnetic fields $A^a$. It is given by the matrix
\be \V =  \left(\begin{array}{cccc} \hspace{2mm}H^{\frac{1}{2}}  \hspace{2mm}& \hspace{2mm} H^{-\frac{1}{2}} \scal{ B - \sfrac{1}{2} [  \baaU , \baaa A] }\hspace{2mm}& \hspace{2mm}  X^{\frac{1}{2}}  \baaU  \hspace{2mm}& \hspace{2mm} X^{-\frac{1}{2}} \scal{ \baaa A + Y  \baaU }  \hspace{2mm}  \\
 \hspace{2mm} 0 \hspace{2mm} &\hspace{2mm} H^{-\frac{1}{2} } \hspace{2mm} & \hspace{2mm}0\hspace{2mm} & \hspace{2mm} 0 \hspace{2mm} \\  0  & - H^{- \frac{1}{2} } \baaa A  & X^{\frac{1}{2} }  & X^{-\frac{1}{2} } Y  \\ 0  &  H^{- \frac{1}{2} }  \baaU  & 0  & X^{-\frac{1}{2} } \end{array} \right) \ .\ee 
 The component of $\V^{-1} d \V$ lying in the orthogonal complement of  $\mathfrak{spin}(2,6) \oplus \mathfrak{so}(2) $ inside $\mathfrak{spin}(2,8)$ is given by 
 \begin{multline}  2 P \equiv \V^{-1} d \V + C ( \V^{-1} d \V )^t C = \\ \left(\begin{array}{cc} \hspace{2mm}H^{-1} d H   \hspace{2mm}& \hspace{2mm} H^{-1}  \scal{ d B  + \sfrac{1}{2} \{  \baaU , d \baaa A \} - \sfrac{1}{2} \{ \baaa A , d  \baaU \} }\hspace{2mm} \\ \hspace{2mm} H^{-1}  \scal{ d B  + \sfrac{1}{2} \{  \baaU , d \baaa A \} - \sfrac{1}{2} \{ \baaa A , d  \baaU \} } \hspace{2mm} &\hspace{2mm}- H^{-1} d H  \hspace{2mm} \\ -  \left(\frac{X}{H}\right)^{\frac{1}{2}} d  \baaU   & - (HX )^{- \frac{1}{2} } \scal{ d \baaa A + Y d  \baaU}   \\  - ( H X )^{-\frac{1}{2}} \scal{ d \baaa A + Y d  \baaU }  &  \left(\frac{X}{H}\right) ^{\frac{1}{2} } d  \baaU     \end{array} \right . \hspace{10mm} \\*
 \hspace{20mm} \left .  \begin{array}{cc} 
  \hspace{2mm}  \left(\frac{X}{H}\right)^{\frac{1}{2}} d  \baaU  \hspace{2mm}& \hspace{2mm} ( H X )^{-\frac{1}{2}} \scal{ d \baaa A + Y d  \baaU }  \hspace{2mm}  \\
   \hspace{2mm} ( H X )^{-\frac{1}{2}} \scal{ d \baaa A + Y d  \baaU } \hspace{2mm} & \hspace{2mm} -  \left(\frac{X}{H}\right)^{\frac{1}{2}} d  \baaU  \hspace{2mm} \\  
   X^{-1} d X   & X^{- 1 } d Y  \\
   X^{- 1 } d Y   & -  X^{-1} d X 
 \end{array}  \right)  \end{multline} 
in such a way that the action (\ref{n4action}) is given by
\be 
\int dx^2 \biggl( - 2 \partial^\alpha \sigma \partial_\alpha \rho  
+ \rho \trace P_\alpha P^\alpha \biggr) \ .
\ee

\subsection{$SO(2,6)\times SO(2)$-orbits of solutions}
The simplest Reissner--Nordstr\"{o}m like solutions of the $\N = 4$ 
theory are the ones for which the axion field is identically zero 
\cite{Cosmic}. The vector source term for the axion field then obeys 
\be 
\{ \partial^\alpha  \baaU , \partial_\alpha \baaa A \}- \{\partial^\alpha  
\baaa A , \partial_\alpha  \baaU \}= 0 \ .
\ee
These solutions have electric and magnetic charge vectors which are 
orthogonal in $\mathds{R}^6$. They can be obtained from the 
Schwarzschild solution by the following $Spin(2,6)$ transformation 
\be 
u(\ba p,\ba q) = \frac{1}{\sqrt{ (1 - p^2 ) ( 1- q^2)}} 
\left(\begin{array}{cccc} \, 1 \, & \, - \ba q \ba p \,  & \, 
\ba q\, & \, \ba p \, \\
 \, \ba q \ba p \, & 1 & - \ba p & \ba q \\
 \ba q & - \ba p & 1 & \, \ba q \ba p  \, \\ \ba p & \ba q & 
\, -\ba q \ba p \, & 1 \end{array}\right) 
\ee
with $\{ \ba q , \ba p \} = 0$. The action of $u(\ba p,\ba q)$ on the Schwarzschild matrix $v_0$ of mass 
$m=c$ gives the dilaton black hole matrix $\V$ through 
\be 
\V = u(\ba p, \ba q) \, v_0 \, u\left(  - \sqrt{\sfrac{r-c}{r+c} }\,  \ba p  
,  - \sqrt{\sfrac{r-c}{r+c}}\,   \ba q\right) \ .
\ee   
The dilaton black hole then has mass $M = \frac{1 - q^2 p^2 }{(1- q^2)
( 1 - p^2)} c$, electric charge $Q^a = \frac{q^a }{1-  q^2} c$ and magnetic 
charge $P^a = \frac{ p^a }{1- p^2} c$ with $P_a Q^a =0$, and dilaton charge 
$\Sigma = \frac{ \batP^2 - \batQ^2}{M}$. The BPS parameter is given by 
the formula
\be 
c^2 = M^2 - 2\,  \baa Q^2 - 2\,   \baaaP^2  + \Sigma^2 \ ,
\ee
while the coset representative $\V$ is
\be 
\V =  \left(\begin{array}{cccc} \vspace{2mm} \hspace{2mm}\sqrt{\frac{r^2 - c^2}{(r + M)^2  - \Sigma^2}} \hspace{2mm}& \hspace{2mm} \frac{ - 2 [ \, \, \baaQ, \, \, \baaP ]}{\sqrt{r^2 - c^2}\sqrt{ (r + M)^2 - \Sigma^2}} \hspace{2mm}& \hspace{2mm} \frac{2 \, \, \baaQ}{\sqrt{ (r + M)^2 - \Sigma^2}}  \hspace{2mm}& \hspace{2mm}  \frac{2 \, \, \baaP}{\sqrt{ (r + M)^2 - \Sigma^2}}   \hspace{2mm}  \\ \vspace{2mm}
 \hspace{2mm} 0 \hspace{2mm} &\hspace{2mm} \sqrt{\frac{(r + M)^2  - \Sigma^2}{r^2 - c^2}}  \hspace{2mm} & \hspace{2mm}0\hspace{2mm} & \hspace{2mm} 0 \hspace{2mm} \\ \vspace{2mm} 0  & -\frac{2 \, \, \baaP }{\sqrt{r^2 - c^2}} \sqrt{ \frac{ r + M - \Sigma}{r + M + \Sigma}}    &  \sqrt{ \frac{ r + M - \Sigma}{r + M + \Sigma}}  & 0  \\  \vspace{2mm}0  &  \frac{2 \, \, \baaQ }{\sqrt{r^2 - c^2}} \sqrt{ \frac{ r + M + \Sigma}{r + M - \Sigma}}  & 0  &\sqrt{ \frac{ r + M + \Sigma}{r + M - \Sigma}}  \end{array} \right) \ .
\ee 
The non-linear $SO(2)$ action of the $SL(2,\mathds{R})$ dilaton-axion sigma 
model permits one to obtain the general solution for arbitrary electric and 
magnetic charges and with a non-trivial axion field. The non-linear $SO(2)$ of 
the pure gravity $SL(2,\mathds{R})$ sigma model turns on the NUT charge, 
just as for pure gravity. For general solutions   \cite{N4general}, the dilaton and 
axion charges $\Sigma$ and $\Xi$ are given by\footnote{Recall that we 
  identify the elements of $Cl(6,\mathds{R})$ proportional to the unit matrix
  $\mathds{1}$ with ordinary real numbers.} 
\be 
\Sigma = \frac{(  \baaaP^2 - \, \baa Q^2 ) M + \{\, \baa Q ,  \baaaP \} N }{M^2 + N^2} \hspace{10mm} \Xi = \frac{ \{ \, \baa Q ,  \baaaP \} M - ( \baaaP^2 - \, \baa Q^2 ) N }{ M^2 + N^2} \label{N4charges} \ .
\ee 
The $\mathfrak{so}(2,8) \ominus ( \mathfrak{so}(2)\oplus\mathfrak{so}(2,6))$ 
charge matrix $\C$ is then given by 
\be 
\C= \left( \begin{array}{cccc} \,  M \, & \, -  N \, & \, -  
\, \baa Q \, & \, -   \baaaP \, \\
\, -  N \, & -  M & -  \baaaP &  \, \baa Q \\  \, \baa Q &   
\baaaP &  \Sigma & -  \Xi \\
   \baaaP & -  \, \baa Q & -  \Xi & -  \Sigma \end{array} \right) 
\ee 
(thus justifying our definition of the electric and magnetic charges including 
a factor $\sqrt{2}$ with respect to the usual one \cite{Cosmic}). 
The BPS parameter in the $SO(2,6)$ basis is given by 
\be 
c^2 \equiv \frac{1}{16}Ê\trace \C^2 =  M^2 + N^2 - 2 \, \baa Q^2 - 2  
\baaaP^2 +  \Sigma^2 + \Xi^2 \ .
\ee 
Using the explicit form of the charge matrix, the cubic equation 
$\C^3 = c^2 \C$ is perfectly equivalent to the complex equation 
\be  
( M + i N ) ( \Sigma + i \Xi ) = (  \baaaP + i \, \baa Q ) ^2 \ , \label{CS} 
\ee
from which the expression (\ref{N4charges}) for the scalar charges can be 
derived. To establish a link with the notation of the preceding sections, 
one must use the isomorphism $Spin(6) \cong SU(4)$ and the fact that the 
matrices $i [ÊC \gamma^a]_{ij}$ define a basis for the complex self-dual 
antisymmetric tensors of $SU(4)$.  We have that $Z_{ij} \equiv [ÊÊC ( 
\baaaP + i \, \baa QÊÊ)]_{ij}$ and $\Sigma_{ijkl} \equiv \frac{1}{4!} 
\varepsilon_{ijkl} \scal{Ê\Sigma + i \Xi}$. As explained in Section 
\ref{PureSpinorS}, equation (\ref{CS}) is in fact the $Spin^*(8)$ pure 
spinor equation, which corresponds within $U(1) \times SO(6,2)$ to the 
fact that the charge matrix defines a complex null vector. 

Because (\ref{CS}) is invariant under the action of $Spin(2,6) \times SO(2)$,
its solutions define non-linear representations of $SO(2,6) \times SO(2)$. 
The maximal compact subgroup $ U(4) \times SO(2) $ is linearly realised 
on $M,\, N,\, \, \baa Q$ and $ \baaaP$.  $U(4)$ acts only on 
$\, \baa Q + i  \baaaP$ as it does  on the vector fields, and $SO(2)$ 
rotates $ \baaaP$ into $\, \baa Q$ and $M$ into $N$ with doubled weight 
for the latter. The non-compact elements act non-linearly in the 
following way
\be
\begin{split}
M (q)  &= \frac{1}{ 1 - q^2} M -   \frac{q^2}{ 1 - q^2} \Sigma + \frac{1}{1 - q^2} \{ \ba q , \, \baa Q \} \\
N (q) &= \frac{1}{ 1 - q^2} N -   \frac{q^2}{ 1 - q^2} \Xi - \frac{1}{1 - q^2} \{ \ba q ,  \baaaP \} \\
M (p) &= \frac{1}{ 1 - p^2} M +   \frac{p^2}{ 1 - p^2} \Sigma + \frac{1}{1 - p^2} \{ \ba p ,  \baaaP \} \\
 N  (p)&= \frac{1}{ 1 -  p^2} N  +   \frac{p^2}{ 1 - p^2} \Xi +  \frac{1}{1 - p^2} \{ \ba p , \, \baa Q \}
    \end{split}\hspace{10mm}\begin{split}
 \, \baa Q (q) &= \frac{\, \baa Q + \ba q \, \baa Q\ba q }{ 1 - q^2} + \frac{ \ba q }{ 1 - q^2} ( M - \Sigma ) \\
     \baaaP(q)  &= \frac{ \baaaP + \ba q  \baaaP\ba q }{ 1 - q^2} - \frac{ \ba q }{ 1 - q^2} ( N - \Xi ) 
 \\ \, \baa Q  (p)&= \frac{\, \baa Q + \ba p \, \baa Q\ba p }{ 1 - p^2} + \frac{ \ba p }{ 1 - p^2} ( N + \Xi  ) \\
    \baaaP (p) &= \frac{ \baaaP + \ba p  \baaaP\ba p }{ 1 - p^2} + \frac{ \ba p }{ 1 - p^2} (M + \Sigma ) \ .
    \end{split}
\ee
For a non-zero fixed value of the BPS parameter $c$, this gives an 
irreducible representation of $SO(2,6) \times SO(2)$ on which this group 
acts transitively. One can see explicitly that the moduli spaces of 
spherically symmetric $\ft14$ BPS and $\ft12$ BPS Taub--NUT black holes  
(\ie $\mathcal{M}_1$  and  $\mathcal{M}_2$) define distinct 
$SO(2,6) \times SO(2)$-orbits from the factorisation of the BPS parameter 
square $c^2$ into
\be 
c^2 = \left( \sqrt{ M^2 + N^2}  - \frac{\, \baa Q^2 +  \baaaP^2 + 
\sqrt{- [\, \baa Q,  \baaaP]^2}}{\sqrt{ M^2 + N^2} } \right) \left(  
\sqrt{ M^2 + N^2}  - \frac{\, \baa Q^2 +  \baaaP^2 - \sqrt{ -[\, \baa Q,  
\baaaP]^2}}{\sqrt{ M^2 + N^2} }\right) \ .
\ee
If only one of these factors is zero, the solution becomes $\ft14$ BPS, and 
if both of them are zero (without the solution being trivial), it becomes one-half 
BPS. For the $\ft14$ BPS case, we consider in fact only the situation where the 
smaller factor is vanishing, so as to respect the positivity 
of the Bogomolny bounds. The compact subgroup $U(4) \times SO(2)$ leaves 
invariant each of these factors. Since the linear $SO(2) \times SO(2)$ 
acts freely on $\Sigma + i \Xi$ and $M + i N$, one can restrict oneself to 
the action of the non-compact generators for a dilaton black hole with 
$N = \Xi = 0$. In this case, one can write $ \baaaP$ and $\, \baa Q$ as 
numbers, and the non-compact generators then act non trivially only if $q^a$ 
is in the direction of $ Q^a$ and respectively if $p^a$ is in the direction 
of $P^a$. In this case, the Lie algebra action for the generators $\bf{p}$
and $\bf{q}$ on the two factors is 
\bea 
\mathbf{q}\;\; &:& \qquad \delta\left( M - \frac{(Q \pm P)^2}{M} \right) 
 = \mp \frac{2P}{M}   \left( M - \frac{(Q \pm P)^2}{M} \right) \CR 
\mathbf{p}\;\; &:& \qquad  \delta \left( M - \frac{(Q \pm P)^2}{M} \right) 
= \mp \frac{2Q}{M}   \left( M - \frac{(Q \pm P)^2}{M} \right) \ .
\eea
We see that the action of $SO(2,6) \times SO(2)$ on the two factors is a 
non-linear rescaling. Thus, this action leaves invariant the number of 
preserved supersymmetry charges of a given solution. We conclude that 
the irreducible representation of  $SO(2,6) \times SO(2)$  for a non-zero value of
$c$ decomposes for vanishing $c$ into three irreducible representations, 
which are the $\ft14$ BPS set, the $\ft12$ BPS set and the fully 
BPS Minkowski singlet, as stated in Section \ref{SpinOrbits}. 

Let us now describe the coset decomposition of the space of solutions. The product group of the trombone symmetry $\mathds{R}_+^*$ and $SO(2,6) \times SO(2)$ acts transitively on non-BPS solutions for a fixed value of $\frac{a}{c}$. Since the subgroup leaving a pure gravity solution invariant is the four-dimensional duality group, such an orbit takes the form $\scal{Ê\mathds{R}_+^* \times SO(2,6) \times SO(2) } / SO(2) \times SO(6)$. There are actually non-BPS solutions with a positive value of $c^2$ that do not lie on the Schwarzschild orbit. These can be obtained from the orbit of a purely dilatonic solution for which all the charges are zero except for the dilaton charge $\Sigma $. Such a charge obviously satisfies (\ref{CS}). The metric is then given by
\be ds^2  = \frac{r^2 - \Sigma^2}{r_+ r_-} \scal{Êdz^2 + d\rho^2} + \rho^2 d \varphi^2 - dt^2 \ee
and the associated Ricci scalar is $ \mathcal{R} = \frac{2 \Sigma^2}{(r^2 - \Sigma^2)^2} $.
Note that this solution has a naked singularity. In fact, all the solutions of the corresponding $SO(2,6) \times SO(2)$-orbit violate simultaneously the two Bogomolny bounds and  so this orbit consists entirely of badly behaved solutions and will be disregarded.

For BPS solutions, the action of the trombone is identified with the action of one of the generators of $SO(2,6) \times SO(2)$. It is enough to compute the isotropy subgroup for a particular solution. Starting from a $\ft14$ BPS solution with $\{ \, \baa Q ,  \baaaP \} = 0$ and $N=0$, the $ \mathfrak{spin}(2,6) \oplus \mathfrak{so}(2)$ elements commuting with the charge matrix take the following form\footnote{Note that $\, \baa Q$ and $ \baaaP$ are  both necessarily non-zero for a strictly $\ft14$ BPS solution.} 
\be \left(\begin{array}{cccc} \vspace{2mm} \hspace{2mm}\ba \upsilon \hspace{2mm}& \hspace{2mm} {\scriptstyle a +  \frac{ \{ \,\baQ , [\baupsilon , \baP] \} - 4 ( \,\baQ^2 + \baP^2 )a }{4  QP}}Ê\hspace{2mm}& \hspace{2mm} - \frac{ 2 a \baaP + [\baupsilon , \,\baQ] }{2Q} \hspace{2mm}& \hspace{2mm}   \frac{2 a \,\baaQ - [\baupsilon ,\, \baaP] }{2P}   \hspace{2mm}  \\ \vspace{2mm}
 \hspace{2mm}{\scriptstyle -  a -  \frac{ \{ \, \baQ, [\baupsilon , \baP] \} - 4 ( \, \baQ^2 + \baP^2 )a }{4  QP}Ê} \hspace{2mm} &\hspace{2mm}\ba \upsilon   \hspace{2mm} & \hspace{2mm} - \frac{2a \,\baaQ - [\baupsilon ,\, \baaP] }{2P}   \hspace{2mm} & \hspace{2mm} - \frac{ 2 a \baaP + [\baupsilon , \,\baaQ] }{2Q} \hspace{2mm} \\ \vspace{2mm} - \frac{ 2 a \baaP + [\baupsilon , \,\baaQ] }{2Q}  & - \frac{2 a \,\baaQ - [\baupsilon ,\, \baaP] }{2P}    & \ba \upsilon  &  {\scriptstyle  a -  \frac{ \{ \,\baQ , [\baupsilon , \baP] \} - 4 ( \,\baQ^2 + \baP^2 )a }{4 \,  Q P}}Ê\\  \vspace{2mm} \frac{2 a \,\baaQ - [\baupsilon ,\,\baaP] }{2P}    & - \frac{2 a \baaP  +  [\baupsilon , \,\baaQ] }{2Q}&{\scriptstyle - a+  \frac{ \{ \,\baQ , [\baupsilon , \baP] \} - 4 ( \,\baQ^2 + \baP^2 )a }{4   Q P}}Ê &\ba \upsilon  \end{array} \right) \ .\ee 
We define the indices $i,\, j,\, \cdots$ as the $SO(6)$ indices orthogonal to both $\, \baa Q$ and $ \baaaP$, and we take $1$ and $2$ as the index values for these directions. With the redefinitions
\be  {\upsilon^\alpha}_\beta \equiv {\varepsilon^\alpha}_\beta a 
\hspace{10mm} z \equiv \upsilon_{12}  \hspace{10mm} x^\alpha_i \equiv (\upsilon_{i1} , \upsilon_{i2})
\ee
where $\varepsilon_{\alpha\beta}$ is the $SO(2)$ antisymmetric invariant tensor, the corresponding generators have the following non-vanishing commutators
 \begin{gather}  [ {\upsi_i}^j , {\upsi_k}^l ] = 2 \delta_{[k}^{[j} {\upsi_{i]}}^{l]} \nonumber \\
  [Ê{\upsi^\alpha}_\beta , \x_i^\gamma ] = \delta_\beta^\gamma \x^\alpha_i 
 \hspace{10mm} [ {\upsi_i}^j , \x_k^\alpha ] = \delta_k^j \x^\alpha_i \nonumber \\
  [\x_i^\alpha , \x_j^\beta ] = \delta_{ij} \varepsilon^{\alpha\beta} \z\ .
 \end{gather}
We will call this algebra $\mathfrak{ic}(\mathfrak{so}(2)\oplus \mathfrak{so}(4))$, \ie this is the Poincar\'{e}-like algebra $\mathfrak{i}(\mathfrak{so}(2)\oplus \mathfrak{so}(4))$ with a central charge, and with corresponding group $I\hspace{-0.6mm}c(SO(2) \times SO(4))$. 

Purely electric dilatonic $\ft12$ BPS black holes have a charge matrix of the form
 \be \C= \left( \begin{array}{cccc} \,  Q \, & \, 0  \, & \, -  \, \baa Q \, & \, 0 \, \\
 \, 0 \, & -  Q  & 0  &  \, \baa Q \\  \, \baa Q &  0 &  -Q  &0 \\
  0  & -  \, \baa Q & 0  & Q  \end{array} \right) \ .\ee 
One can easily check that this matrix satisfies $\C^2 = 0$. The $ \mathfrak{spin}(2,6) \oplus \mathfrak{so}(2)$ elements that commute with this charge matrix are of the following form
\be \left(\begin{array}{cccc} \vspace{2mm} \hspace{2mm}\ba \upsilon \hspace{2mm}& \hspace{2mm} - \frac{\{ \bap , \, \baaQ \} }{2Q} Ê\hspace{2mm}& \hspace{2mm}- \frac{ [\baupsilon , \, \baaQ] }{2 Q}  \hspace{2mm}& \hspace{2mm} \ba p   \hspace{2mm}  \\ \vspace{2mm}
 \hspace{2mm}\frac{\{ \bap , \, \baaQ \} }{2Q} Ê \hspace{2mm} &\hspace{2mm} \ba \upsilon   \hspace{2mm} & \hspace{2mm} -\ba p   \hspace{2mm} & \hspace{2mm} - \frac{ [\baupsilon , \, \baaQ] }{2 Q}  \hspace{2mm} \\ \vspace{2mm} - \frac{ [\baupsilon , \, \baaQ] }{2 Q}   & -\ba p  &  \ba \upsilon  &   \frac{\{ \bap , \, \baaQ \} }{2Q} Ê \\  \vspace{2mm} \ba p  &- \frac{ [\baupsilon , \, \baaQ] }{2 Q} & - \frac{\{ \bap , \, \baaQ \} }{2Q} Ê  &\ba \upsilon  \end{array} \right) \ .\ee 
These elements generate the six-dimensional Poincar\'e algebra $\mathfrak{iso}(1,5)$, where the components of $\upsi$ and $\boldsymbol{p}$ orthogonal to $\, \baa Q$ generate $\mathfrak{so}(1,5)$ and their components collinear to $\, \baa Q$ generate the abelian subalgebra $\mathds{R}^6$. 

Finally, the space of asymptotically flat particle-like stationary solutions has the following decomposition into $SO(2,6) \times SO(2)$-orbits 
\be [-1,1]  \times \frac{\mathds{R}_+^* \times SO(2,6) \times SO(2)}{SO(2) \times SO(6)} \cup \frac{SO(2,6) \times SO(2)} { I\hspace{-0.6mm}c(SO(2)\times SO(4))} \cup  \frac{SO(2,6) \times SO(2)} { ISO(1,5)} \cup \{ 0 \} \ee
where $[-1,1]$ stands for the angular momentum per unit of mass, in perfect agreement with the results of Section \ref{SpinOrbits}.

\subsection{$\N=4$ supergravity coupled to $n$ vector multiplets}
Let us consider briefly the more general case of $\N=4$ supergravity coupled to $n$ vector multiplets. We will just give here the main results without explaining the full details. The scalar fields of the corresponding non-linear sigma model lie in the coset space $Spin(8,2+n)/(SO(6,2) \times SO(2,n))$ and the charge matrix $\C$ can be represented as a Majorana--Weyl chiral spinor of $Spin^*(8) \cong Spin(6,2)$ valued in the vector representation of $SO(2,n)$
\be 
| \C \rangle \equiv \left( \begin{array}{r}
\big(\w \  + Z_{ij} \ \, a^i a^j + \frac{1}{4} \varepsilon_{ijkl} \, \Sigma \ \,  
a^i a^j a^k a^l \big)|0\rangle  \vspace{2mm} \\*     
\big(Êz^A  + \Sigma_{ij_+}^A a^i a^j + \frac{1}{4} \varepsilon_{ijkl} \, \bar z^A  \, 
a^i a^j a^k a^l \big) |0,A\rangle  \end{array} \right)  \ ,
\ee
where the index $A$ lies in the vector representation of $SO(n)$.
Note that only the $SO(n)$ vector components obey the $Spin^*(8)$
self-duality constraint, while the first two components of the 
$SO(2,n)$ vector have been combined into a complex state. 
The `Dirac equation' (\ref{susy1}) gives the same constraints 
on $\w, \, Z_{ij}$ and $\Sigma$ as in the pure supergravity case 
and furthermore we have it that
\be 
\Scal{Ê\Sigma_{ij_+}^A - \frac{ z^A Z_{ij} }{\w} } \epsilon^j_\alpha = 0 
\ee
from which one can derive the $\ft14$ and the $\ft12$ BPS conditions.

It follows from the $3$-graded decomposition of the spinor representation of $\mathfrak{spin}(8,2+n)$ that the cubic constraint, $\C^3 =c^2 \C$, must be satisfied in the spinor representation, which implies its validity in the vector representation. Its components bilinear in the gamma matrices of $\mathfrak{spin}(6,2)$ and $\mathfrak{spin}(2,n)$ yield a component of $\C \otimes \C$ in the symmetric traceless rank two tensor representation of $SO(2,n)$ which vanishes, and its component bi-linear in the antisymmetric product of three gamma matrices of $\mathfrak{spin}(6,2)$ and $\mathfrak{spin}(2,n)$ yields a component of $\C \otimes \C \otimes \C$ in the product of the antisymmetric rank three tensor representation of $SO(6,2)$ times the antisymmetric rank three tensor representation of $SO(2,n)$ which vanishes too, \ie 
\be \eta_{\mathcal{I}\mathcal{J}} \, \C^{\mathcal{I}}_{\mathcal{A}} \C^{\mathcal{J}}_{\mathcal{B}} = \frac{1}{8} \eta_{\mathcal{A} \mathcal{B}} \, \eta^{\mathcal{C}\mathcal{D}} \,  \eta_{\mathcal{I}\mathcal{J}} \, \C^{\mathcal{I}}_{\mathcal{C}} \C^{\mathcal{J}}_{\mathcal{D}}  \hspace{15mm}Ê\C^{[\mathcal{I}}_{[\mathcal{A}} \C^{\mathcal{J}}_{\mathcal{B}} \C^{\mathcal{K}]}_{\mathcal{C}]} = 0 \ .\label{CubicMatter} \,\ee
where $\mathcal{I},\, \mathcal{J} ,\, \cdots $ and $\mathcal{A},\, \mathcal{B},\, \cdots $ lie in the vector representation of $SO(6,2)$ and $SO(2,n)$, respectively, and $\eta_{\mathcal{I}\mathcal{J}}$ and $ \eta_{\mathcal{A} \mathcal{B}}$ are the corresponding invariant tensors. The general solution is a non-rational function of $\w$, $Z_{ij}$ and $z^A$, but one can nevertheless determine the general solution by using the transitivity of $SO(6,2) \times SO(2,n)$ on non-extremal solutions. A general non-extremal solution ($c^2>0$) can indeed be obtained by acting with a general $SO(2,n)$ element on a general $Spin^*(8)$ pure spinor
\be | \C \rangle \equiv \left( \begin{array}{c}
\big({\rm x}  + X_{ij} a^i a^j + \frac{1}{2 {\rm x}} X_{ij} X_{kl}  a^i a^j a^k a^l  \big)|0\rangle
\vspace{2mm} \\*     0  \end{array} \right)  \ .
\ee
The $SO(2,n)$ element can be chosen to be the product of an $SO(2)$ rotation and transformations generated by two orthogonal non compact generators. It gives the general non-extremal solution of (\ref{CubicMatter}) as
\bea \w &=& e^{i\alpha} \left( \frac{ \cosh u + \cosh v }{2}  {\rm x} + \frac{ \cosh u - \cosh v }{2} \frac{1}{2 \bar {\rm x} } \varepsilon_{ijkl} X^{ij} X^{kl} \right) \CR
Z_{ij} &=& e^{i\alpha} \left( \frac{ \cosh u + \cosh v }{2} X_{ij}  + \frac{ \cosh u - \cosh v }{2} \frac{1}{2}  \varepsilon_{ijkl}  X^{kl} \right) \CR
\Sigma &=& e^{i\alpha} \left( \frac{ \cosh u + \cosh v }{2} \frac{1}{2  {\rm x} } \varepsilon^{ijkl} X_{ij} X_{kl}   + \frac{ \cosh u - \cosh v }{2} {\bar {\rm x}} \right) \CR
z^A &=&\frac{1}{2}Ê \hat{u}^A \sinh u \Scal{Ê{\rm x} + \frac{1}{2 \bar {\rm x} } \varepsilon_{ijkl} X^{ij} X^{kl} } + \frac{i}{2} \hat{v}^A \sinh v  \Scal{Ê{\rm x} - \frac{1}{2 \bar {\rm x} } \varepsilon_{ijkl} X^{ij} X^{kl} } \CR
\Sigma_{ij_+}^A &=&  \hat{u}^A \sinh u X_{ij_+} + i  \hat{v}^A \sinh v  X_{ij_-} \ ,\eea
where $ \hat{u}^A$ and $ \hat{v}^A $ are real orthogonal $SO(n)$ vectors of norm one. Non-BPS extremal solutions correspond to the limit where the $SO(2,n)$ element goes to the $SO(2,n)$ boundary, that is when either $u$, or $v$, or both go to infinity. The generic case corresponds to the limit where both go to infinity in such a way that $e^u - e^v$ remains finite.  The corresponding non-BPS extremal solutions satisfy
\be \Sigma =\frac{ \bar z^A \bar z_A}{ \bar \w} = \frac{1}{2 \w} \varepsilon^{ijkl} Z_{ij} Z_{kl}  \hspace{10mm}Ê \hspace{10mm} \bar z_A z^A = |\w|^2 + |\Sigma|^2 \ . \ee
Finally, there are two distinguished cases, either where $ |Êz^A z_A | < |\w|^2 $, in which case the solution remains  rather complicated in general, or where $ | z^A z_A| = | \w|^2 $, in which case
\be \Sigma^A_{ij_+} = \frac{z^A Z_{ij}}{\w}   =  \frac{1}{2} \varepsilon_{ijkl} \frac{\bar z^A Z^{kl}}{\bar \w} \ .\ee 
The strata are\footnote{Note, however, that $\mathcal{M}_\gra{0}{1}$,  $\mathcal{M}_{\gra{0}{1}^\circ} $ and $\mathcal{M}_\gra{1}{1}$ are empty in the case $n=1$, and note that $ISO(1)$ must be understood as the abelian translation group $\mathds{R}$. $\Ic(SO(n-2) \times SO(2))$ is $SO(2)\times \mathds{R}$ for $n=2$ and is $\Ic SO(2)$ for $n=3$. Moreover, $\mathcal{M}_\gra{0}{1}$,  $\mathcal{M}_{\gra{0}{1}^\circ} $ and $\mathcal{M}_\gra{1}{1}$ have two connected components in the case $n=2$, which can be transformed into one another by $O(2,2)$ parity.} 
\bea
&     \hspace{10mm} \mathcal{M}_\gra{0}{0} \cong  \mathds{R}_+^* \times \frac{ SO(6,2) \times SO(2,n) }{ SO(2) \times SO(6)  \times SO(n)} \CR
& \mathcal{M}_\gra{1}{0}  \cong \frac{ SO(6,2) \times SO(2,n) }{\Ic ( SO(4) \times SO(2) ) \times SO(n)} \hspace{20mm} \mathcal{M}_\gra{0}{1} \cong   \frac{ SO(6,2) \times SO(2,n) }{ SO(6) \times \Ic ( SO(n-2) \times SO(2) ) } \CR
& \mathcal{M}_{\gra{1}{0}^\circ}   \cong \frac{ SO(6,2) \times SO(2,n) }{\left( \mathds{R}_+^* \times SO(4) \times SO(n-1) \right) \ltimes \left( ({\bf 1} \oplus {\bf 4} \oplus {\bf n-1})^\ord{1} \oplus {\bf 4}^\ord{2} \oplus {\bf 1}^\ord{3} \right)} \hspace{60mm} \CR
&\hspace{60mm} \mathcal{M}_{\gra{0}{1}^\circ}  \cong   \frac{ SO(6,2) \times SO(2,n) }{ \left( \mathds{R}_+^* \times SO(5) \times SO(n-2) \right) \ltimes \left( ({\bf 1} \oplus {\bf 5} \oplus {\bf n-2})^\ord{1} \oplus {\bf n-2}^\ord{2} \oplus {\bf 1}^\ord{3} \right) } \CR
 &     \hspace{10mm} \mathcal{M}_\gra{1}{1}  \cong  \frac{ SO(6,2) \times SO(2,n) }{  ( GL(2,\mathds{R}) \times SO(4) \times SO(n-2) ) \ltimes (  {\bf 1}^\ord{-2} \oplus  {\bf \bar 2}^\ord{-1} \otimes {\bf 4}Ê\oplus  {\bf  2}^\ord{1}\otimes   {\bf n-2}  \oplus {\bf 1}^\ord{2} )}  \CR
&  \mathcal{M}_\gra{2}{0} \cong \frac{ SO(6,2) \times SO(2,n) }{ ISO(5,1)  \times SO(1,n)} 
\hspace{20mm}Ê \mathcal{M}_\gra{0}{2} \cong \frac{ SO(6,2) \times SO(2,n) }{SO(6,1) \times ISO(1,n-1)} 
\CR
&\hspace{10mm} \mathcal{M}_\gra{2}{2} \cong \frac{ SO(6,2) \times SO(2,n) }{ ( SO(1,1) \times SO(5,1) \times SO(1,n-1) ) \ltimes ( {\bf 6}^\ord{-1} \oplus  {\bf n }^\ord{1} )  }
 \eea
 where the stratum $\mathcal{M}_\gra{p}{q}$ corresponds to solutions which are $\ft p 4$ BPS, and $\mathcal{M}_\gra{p}{q} \subset \overline{\mathcal{M}}_\gra{r}{s}$ if and only if both $p\ge r$ and $q\ge s$ (with, in addition, $\partial \mathcal{M}_\gra{1}{0} = \overline{\mathcal{M}}_{\gra{1}{0}^\circ} $ and $\partial \mathcal{M}_\gra{0}{1} = \overline{\mathcal{M}}_{\gra{0}{1}^\circ} $). The properties of the strata are summarised in the Table below 
\begin{gather}
\begin{array}{|c|c|c|c|}
\hline
 & \,\,  \mbox{dim} \,\, & \,\,  \mbox{nilpotency} \,\, &  \,\,  \mbox{Horizon area} \,\, \\*
 & & &\vspace{-4mm} \\*
 \hline
 & & &\vspace{-4mm} \\*
\hspace{2mm} \mathcal{M}_\gra{0}{0}   \hspace{2mm}&\hspace{2mm}14+2n \hspace{2mm} & \hspace{2mm} \C^3 = c^2 \C  \hspace{7mm} [ \C , [ \C , [ \C ,   \Upgamma^{\mathpzc{M}} ]]] = c^2 [ \C ,   \Upgamma^{\mathpzc{M}} ]  \hspace{2mm}  & \hspace{2mm}   A >0 \hspace{2mm} \\*
\ & & &\vspace{-4mm} \\*
 \hline
 & & &\vspace{-4mm} \\*
\hspace{2mm} \mathcal{M}_\gra{1}{0} ,\,  \mathcal{M}_\gra{0}{1}    \hspace{2mm}&\hspace{2mm}13+2n \hspace{2mm} & \hspace{2mm} \C^3 = 0 \hspace{5mm} [ \C , [ \C , [ \C ,   \Upgamma^{\mathpzc{M}} ]]] = 0 \hspace{5mm} {\ad_\C}^5 = 0   \hspace{2mm}  & \hspace{2mm}   A >0 \hspace{2mm}\\*
 & & & \vspace{-4mm} \\*
 \hline
 & & &\vspace{-4mm} \\*
\hspace{2mm} \mathcal{M}_{\gra{1}{0}^\circ}  ,\,  \mathcal{M}_{\gra{0}{1}^\circ}     \hspace{2mm}&\hspace{2mm}12+2n \hspace{2mm} & \hspace{2mm} \C^3 = 0 \hspace{5mm} [ \C , [ \C , [ \C ,   \Upgamma^{\mathpzc{M}} ]]] = 0 \hspace{5mm} {\ad_\C}^4 = 0   \hspace{2mm} & \hspace{2mm}   A = 0 \hspace{2mm} \\*
 & & &\vspace{-4mm} \\*
 \hline
 & & &\vspace{-4mm} \\*
 \hspace{2mm} \mathcal{M}_\gra{1}{1}   \hspace{2mm}&\hspace{2mm}10+2n \hspace{2mm} & \hspace{2mm}  \C^3 = 0 \hspace{5mm}  [ \C , [ \C ,   \Upgamma^{\mathpzc{M}} ]] = 0 \hspace{5mm} {\ad_\C}^3 = 0 \hspace{2mm} & \hspace{2mm}   A = 0 \hspace{2mm} \\*
 & & & \vspace{-4mm} \\*
 \hline
 & & &\vspace{-4mm} \\*
\hspace{2mm} \mathcal{M}_\gra{2}{0} ,\, \mathcal{M}_\gra{0}{2}   \hspace{2mm}&\hspace{2mm}8+n \hspace{2mm} & \hspace{2mm}  \C^2 = 0 \hspace{5mm} [ \C , [ \C , [ \C ,   \Upgamma^{\mathpzc{M}} ]]] = 0 \hspace{5mm} {\ad_\C}^3 = 0 \hspace{2mm} & \hspace{2mm}   A = 0 \hspace{2mm} \\*
 & & &\vspace{-4mm} \\*
 \hline
 & & &\vspace{-4mm} \\*
 \hspace{2mm} \mathcal{M}_\gra{2}{2}  \hspace{2mm}&\hspace{2mm}7+n \hspace{2mm} & \hspace{2mm} \C^2 = 0 \hspace{9mm} [ \C , [ \C ,   \Upgamma^{\mathpzc{M}} ]] = 0 \hspace{9mm} {\ad_\C}^3 = 0  \hspace{2mm} & \hspace{2mm}   A = 0 \hspace{2mm} \vspace{-4mm} \\*
 & & & \\*
 \hline
\end{array} \nonumber
\end{gather}
\begin{centerline} {\small Table VII : dimension of strata in $\N=4$ supergravity with $n$ vector multiplets} \end{centerline}
This stratification is in agreement with the stratification of the nilpotent orbits $\mathfrak{N}_{SO(8,2+n)}$ of $SO(8,2+n)$ as described in \cite{SOpqstrat,SOpqstrat2}. Nevertheless, for $n\ge 2$, the stratification of  $\mathfrak{N}_{SO(8,2+n)}$ suggests that there is an additional stratum of charge matrices which correspond to extremal black holes without any saturated central charge
\be \mathcal{M}_{\gra{0}{0}^\circ} \cong \frac{ SO(6,2) \times SO(2,n) }{  ISO(5)  \times ISO(n-1) \times \mathds{R} } \ee
which satisfies the ordering $\mathcal{M}_{\gra{1}{0}^\circ} \cup \mathcal{M}_{\gra{0}{1}^\circ} \subset \overline{\mathcal{M}}_{\gra{0}{0}^\circ} \subset \overline{\mathcal{M}}_\gra{0}{0}$.   Such solutions do indeed exist, some examples of which having been found within the $STU$ model \cite{nonBPSextrem}.

The fact that spherically symmetric extremal solutions of $\N=4$ supergravity are associated to nilpotent orbits of $SO(8,2+n)$ has already been discussed in \cite{Rousset}. Note that although the $\ft14$ BPS solutions are naturally related to the complex geometry of twistor spaces, this is not necessarily the case for the $\ft12$ BPS ones. For instance, our analysis (although not yet complete) leads us to believe that the general $\ft12$ BPS solutions of $\N=4$ supergravity coupled to $n$ vector multiplets depend on $2+n$ harmonic functions (instead of $4+2n$ harmonic functions for the general $\ft14$ BPS solutions).\footnote{The number of harmonic functions is the dimension of the maximal vector space lying inside the relevant stratum (\ie $\mathds{R}^{4+2n} \subset \mathcal{M}_\gra{1}{0}$ and $\mathds{R}^{2+n} \subset \mathcal{M}_\gra{2}{0}$ for the $\ft14$ and $\ft12$ BPS solutions respectively).} Roughly speaking, the $\ft14$ BPS constraints are holomorphic in the complex charges $\w,\, Z_{ij}$ and $z^A$, whereas the $\ft12$ BPS constraints involve a reality condition coming from the complex self-duality of the vector multiplets.

The embedding $SO(8,2+n) \subset E_{8(8)}$ for $n\le 6$ implies that $\N=4$ supergravity coupled to  $n\le 6$ vector multiplets is a consistent truncation of $\N=8$ supergravity. The solutions lying inside $\mathcal{M}_\gra{p}{q}$ are then $\frac{p+q}{8}$ BPS in $\N=8$ supergravity. Note that the charge matrices which lie in the minimal adjoint orbit of $SO(6,2+n)$ correspond to $\ft12$ BPS solutions in maximal supergravity. This suggests the existence of an intriguing link between minimal adjoint orbits and maximally supersymmetric black holes.

\section{Conclusion}
\label{Concl}

In this paper, we have characterised in depth the stationary asymptotically flat solutions of $D=4$ supergravities by a detailed analysis of the duality orbits of the corresponding timelike-reduced Euclidean-signature $D=3$ supergravities. This proceeds initially by analogy with the classification \cite{3Dclass} of solutions to three-dimensional supergravities obtained via a spacelike dimensional reduction. A special feature of these Euclidean stationary solution orbits, however, is the noncompact nature of the isotropy group $\H^*$, which appears upon making a timelike dimensional reduction to $D=3$. For $\N$-extended supergravity, the group $\H^*$ is the product of $Spin^*(2\N)_{\rm c} \cong Spin^*(2\N) / \ker(S_+)$ (with $\ker(S_+)$ being the kernel of the $Spin^*(2\N)$ chiral Weyl spinor representation) with a group determined by the matter content of the theory.

Rejecting orbits that contain only solutions with naked singularities led us to the quintic characteristic equation (\ref{polynom}) for the charge matrix $\C$ (\ref{Charge}). In all but two exceptional cases where the $D=3$ symmetry groups are $E_{8(8)}$ or $E_{8(-24)}$, this characteristic equation is strengthened to a cubic equation (\ref{cubic}) for the charge matrix. The $D=3$ charge matrix $\C$ is the Noether charge for the $D=3$ duality symmetry; the characteristic equations determine its values in terms of the smaller number of $D=4$ charges of the same theory (\ie the gravitational mass and NUT charge and the various electric and magnetic charges of the vector field species). This analysis works for rotating as well as non-rotating solutions; the characteristic equation guarantees that each acceptable orbit passes through some Kerr solution. For pure $\N$-extended supergravity with $\N\le 5$, the characteristic is equivalent to the Cartan pure spinor condition on the Weyl $Spin^*(2\N)$ spinor $|\C\rangle$.   

The characteristic equations involve the BPS parameter $c^2 = \frac{1}{k} \trace \C^2$ (\ref{bpsparam}). Extremal rotating solutions have $c^2=a^2$, where $a$ is the angular momentum parameter. Non-rotating extremal solutions thus have $c^2=0$, leading to a key algebraic feature of the extremal solution suborbits: the charge matrix becomes nilpotent -- cubic in most cases, quintic in the two $E_8$ exceptional cases. This allowed us to make contact with  extensive studies of nilpotent orbits of noncompact groups in the mathematical literature \cite{Sekiguchi,homeoKS,E7strat,MagicStrat,E8strat}.

The extremality condition is not always synonymous with the BPS condition, however. For pure $\N\leq 5$ supergravities, the two conditions are synonymous, but not for $\N=6$ or $\N=8$ or any supergravity coupled to vector supermultiplets. Algebraic analysis of the BPS solutions led us to the `Dirac equation' condition (\ref{susy}) in which the charge matrix $\C$ is given an interpretation as a $Spin^*(2\N)$ Weyl spinor, using a creation/annihilation operator construction for the $\mathfrak{so}^*(2\N)$ generators. This `Dirac equation' allows the charge matrix $\C$ to be solved for explicitly in terms of a simple rational function of the $D=4$ charges.

Having established the relevant families of stationary supergravity solutions, we extended the $D\ge 4$ analysis \cite{Active} of active duality transformations (\ie transformations that leave the asymptotic values of all fields unchanged) to the action of the three-dimensional duality group $\G$ on these solution families. As in the higher-dimensional cases, in order to preserve the asymptotic values of the fields, the active realisations operate via the quotient of $\G$ by $\P_0$, the quotient group of its maximal parabolic subgroup $\P$ by its defining $\mathds{R}_+^*$ subgroup. Here, a peculiarity of the non-compact nature of the $D=3$ scalar isotropy group $\H^*$ plays a key r\^ole: although the Iwasawa decomposition remains valid almost everywhere in the moduli space of solutions, it fails precisely on the subspace of extremal solutions. The Iwasawa failure set is not in general homeomorphic to the moduli space of spherically symmetric extremal solutions, however. As a result, there is not in general a well-defined active group action on the whole stationary solution space -- some $\G$ transformations become singular as one approaches the extremal strata. As a result, one has to speak of an `almost group action' of active transformations on the solution space. We speculate that this curious problem may be resolved in cases where the isotropy group $\H^*$ is semi-simple, in particular for the $\N=8$ theory.

The results of this $D=3$ duality group analysis should have a bearing on the debate, continuing since the appearance of reference \cite{HT}, about the extent to which continuous duality symmetries of lower-spacetime-dimensional classical supergravity theories should be replaced by arithmetic subgroups such as $E_8(\Z)$ at the quantum level. Although such subgroups certainly exist in the abstract, their concrete realisation as quantum symmetries is problematical because there does not appear to be any way in which the concrete active realisations ({\it c.f.}\ (\ref{nonlinear}) for an $SL(2,\mathds{R})$ example) that we have found for the action of the $D=3$ duality groups $\G$ on the $D=3$ charges might be consistent with a Dirac quantisation rule.


\section*{Appendices}
\appendix 
 
\section{Simple duality groups and their five-graded decomposition}
\label{groups}

Let us review briefly the various simple duality groups in three dimensions which occur in time-like dimensionally reduced four-dimensional theories. We will see that excepted for $E_8$, all these groups have a five-graded decomposition with respect to which their fundamental representation admits a three-graded decomposition. 

Most of these theories can be embedded into supergravity theories. Whenever the symmetric space in which the four-dimensional scalars lie is K\"{a}hler, the theory can be embedded into an $\N=1$ supergravity. When the symmetric space is furthermore special K\"{a}hler, the theory can moreover be embedded into an $\N=2$ supergravity coupled to several vector multiplets. An $\N=2$ supergravity theory with hypermultiplets always leads to a three-dimensional theory with a reducible symmetric space of scalars, and we do not consider such cases in the present publication. The homogenous special K\"{a}hler spaces have been classified in \cite{Cremmer}. See \cite{Ferrara} for a complete classification.  
\subsection*{a) $SL(2+n,\mathds{R})/SO(2,n)$}
This coset space corresponds to the dimensional reduction of pure gravity in $4+n$ dimensions. The scalar fields of the four-dimensional theory lie in the coset $GL(n,\mathds{R})/SO(n)$ and the five-graded decomposition of $\mathfrak{sl}(2+n,\mathds{R})$ is as follows 
\be \mathfrak{sl}(2+n,\mathds{R})   \cong  {\bf 1}^{\ord{-2}} \oplus ({{ \Yboxdim8pt  \yng(1)}^\ord{2} \oplus \overline{ { \Yboxdim8pt  \yng(1)}}^\ord{-2}} )^{\ord{-1}} \oplus {\bf 1}^\ord{0} \oplus \scal{Ê \mathfrak{gl}(1,\mathds{R}) \oplus \mathfrak{sl}(n,\mathds{R})}^\ord{0}  \oplus ({{ \Yboxdim8pt  \yng(1)}^\ord{2} \oplus \overline{ { \Yboxdim8pt  \yng(1)}}^\ord{-2}} )^\ord{1} \oplus {\bf 1}^\ord{2} \ .\ee
The fundamental representation decomposes as 
\be \boldsymbol{n+2}  \cong  ({\bf 1}^\ord{-1})^{\ord{-1}} \oplus ({{ \Yboxdim8pt  \yng(1)}^\ord{1} })^\ord{0} \oplus  ({\bf 1}^\ord{-1})^\ord{1} \ .\ee
\subsection*{b) $SU(1+m,1+n)/S(U(m,1) \times U(1,n))$}
The corresponding four-dimensional theory is the bosonic sector of an $\N=1$ supergravity coupled to $m+n$ abelian vector supermultiplets and $mn$ scalar supermultiplets. In the special case $m=0,\, n=1$, this theory is Maxwell--Einstein theory, which is also the bosonic sector of $\N=2$ pure supergravity. For $m=1$ it is the bosonic sector of an $\N=2$ supergravity coupled to $n$ abelian vector supermultiplets. For $m=3$ it is the bosonic sector of $\N=3$ supergravity theory coupled to $n$ abelian vector supermultiplets. The scalar fields of the four-dimensional theory lie in the K\"{a}hler coset $U(m,n)/(U(m) \times U(n))$ and the five-graded decomposition of $\mathfrak{su}(1+m,1+n)$ is as follows 
\be \mathfrak{su}(1+m,1+n)  \cong  {\bf 1}^{\ord{-2}} \oplus ({{ \Yboxdim8pt  \yng(1)}^\ord{2} \oplus \overline{ { \Yboxdim8pt  \yng(1)}}^\ord{-2}} )^{\ord{-1}} \oplus {\bf 1}^\ord{0} \oplus  (\mathfrak{u}(1) \oplus \mathfrak{su}(m,n))^\ord{0}  \oplus ({{ \Yboxdim8pt  \yng(1)}^\ord{2} \oplus \overline{ { \Yboxdim8pt  \yng(1)}}^\ord{-2}} )^\ord{1} \oplus {\bf 1}^\ord{2} \ .\ee
The complex fundamental representation decomposes as 
\be \boldsymbol{m+n+2}  \cong  ({\bf 1}^\ord{-1}_\mathds{C})^{\ord{-1}} \oplus ({{ \Yboxdim8pt  \yng(1)}^\ord{1} \oplus \overline{ { \Yboxdim8pt  \yng(1)}}^\ord{1}} )^\ord{0} \oplus  ({\bf 1}_\mathds{C}^\ord{-1})^\ord{1} \ .\ee
\subsection*{c) $SO(2+m,2+n)/(SO(m,2) \times SO(2,n))$}
For $m=2$ the corresponding four-dimensional theory is the bosonic sector an $\N=2$ supergravity coupled to $1+n$ abelian vector supermultiplets. In the case $m=6$, this is the bosonic sector of $\Nquatre$ supergravity coupled to $n$ abelian vector supermultiplets. The scalar fields lie in the coset $SO(2,1)/ SO(2) \times SO(m,n)/(SO(m) \times SO(n))$ and the five-graded decomposition of $\mathfrak{so}(2+m,2+n)$ is as follows 
\be \mathfrak{so}(2+m,2+n)  \cong  {\bf 1}^{\ord{-2}} \oplus ({{ \Yboxdim8pt  \yng(1)} \otimes  { \Yboxdim8pt  \yng(1)}} )^{\ord{-1}} \oplus {\bf 1}^\ord{0} \oplus  \scal{Ê\mathfrak{sl}(2,\mathds{R}) \oplus \mathfrak{so}(m,n)}^\ord{0}  \oplus ({{ \Yboxdim8pt  \yng(1)} \otimes  { \Yboxdim8pt  \yng(1)}} )^\ord{1} \oplus {\bf 1}^\ord{2} \ .\ee
It is convenient to consider the irreducible spinor representations $S^\pm$ of $Spin(2+m,2+n)$ and $Spin(m,n)$, for which we get the decomposition
\be S^\pm  \cong  ({\boldsymbol{1} \otimes S^\mp})^{\ord{-1}} \oplus ({{ \Yboxdim8pt  \yng(1)}} \otimes S^\pm )^\ord{0} \oplus  ({{\boldsymbol{1} \otimes S^\mp}})^\ord{1} \ .\ee
The vector representation decomposes as
\be V \cong ({ \Yboxdim8pt  \yng(1)}  \otimes {\bf 1})^{\ord{-1}} \oplus ( {\bf 1}Ê\otimes { \Yboxdim8pt  \yng(1)}  )^\ord{0} \oplus  ({ \Yboxdim8pt  \yng(1)}  \otimes {\bf 1} )^\ord{1} \ .\ee
\subsection*{d) $SO^*(4 + 2n)/U(2,n)$}
For $n=0$, $Spin^*(4) \cong SU(2) \times SL(2,\mathds{R})$ and the corresponding four-dimensional theory is Einstein theory, \ie the bosonic sector of pure $\N=1$ supergravity. In the case $n=1$, $Spin^*(6) \cong SU(1,3)$ and the corresponding four-dimensional theory is the above-discussed bosonic sector of an $\N=1$ supergravity coupled to $2$ vector supermultiplets. In general, the corresponding four-dimensional theory is the bosonic sector of an $\N=1$ supergravity coupled to $2n$ abelian vector supermultiplets and $\frac{n(n-1)}{2}$ scalar supermultiplets. The scalar fields of the latter lie in the K\"{a}hler coset $SU(2)/ SU(2) \times SO^*(2n)/ U(n) $, and the five-graded decomposition of $\mathfrak{so}^*(4 + 2n)$ is as follows
\be \mathfrak{so}^*(4 + 2n)  \cong  {\bf 1}^{\ord{-2}} \oplus ({{ \Yboxdim8pt  \yng(1)} \otimes_\mathds{C}   { \Yboxdim8pt  \yng(1)}} )^{\ord{-1}} \oplus {\bf 1}^\ord{0} \oplus  \scal{Ê\mathfrak{su}(2) \oplus \mathfrak{so}^*(2n)}^\ord{0}  \oplus ({{ \Yboxdim8pt  \yng(1)} \otimes_\mathds{C}  { \Yboxdim8pt  \yng(1)}} )^\ord{1} \oplus {\bf 1}^\ord{2} \ .\ee
It is convenient to consider the irreducible spinor representations $S^\pm$ of $Spin^*(4 + 2n)$ and $Spin^*(2n)$, for which we get the decomposition
\be S^\pm  \cong  ({\boldsymbol{1} \otimes  S^\pm})^{\ord{-1}} \oplus ({{ \Yboxdim8pt  \yng(1)}} \otimes_\mathds{C} S^\mp )^\ord{0} \oplus  ({{\boldsymbol{1} \otimes S^\pm}})^\ord{1} \ .\ee
\subsection*{e) $Sp(2 + 2n,\mathds{R})/ U(1,n)$}
The corresponding four-dimensional theory is the bosonic sector of an $\N=1$ supergravity coupled to $n$ abelian vector supermultiplets and $\frac{n(n+1)}{2}$ scalar supermultiplets. The scalar fields of the latter lie in the K\"{a}hler coset $Sp(2n,\mathds{R})/ U(n)$ and the five-graded decomposition of $\mathfrak{sp}(2+2n,\mathds{R})$ is as follows 
\be \mathfrak{sp}(2+2n,\mathds{R})  \cong  {\bf 1}^{\ord{-2}} \oplus {{ \Yboxdim8pt  \yng(1)}}^{\ord{-1}} \oplus {\bf 1}^\ord{0} \oplus  \mathfrak{sp}(2n,\mathds{R}) ^\ord{0}  \oplus {{ \Yboxdim8pt  \yng(1)}}^\ord{1} \oplus {\bf 1}^\ord{2} \ .\ee
The fundamental representation decomposes as 
\be \boldsymbol{2 +2n}  \cong  {\bf 1}^{\ord{-1}} \oplus {{ \Yboxdim8pt  \yng(1)}}^\ord{0} \oplus  {\bf 1}^\ord{1} \ .\ee
\subsection*{f) $G_{2(2)}/ (SU(1,1)\times SU(1,1))$}
The corresponding four-dimensional theory is the bosonic sector of an $\N=2$ supergravity theory coupled to one vector supermultiplet, which corresponds itself to the dimensional reduction of minimal supergravity in five dimensions. The scalar fields of the four-dimensional theory lie in the special K\"{a}hler coset $SU(1,1) / U(1)$ and the five-graded decomposition of $\mathfrak{g}_{2(2)}$ is as follows 
\be \mathfrak{g}_{2(2)} \cong  {\bf 1}^{\ord{-2}} \oplus {{ \Yboxdim8pt  \yng(3)}}^{\ord{-1}} \oplus {\bf 1}^\ord{0} \oplus  \mathfrak{su}(1,1)^\ord{0}  \oplus {{ \Yboxdim8pt  \yng(3)}}^\ord{1} \oplus {\bf 1}^\ord{2} \ .\ee
The fundamental representation decomposes as 
\be \boldsymbol{7}  \cong {{ \Yboxdim8pt  \yng(1)}}^\ord{-1} \oplus {{ \Yboxdim8pt  \yng(2)}}^\ord{0} \oplus  {{ \Yboxdim8pt  \yng(1)}}^\ord{1} \ .\ee
\subsection*{g) $F_{4(4)}/ (SU(1,1) \times Sp(6,\mathds{R}))$}
The corresponding four-dimensional theory is the bosonic sector of the real magic $\N=2$ supergravity, which admits $6$ abelian vector supermultiplets. The scalar fields of the latter lie in the special K\"{a}hler coset $Sp(6,\mathds{R})/ U(3)$ and the five-graded decomposition of $\mathfrak{f}_{4(4)}$ is as follows 
\be \mathfrak{f}_{4(4)} \cong  {\bf 1}^{\ord{-2}} \oplus {{ \Yboxdim8pt  \yng(1,1,1)}}^{\ord{-1}} \oplus {\bf 1}^\ord{0} \oplus  \mathfrak{sp}(6,\mathds{R})^\ord{0}  \oplus {{ \Yboxdim8pt  \yng(1,1,1)}}^\ord{1} \oplus {\bf 1}^\ord{2} \ .\ee
The fundamental representation decomposes as 
\be \boldsymbol{26}  \cong {{ \Yboxdim8pt  \yng(1)}}^\ord{-1} \oplus {{ \Yboxdim8pt  \yng(1,1)}}^\ord{0} \oplus  {{ \Yboxdim8pt  \yng(1)}}^\ord{1} \ .\ee
\subsection*{h) $E_{6(6)}/ Sp(8,\mathds{R})$}
The scalar fields of the corresponding four-dimensional theory lie in the coset $SL(6,\mathds{R})/ SO(6)$ and the five-graded decomposition of $\mathfrak{e}_{6(6)}$ is as follows 
\be \mathfrak{e}_{6(6)} \cong  {\bf 1}^{\ord{-2}} \oplus {{ \Yboxdim8pt  \yng(1,1,1)}}^{\ord{-1}} \oplus {\bf 1}^\ord{0} \oplus  \mathfrak{sl}(6,\mathds{R})^\ord{0}  \oplus {{ \Yboxdim8pt  \yng(1,1,1)}}^\ord{1} \oplus {\bf 1}^\ord{2} \ .\ee
The fundamental representation decomposes as 
\be \boldsymbol{27}  \cong \overline{{{ \Yboxdim8pt  \yng(1)}}}^\ord{-1} \oplus {{ \Yboxdim8pt  \yng(1,1)}}^\ord{0} \oplus  \overline{{{ \Yboxdim8pt  \yng(1)}}}^\ord{1} \ .\ee
\subsection*{i) $E_{6(2)}/ (SU(1,1)\times SU(3,3))$}
The corresponding four-dimensional theory is the bosonic sector of the complex magic $\N=2$ supergravity, which admits $9$ abelian vector supermultiplets. The scalar fields of the latter lie in the special K\"{a}hler coset $SU(3,3)/ S(U(3)\times U(3))$ and the five-graded decomposition of $\mathfrak{e}_{6(2)}$ is as follows 
\be \mathfrak{e}_{6(2)} \cong  {\bf 1}^{\ord{-2}} \oplus {{ \Yboxdim8pt  {\yng(1,1,1)}}_+}^{\ord{-1}} \oplus {\bf 1}^\ord{0} \oplus  \mathfrak{su}(3,3)^\ord{0}  \oplus {{ \Yboxdim8pt  {\yng(1,1,1)}}_+}^\ord{1} \oplus {\bf 1}^\ord{2} \ ,\ee
where the $_+$ subscript states for complex-self-duality. The complex fundamental representation decomposes as 
\be \boldsymbol{27}  \cong \overline{{{ \Yboxdim8pt  \yng(1)}}}^\ord{-1} \oplus {{ \Yboxdim8pt  \yng(1,1)}}^\ord{0} \oplus  \overline{{{ \Yboxdim8pt  \yng(1)}}}^\ord{1} \ .\ee
\subsection*{j) $E_{6(-14)}/ (U(1) \times SO^*(10) )$}
The corresponding  four-dimensional theory is the bosonic sector of $\N=5$ supergravity. The scalar fields lie in the coset $SU(5,1)/ U(5)$ and the five-graded decomposition of $\mathfrak{e}_{6(-14)}$ is as follows 
\be \mathfrak{e}_{6(-14)} \cong  {\bf 1}^{\ord{-2}} \oplus{{ \Yboxdim8pt  {\yng(1,1,1)}}_+}^{\ord{-1}} \oplus {\bf 1}^\ord{0} \oplus  \mathfrak{su}(5,1)^\ord{0}  \oplus {{ \Yboxdim8pt  {\yng(1,1,1)}}_+}^\ord{1} \oplus {\bf 1}^\ord{2} \ .\ee
The complex fundamental representation decomposes as 
\be \boldsymbol{27}  \cong \overline{{{ \Yboxdim8pt  \yng(1)}}}^\ord{-1} \oplus {{ \Yboxdim8pt  \yng(1,1)}}^\ord{0} \oplus  \overline{{{ \Yboxdim8pt  \yng(1)}}}^\ord{1} \ .\ee
\subsection*{k) $E_{7(7)}/ SU(4,4)$}
The scalar fields of the corresponding four-dimensional theory lie in the coset $SO(6,6)/ (SO(6)\times SO(6))$ and the five-graded decomposition of $\mathfrak{e}_{7(7)}$ is as follows 
\be \mathfrak{e}_{7(7)} \cong  {\bf 1}^{\ord{-2}} \oplus S_+^{\ord{-1}} \oplus {\bf 1}^\ord{0} \oplus  \mathfrak{spin}(6,6)^\ord{0}  \oplus S_+^\ord{1} \oplus {\bf 1}^\ord{2} \ee
The fundamental representation decomposes as 
\be \boldsymbol{56}  \cong V^\ord{-1} \oplus S_-^\ord{0} \oplus  V^\ord{1} \ ,\ee
where $S_\pm$ are the 32-dimensional Majorana--Weyl representations of $Spin(6,6)$ and $V$ is the vector representation of $SO(6,6)$. 
\subsection*{l) $E_{7(-5)}/ (SU(1,1) \times SO^*(12))$}
The corresponding four-dimensional theory is the bosonic sector of both $\N=6$ supergravity and of the quaternionic magic $\N=2$ supergravity, which admits $15$ abelian vector supermultiplets. The scalar fields lie in the special K\"{a}hler coset $SO^*(12)/ U(6)$ and the five-graded decomposition of $\mathfrak{e}_{7(-5)}$ is as follows 
\be \mathfrak{e}_{7(-5)} \cong  {\bf 1}^{\ord{-2}} \oplus S_+^{\ord{-1}} \oplus {\bf 1}^\ord{0} \oplus  \mathfrak{spin}^*(12)^\ord{0}  \oplus S_+^\ord{1} \oplus {\bf 1}^\ord{2} \ .\ee
The complex fundamental representation decomposes as 
\be \boldsymbol{56}  \cong V^\ord{-1} \oplus S_-^\ord{0} \oplus  V^\ord{1} \ ,\ee
where $S_+$ is the Majorana--Weyl representation of $Spin^*(12)$, whereas $V$ and $S_-$ are complex,  respectively vector and Weyl spinor, representations of $Spin^*(12)$.
\subsection*{m) $E_{7(-25)}/ (SO(2) \times E_{6(-14)} )$}
The corresponding four-dimensional theory is an $\N=1$ supergravity coupled to $16$ abelian vector supermultiplets and $10$ scalar supermultiplets. The scalar fields of the latter lie in the K\"{a}hler coset $SO(2,10)/ (SO(2)\times SO(10))$ and the five-graded decomposition of $\mathfrak{e}_{7(-25)}$ is as follows 
\be \mathfrak{e}_{7(-25)} \cong  {\bf 1}^{\ord{-2}} \oplus S_+^{\ord{-1}} \oplus {\bf 1}^\ord{0} \oplus  \mathfrak{spin}(2,10)^\ord{0}  \oplus S_+^\ord{1} \oplus {\bf 1}^\ord{2} \ .      \ee
The fundamental representation decomposes as 
\be \boldsymbol{56}  \cong V^\ord{-1} \oplus S_-^\ord{0} \oplus  V^\ord{1} \ ,\ee
where $S_\pm$ are the 32-dimensional Majorana--Weyl representations of $Spin(2,10)$ and $V$ is the vector representation of $SO(2,10)$. 
\subsection*{n) $E_{8(8)}/ SO^*(16)$}
The corresponding four-dimensional theory is the bosonic sector of $\N=8$ supergravity. The scalar fields of the latter lie in the coset $E_{7(7)}/ ( SU(8) / \mathds{Z}_2 ) $ and the five-graded decomposition of $\mathfrak{e}_{8(8)}$ is as follows 
\be \mathfrak{e}_{8(8)} \cong  {\bf 1}^{\ord{-2}} \oplus {\bf 56}^{\ord{-1}} \oplus {\bf 1}^\ord{0} \oplus  \mathfrak{e}_{7(7)}^\ord{0}  \oplus {\bf 56}^\ord{1} \oplus {\bf 1}^\ord{2} \ .\ee
The fundamental is the adjoint, and the $\bf 3875$ representation is also five-graded, 
\be {\bf 3875} \cong {\bf 133}^\ord{-2} \oplus {\bf 56}^\ord{-1} \oplus {\bf 912}^\ord{-1} \oplus {\bf 1}^\ord{0} \oplus {\bf 133}^\ord{0} \oplus {\bf 1539}^\ord{0} \oplus {\bf 56}^\ord{1} \oplus {\bf 912}^\ord{1} \oplus {\bf 133}^\ord{2} \ .\ee
\subsection*{o) $E_{8(-24)}/ (SU(1,1) \times E_{7(-25)})$}
The corresponding four-dimensional theory is the bosonic sector of the octonionic magic $\N=2$ supergravity, which admits $27$ abelian vector supermultiplets.  The scalar fields of the latter lie in the special K\"{a}hler coset  $E_{7(-25)}/ (U(1) \times E_{6(-78)}) $ and the five-graded decomposition of $\mathfrak{e}_{8(-24)}$ is as follows 
\be \mathfrak{e}_{8(-24)} \cong  {\bf 1}^{\ord{-2}} \oplus {\bf 56}^{\ord{-1}} \oplus {\bf 1}^\ord{0} \oplus  \mathfrak{e}_{7(-25)}^\ord{0}  \oplus {\bf 56}^\ord{1} \oplus {\bf 1}^\ord{2} \ .\ee
The fundamental is the adjoint, and the $\bf 3875$ representation is also five-graded, 
\be {\bf 3875} \cong {\bf 133}^\ord{-2} \oplus {\bf 56}^\ord{-1} \oplus {\bf 912}^\ord{-1} \oplus {\bf 1}^\ord{0} \oplus {\bf 133}^\ord{0} \oplus {\bf 1539}^\ord{0} \oplus {\bf 56}^\ord{1} \oplus {\bf 912}^\ord{1} \oplus {\bf 133}^\ord{2} \ .\ee


\section{$Spin^*(2\N)$ and its representations}  
\label{Spindix}
In this appendix we summarise some pertinent results concerning the group
$Spin^*(2\N)$ and its spinorial representations, comparing them to 
the corresponding representations of the compact group $Spin(2\N)$. We also refer to Ref.\ \cite{SO2N} for a detailed discussion of the algebra $\mathfrak{so}^*(2\N)$. These two
groups are different real forms of the same complex Lie group 
$Spin(2\N, \mathds{C})$, with the compact $U(\N)$ group as their intersection.
Because their complex representations are thus the same, it will be convenient 
to analyse these representations in the basis 
$\oplus_n \bigwedge^n \mathds{C}^\N$. For this purpose, we will make use 
of the fermionic creation and annihilation operators $a_i$ and 
$a^i\equiv (a_i)^\dagger$ already introduced in Section~1.2 
(with $i,j, \dots\in \{1,\dots\N\}$) 
\be
\{ a_i , a_j\} = \{a^i , a^j \} = 0 \quad , \qquad
\{a_i , a^j \} = \delta^j_i\ .
\ee
Since all the generators of both $Spin^*(2\N)$ and $Spin(2\N)$ commute with 
the diagonal matrix $(-1)^n$, the spinor representations decompose into 
chiral and anti-chiral Weyl spinor representations
$\oplus_p \bigwedge^{2p} \mathds{C}^\N$ and 
$\oplus_p \bigwedge^{2p+1} \mathds{C}^\N$, respectively. These representations
can thus be obtained by acting with an even or an odd number of creation
operators on the vacuum $|0\rangle$, that is, we have
\be
|\C\rangle = \Big(\w + Z_{ij} a^i a^j + \Sigma_{ijkl} a^i a^j a^k a^l
             + \dots \Big) |0\rangle
\ee
for the chiral and
\be
|\C\rangle = \Big( \psi_i a^i + \chi_{ijk} a^i a^j a^k  
              + \dots \Big) |0\rangle
\ee
for the antichiral representations, respectively.

The groups $Spin(2\N)$ and $Spin^*(2\N)$ are respectively the two 
real forms of $Spin(2\N,\mathds{C})$ defined by the conditions 
\be 
U^\dagger  = U^{-1} \;\;\; \mbox{[for $Spin(2\N)$]}
\qquad \textrm{and} \qquad 
U^\dagger = \beta U^{-1} \beta \;\;\; \mbox{[for $Spin^*(2\N)$]}
\ee
where the matrix $\beta$ is defined to act on both  
$\oplus_p \bigwedge^{2p} \mathds{C}^\N$ and 
$\oplus_p \bigwedge^{2p+1} \mathds{C}^\N$ as $(-1)^p$. The generators 
of the $\mathfrak{u}(\N)$ maximal subalgebra of both algebras are defined 
in terms of the anti-Hermitean parameters ${\Lambda_i}^j = -{\Lambda^j}_i$ as
\be 
K(\Lambda) = \frac12 \, {\Lambda_i}^j \, [ a^i , a_j ] \quad\Rightarrow\qquad 
K(\Lambda)^\dagger = - K(\Lambda) 
\ee
The remaining generators depend on the antisymmetric tensors $\Lambda_{ij}$ 
of $U(\N)$: for $Spin(2\N)$ we have the anti-Hermitean generators 
\be 
T(\Lambda) = \Lambda_{ij} a^i a^j + \Lambda^{ij} a_i a_j 
\quad\Rightarrow \qquad T(\Lambda)^\dagger = - T(\Lambda) \ ,
\ee
whereas for the non-compact real form $Spin^*(2\N)$ we have
\be  
T^*(\Lambda) = \Lambda_{ij} a^i a^j - \Lambda^{ij} a_i a_j 
\quad\Rightarrow\qquad 
T^*(\Lambda)^\dagger =  T^*(\Lambda) \ .
\ee
With the above definition of $\beta$ it follows that
\be
\beta  G(\Lambda) \beta =  - G(\Lambda)^\dagger
\ee
for both $G=K$ and $G=T^*$. From these formulas we see that the conjugate
of a spinor $|\lambda\rangle$ must be defined as
\be 
\left< \lambda \right | \equiv \scal{ \left| \lambda \right > }^\dagger 
\;\;\; \mbox{[for $Spin(2\N)$]} \qquad \textrm{and} \qquad
\bras \lambda \right | \equiv \scal{ \left| \lambda \right > }^\dagger \beta 
\;\;\; \mbox{[for $Spin^*(2\N)$]} \ .
\ee
Let us also record the expression for the $\mathfrak{u}(1)$ generator 
of $\mathfrak{u}(\N)$ in terms of oscillators, \viz
\be
J\equiv \frac12 \, [a^i , a_i] = a^i a_i - \frac12 \N \ ,                      
\ee
which permits one to re-express $\beta$ as
\be 
\left . \beta \right |_{ \oplus_p \bigwedge^{2p} \mathds{C}^\N }  
\equiv (-1)^{\frac{J}{2} + \frac{\N}{4}  } \hspace{10mm} 
\left . \beta \right |_{ \oplus_p \bigwedge^{2p-1} \mathds{C}^\N }  
\equiv (-1)^{\frac{J}{2} + \frac{\N}{4}  + \frac{1}{2} }
\ee
for the chiral and the antichiral Weyl spinors respectively. 

As for $Spin(2\N)$, the centre of $Spin^*(2\N)$ is generated by the 
group elements $e^{i\pi J }$ and $- \mathds{1}$. For odd $\N$, we have 
$\scal{ e^{i\pi J } }^2 = - \mathds{1}$ and the centre is $\mathds{Z}_4$. 
For even $\N$, $\scal{ e^{i\pi J } }^2 = \mathds{1}$ and the centre is 
$\mathds{Z}_2 \times \mathds{Z}_2$. In the latter case, the $\mathds{Z}_2$ 
subgroup generated by the group element $e^{i\pi \scal{Ê  J + \frac{\N}{2}}}$ 
acts trivially on the chiral Weyl spinor representation, whereas it acts 
as $- \mathds{1}$ on the anti-chiral Weyl spinor representation and the 
vector representation. The chiral Weyl spinor representation is thus a 
representation of the group $Spin^*(2\N) / \mathds{Z}_2$, and it is 
this latter which appears in the definition of the scalar-field 
coset space, \ie
\be 
U(1) \times_{\mathds{Z}_2} \frac{ Spin^*(8)}{\mathds{Z}_2} \cong 
SO(2) \times_{\mathds{Z}_2} SO(2,6) , \hspace{5mm} 
SU(1,1) \times_{\mathds{Z}_2} \frac{ Spin^*(12)}{\mathds{Z}_2} ,  
\hspace{5mm} \frac{ Spin^*(16)}{\mathds{Z}_2} \nn 
\ee
for $\N= 4,\, 6$ and $8$ respectively.

The (anti-)chiral representations given above are not always irreducible.
To analyse the values of $\N$ for which this happens, we first note that one can 
define certain anti-involutions or pseudo-anti-involutions for both 
$Spin(2\N)$ and $Spin^*(2\N)$ by making use of the $SU(\N)$--preserving 
Hodge star operator $\star$ which maps $\oplus_n \bigwedge^n \mathds{C}^\N$ 
to its conjugate. The Hodge star obeys
\be 
\star^2 = (-1)^{n(\N-n)} \ .
\ee
The definition of the respective (pseudo-)anti-involutions,
which we denote here by $\invo$ and $\invo^*$, respectively, involves 
extra sign factors, as we will explain below. Let us now analyse the
different cases in turn.

For $\N$ odd there is no difference between the spinor representations
of $Spin(2\N)$ and $Spin^*(2\N)$. In this case, 
the (pseudo)-anti-involution  does not commute with $(-1)^n$ 
and therefore the spinor and its conjugate are simply the two inequivalent 
irreducible complex spinor representations, for both $Spin(2\N)$ and 
$Spin^*(2\N)$. For $\N$ even, on the other hand, both $\invo$ and ${\invo}^*$ 
commute with $(-1)^n$ and the Weyl spinor representations become reducible 
if $\invo$ and $\invo^*$ are anti-involutions, that is, if they square to
one on these subspaces. 

For the reader's convenience, we first recall some familiar results 
for the compact real form $Spin(2\N)$ (cf. \cite{Georgi}). For 
$Spin(8M)$, the operation $\invo$ is defined on 
$\oplus_p \bigwedge^{2p} \mathds{C}^{4M}$ as 
\be 
\invo \oplus_{p=1}^{2M} \psi_\ord{2p} \equiv \oplus_{p=1}^{2M} 
\Scal{ (-1)^p \star \overline{\psi_\ord{4M-2p}} } \ .
\ee
Since $\star^2=1$ on even forms, $\invo^2=1$ in this case. On the 
anti-chiral spinor $\oplus_p \bigwedge^{2p-1} \mathds{C}^{4M}$, the formula is 
\be 
\invo \oplus_{p=1}^{2M} \psi_\ord{2p-1} \equiv \oplus_{p=1}^{2M} 
\Scal{ (-1)^p \star \overline{\psi_\ord{4M-2p+1}} } \ .
\ee
Now, $\star^2=-1$ on odd forms in even dimensions, but 
$(-1)^p (-1)^{2M - p +1} = -1$ so that $\invo^2=1$. 
Therefore, in both cases, one can impose the reality condition
$\invo |\lambda\rangle = |\lambda\rangle$, thereby reducing the 
Weyl spinors to Majorana-Weyl spinors.

For $Spin(8M+4)$, $\invo$ is defined to act on 
$\oplus_p \bigwedge^{2p} \mathds{C}^{4M+2}$ as 
\be 
\invo \oplus_{p=1}^{2M+1} \psi_\ord{2p} \equiv \oplus_{p=1}^{2M} 
\Scal{ (-1)^p \star \overline{\psi_\ord{4M+2 -2p}} } \ .
\ee
Although $\star^2=1$ on even forms, $(-1)^p (-1)^{2M +1 -p} = -1$ and  
$\invo^2=-1$ in this case. Similarly, on 
$\oplus_p \bigwedge^{2p-1} \mathds{C}^{4M+2}$, one has 
\be 
\invo \oplus_{p=1}^{2M+1} \psi_\ord{2p-1} \equiv \oplus_{p=1}^{2M} 
\Scal{ (-1)^p \star \overline{\psi_\ord{4M-2p-1}} } \ .
\ee
Because $\star^2=-1$ on odd forms in even dimensions and 
$(-1)^p (-1)^{2M - p} = 1$ one obtains again $\invo^2=-1$. Consequently,
the $Spin(8M+4)$ Weyl spinor representations are irreducible, though 
pseudo-real. Altogether we have thus rederived the well-known result
\be\label{invo1}
\invo\invo = (-1)^\frac{\N}{2} 
\ee 
on $\oplus_n \bigwedge^n \mathds{C}^\N$ for even $\N$. 

For the non-compact real form $Spin^*(4M)$ and its Weyl representations, 
the operation ${\invo}^*\equiv \beta \invo $ is defined on 
$\oplus_p \bigwedge^{2p} \mathds{C}^{2M}$ as 
\be 
\invo^* \oplus_{p=1}^{M} \psi_\ord{2p} \equiv \oplus_{p=1}^{M} 
\Scal{ \star \, \overline{\psi_\ord{2M-2p}} } \ .
\ee
Because $\star^2=1$ on even forms, one gets $\invo^* \invo^* =1$ in this case. 
Similarly, on $\oplus_p \bigwedge^{2p-1} \mathds{C}^{2M}$ one has
\be 
\invo^* \oplus_{p=1}^{M} \psi_\ord{2p-1} \equiv \oplus_{p=1}^{2M} 
\Scal{  \star\,  \overline{\psi_\ord{2M-2p+1}} } \ .
\ee
Now $\star^2=-1$ on odd forms in even dimensions whence $\invo^* \invo^* = -1$
in this case. We thus conclude that the chiral Weyl spinor representation 
of $Spin^*(4M)$ always decomposes into two equivalent Majorana--Weyl 
representations, whereas the anti-chiral Weyl spinor representations of 
$Spin^*(4M)$ are always pseudo-real, hence irreducible. We have thus 
shown that the analogue of (\ref{invo1}) reads, for $\N= 4M$,
\[ \label{invo2}  
\invo^* \invo^* = \left\{ \begin{array}{ll}
                       +1 & \quad \mbox{for chiral spinors} \\
                       -1 & \quad \mbox{for anti-chiral spinors}\ .
                           \end{array}   \right.
\]
These properties are summarised in the following two Tables. For 
$Spin(2\N)$ one has: 

\begin{gather}
\begin{array}{|l|c|c|c|c|c|}
\hline
 & \,\, \mbox{ vector}  \,\, & \, \mbox{ chiral\  spinor}  \,& \,\, \mbox{ antichiral\  spinor} \,\, & \,\,\mbox{ centre} \,\, \\*
\hline
\hspace{2mm}Spin(8M)  \hspace{2mm}&\hspace{2mm} \mbox{ \ real\ } \hspace{2mm} & \hspace{2mm} \mbox{ \ real \ } \hspace{2mm}  & \hspace{2mm} \mbox{ \ real \ }  \hspace{2mm}& \hspace{2mm}\mathds{Z}_2 \times \mathds{Z}_2\hspace{2mm} \\*
\hline
\hspace{2mm}Spin(8M+4)  \hspace{2mm}&\hspace{2mm} \mbox{ \ real\ } \hspace{2mm} & \hspace{2mm} \mbox{ \ pseudo-real \ } \hspace{2mm}  & \hspace{2mm} \mbox{ \ pseudo-real \ }  \hspace{2mm}& \hspace{2mm}\mathds{Z}_2 \times \mathds{Z}_2\hspace{2mm} \\*
\hline
\hspace{2mm}Spin(4M + 2)\hspace{2mm} & \hspace{2mm} \mbox{ \ real\ }  \hspace{2mm} Ê& \hspace{2mm}  \mbox{ complex \ }  \hspace{2mm}& \hspace{2mm}  \mbox{ \ complex \ } \hspace{2mm} Ê &\hspace{2mm} \mathds{Z}_4 \hspace{2mm}\\* \hline
\end{array} \nonumber
\end{gather}

\noindent
The Table for the $Spin^*(2\N)$ spinor representations is\footnote{For 
  $\N=4$ (\ie $M=2$), $Spin^*(8) \cong Spin(2,6)$ and, owing to triality, the complex vector representation 
  of $SO^*(8)$ is isomorphic to the antichiral Weyl spinor representation 
  of $Spin(2,6)$, which leads to the existence of a sixteen real dimensional $Spin(2,6)$ $SU(2)$--Majorana 
  representation.}
\begin{gather}
\begin{array}{|l|c|c|c|c|c|}
\hline
 & \,\, \mbox{ vector}  \,\, & \, \mbox{ chiral\  spinor}  \,& \,\, \mbox{ antichiral\  spinor} \,\, & \,\,\mbox{ centre} \,\, \\*
\hline
\hspace{2mm}Spin^*(4M)  \hspace{2mm}&\hspace{2mm} \mbox{ \ pseudo-real\ } \hspace{2mm} & \hspace{2mm} \mbox{ \ real \ } \hspace{2mm}  & \hspace{2mm} \mbox{ \ pseudo-real \ }  \hspace{2mm}& \hspace{2mm}\mathds{Z}_2 \times \mathds{Z}_2\hspace{2mm} \\*
\hline
\hspace{2mm}Spin^*(4M + 2)\hspace{2mm} & \hspace{2mm} \mbox{ \ pseudo-real\ }  \hspace{2mm} Ê& \hspace{2mm}  \mbox{ complex \ }  \hspace{2mm}& \hspace{2mm}  \mbox{ \ complex \ } \hspace{2mm} Ê &\hspace{2mm} \mathds{Z}_4 \hspace{2mm}\\* \hline
\end{array} \nonumber 
\end{gather}

\vspace{0.3cm}
When $\N = 4M$, the above results for $Spin^*(2\N)$ would seem to pose a 
problem for the boson-fermion balance required by supersymmetry, because 
unlike for $Spin(2\N)$ where both chiral and antichiral spinors share the same 
number of degrees of freedom, the antichiral representation requires twice
as many degrees of freedom as the chiral one. Fortunately, at this point
the presence of the spatial rotation group $SU(2)$ comes to our rescue: 
namely, the spinor fields transform not only under $Spin^*(2\N)$ but 
under $SU(2)\times Spin^*(2\N)$ for any $\N$. The existence of the
$SU(2)$ invariant tensor $\varepsilon_{\alpha\beta}$ allows us to impose
the representation halving condition 
\be
 ( \invo^* |\lambda \rangle )^\alpha = \varepsilon^{\alpha\beta} |\lambda \rangle_\beta
\ee
replacing the Majorana-Weyl condition (which would not work by itself) 
by a {\it symplectic Majorana-Weyl} condition. In this way the 
boson-fermion balance necessary for supersymmetry can be restored.

Let us explain a bit more explicitly how this works for $\N=6$ and $\N=8$.
For simplicity of notation, we will now write $\invo^*\equiv \invo$ and
give all formulas with two signs, the upper ones corresponding to the 
non-compact group $Spin^*(2\N)$, and the lower ones to the compact group
$Spin(2\N)$. For $\N=6$, the chiral spinor can be written as
\be 
|\C\rangle = \Big( \w + Z_{ij} a^i a^j + 
\frac1{4!} \varepsilon_{ijklmn} \Sigma^{ij} a^k a^l a^m a^n +
\frac1{6!} \varepsilon_{ijklmn}Z  a^i a^j a^k a^l a^m a^n \Big) |0\rangle
\ee
on which the coset generators act as 
\be 
\delta |\C\rangle = \Big( \Lambda_{ij} a^i a^j \mp \Lambda^{ij} a_i a_j
       |\C\rangle
\ee
The (pseudo-)anti-involution is defined as follows
\be
\invo |\C\rangle := \Big( \bar Z \pm \Sigma_{ij} a^i a^j + 
\frac1{4!} \varepsilon_{ijklmn} Z^{ij} a^k a^l a^m a^n  \pm
\frac1{6!} \varepsilon_{ijklmn}\bar\w  a^i a^j a^k a^l a^m a^n \Big) |0\rangle
\ee
and is preserved by the transformations 
\be
\begin{split}  \delta \w &= 2 \Lambda^{ij} Z_{ij} \\
\delta Z &= 2 \Lambda_{ij} \Sigma^{ij}Ê
\end{split}\hspace{10mm}\begin{split} 
\delta Z_{ij} &= \Lambda_{ij} \w + \frac12 \varepsilon_{ijklmn} \Lambda^{kl} 
\Sigma^{mn}Ê\\
\delta \Sigma^{ij} &= \Lambda^{ij} Z + \frac12 \varepsilon^{ijklmn} 
\Lambda_{kl} Z_{mn}Ê\ .
\end{split}
\ee
For an antichiral spinor we have
\be 
\left| \chi \right> \equiv \Scal{Ê\psi_i a^i + \chi_{ijk} a^i a^j a^k 
+  \frac{1}{5!} \varepsilon_{ijklmn}Ê\chi^n a^i a^j a^k a^l a^m }Ê
\left| 0 \right > 
\ee
and the (pseudo-)anti-involution reads 
\be 
{\invo}   \left| \chi \right> \equiv \Scal{ \mp \chi_i  a^i + 
\frac{1}{6!} \varepsilon_{ijklmn}\, Ê\chi^{lmn} \, a^i a^j a^k \pm  
\frac{1}{5!} \varepsilon_{ijklmn} \, \psi^n \, a^i a^j a^k a^l a^m }Ê
\left| 0 \right > \ .
\ee

Finally, for maximal supergravity, the relevant group is $Spin^*(16)$, and a 
chiral Weyl spinor can be represented by the state
\begin{multline} 
\left|\C\right> \equiv \Bigl( \w + Z_{ij} a^i a^j + 
\Sigma_{ijkl}Êa^i a^j a^k a^l+Z_{ijklmn}   ÊÊa^i a^j a^k a^l a^m a^n   
\Bigr .  \\ \Bigl .  
+\w_{ijklmnpq}    a^i a^j a^k a^l a^m a^n a^p a^q \Bigr) \left| 0 \right>  \ .
\end{multline}
The anti-involution is then (where the lower sign is for $Spin(16)$)
\begin{multline} {\invo} \left|\C\right> \equiv 
\biggl(  \varepsilon_{ijklmnpq}Ê\, \w^{ijklmnpq}Ê 
\pm \frac{1}{2}\varepsilon_{ijklmnpq}\, ÊZ^{klmnpq } \, a^i a^j\\ 
+ \frac{1}{4!} \varepsilon_{ijklmnpq}Ê\, \Sigma^{mnpq}Ê\, 
a^i a^j a^k a^l \pm Ê \frac{1}{6!} 
\varepsilon_{ijklmnpq}Ê\, Z^{pq} \,   ÊÊa^k a^l a^m a^n a^p a^q   
\biggr .  \\ \biggl .  
+\frac{1}{8!} \varepsilon_{ijklmnpq} \, \bar \w \,   
a^i a^j a^k a^l a^m a^n a^p a^q \biggr) \left| 0 \right>  \ .
\end{multline}
Similarly, for an antichiral spinor one has
\be  
\left| \chi \right> \equiv \Scal{Ê\psi_i a^i + 
\chi_{ijk} a^i a^j a^k +  \chi_{ijklm} a^i a^j a^k a^l  a^m + 
\psi _{ijklmnp}Êa^i a^j a^k a^l  a^m  a^n a^p Ê}Ê\left| 0 \right > \ .
\ee
The anti-involution $\invo$ of $Spin(16)$ corresponds to the 
pseudo-anti-involution of $Spin^*(16)$,
\begin{multline}
Ê {\invo} \left| \chi \right> \equiv 
\biggl( \pm \varepsilon_{ijklmnpq}\, Ê\psi^{jklmnpq}Ê\,  a^i + 
\frac{1}{3!} \varepsilon_{ijklmnpq}\, Ê\chi^{lmnpq} \, a^i a^j a^k 
\biggr . \\ \biggl . 
\pm \frac{1}{5!} \varepsilon_{ijklmnpq}Ê\,  \chi^{npq} 
\, a^i a^j a^k a^l  a^m + 
\frac{1}{7!} \varepsilon_{ijklmnpq}Ê\, \psi^q\, 
Êa^i a^j a^k a^l  a^m  a^n a^pÊ\biggr) \left| 0 \right >  \ .
\end{multline}

\subsection*{Acknowledgements}
We would like thank Ling Bao, P. Breitenlohner, D.~\v{Z}.~\DJo, A.~Kleinschmidt, I.~Melnikov, B.~Pioline,  D.~Vogan and B. de Wit for discussions related to this work. The research of K.S.S.\ was supported in part by the EU under contract MRTN-CT-2004-005104, by the STFC under rolling grant PP/D0744X/1 and by the Alexander von Humboldt Foundation through the award of a Research Prize. K.S.S. would like to thank the Albert Einstein Institute and CERN for hospitality during the course of the work.


\end{document}